\documentclass[nofootinbib
]{revtex4-1}
\usepackage{amssymb}
\usepackage{graphicx}
\usepackage{amsmath,rotating}
\usepackage{color,bm}

\def\ee{\end{eqnarray}}

\def\p{\partial}

\def\underline{\underline}

\def\=:{=\hspace{-.7em}\raisebox{1.1ex}{.}\hspace{.1em}\raisebox{-0.2ex}{.} }

\newcommand {\beq}{\begin{eqnarray}}
\newcommand {\eeq}{\end{eqnarray}}

\begin{document}

\begin{flushright}YGHP-19-03\end{flushright}

\title{
Collision dynamics and reactions of fractional vortex molecules \\
in coherently coupled Bose-Einstein condensates 
}

\author{Minoru Eto$^{1,2}$, Kazuki Ikeno$^3$, Muneto Nitta$^{2,4}$}
\email{meto(at)sci.kj.yamagata-u.ac.jp,\ nitta(at)phys-h.keio.ac.jp}

\affiliation{%
$^1$Department of Physics, Yamagata University, 
Kojirakawa-machi 1-4-12, Yamagata,
Yamagata 990-8560, Japan \\
$^2$Research and Education Center for Natural
Sciences, Keio University, Hiyoshi 4-1-1, Yokohama, Kanagawa 223-8521, Japan\\
$^3$ Adavito Inc., Hongo-cho 4-26-12, Bunkyo-ku, Tokyo, 113-033, Japan\\
$^4$Department of Physics, Keio University, Hiyoshi 4-1-1, Yokohama, Kanagawa 223-8521, Japan
}%

\date{\today}

\begin{abstract}
\ \ \\\ \\

Coherently coupled two-component Bose-Einstein condensates (BEC)
exhibit vortex confinement resembling quark confinement in 
Quantum Chromo Dynamics (QCD).
Fractionally quantized vortices 
winding only in one of two components 
are attached by solitons, and they cannot stably exist  alone. 
Possible stable states are ``hadrons''  
either of mesonic type, {\it i.e.}, molecules made of a vortex and anti-vortex 
in the same component connected by a soliton,
or of baryonic type, {\it i.e.}, molecules made of two vortices winding 
in two different components connected by a soliton.
Mesonic molecules move straight with a constant velocity 
while baryonic molecules rotate.
We numerically simulate collision dynamics of mesonic and baryonic molecules 
and find that the molecules swap a partner in collisions in general like chemical 
and nuclear reactions, 
summarize all collisions as vortex reactions, 
and describe those by Feynman diagrams.
We find a selection rule for final states after collisions of 
vortex molecules, analogous to that for collisions of 
hadrons in QCD.

\end{abstract}

\maketitle

\newpage

\section{Introduction}

Ultracold atomic gases are experimentally controllable systems 
offering setups to simulate various problems in physics 
 \cite{Dalfovo:1999zz,Pitaevskii:2003,Pethick:2008}. 
For instance, 
Bose-Einstein condensates (BECs) of 
Bose gases are superfluids  
admitting quantized vortices, {\it i.e.}, vortices carrying quantized circulations \cite{Fetter} known as global strings in cosmology, 
and nucleation of vortices and detecting 
real time dynamics of them were achieved experimentally recently
\cite{vortex-exp}. 
Among various systems, 
coherently coupled two-component BECs 
realized the JILA group 
 \cite{Hall:1998,coherent}
are one of interesting systems even for high energy physics;
When two components are different hyperfine components of the same atom,
one can introduce a Rabi (Josephson) coupling between them.
Then, they allow vortex molecules, 
fractionally quantized vortices confined by solitons 
(linearly extended objects)
\cite{Son:2001td},
which are suggested to  
share several properties with 
confinement phenomena of Quantum Chromo Dynamics (QCD),
a theory of the strong interaction consisting of quarks and gluons. 
Such fractional vortex molecules have been extensively 
studied theoretically 
\cite{Garcia:2002,Kasamatsu:2004tvg,Kasamatsu:2005,Cipriani:2013nya,
Nitta:2013eaa,Tylutki:2016mgy,Calderaro:2017,Eto:2017rfr,
Kobayashi:2018ezm}  
including 
sine-Gordon like solitons
\cite{Usui:2015,Qu:2016, Qu:2016b, Ihara:2019ihz}. 
Moreover such studies have been extended to three or more components
\cite{Kuopanportti:2011,Eto:2012rc,Eto:2013spa,
Cipriani:2013wia,Dantas:2015fka,Orlova:2016}
as well as spinor BECs \cite{Shinn:2018zde}.
As similarities with QCD, dynamics of a single vortex molecule 
composed of two fractional vortices were studied 
\cite{Tylutki:2016mgy,Eto:2017rfr}, for which 
a vortex in the first component and that in the second component
are confined by a sine-Gordon soliton.
If the two vortices are at the equilibrium distance, the molecule is static. 
On the other hand, if they are more separated or pushed to a short distance,  the molecule rotates clockwise or counterclockwise, respectively.
If they are further pulled to be more separated, the soliton connecting them is broken by creating another pair of fractional vortices, and therefore fractional vortices never been liberated (unless the Rabi coupling is turned off). This situation 
resembles the confinement of quarks in QCD;
QCD vacuum  is considered to be a dual superconductor 
 \cite{Nambu:1974zg,tHooft,Mandelstam:1974pi}, 
where chromo electric fluxes are confined to color electric flux tubes due to magnetic monopole condensations.
Then, quarks are confined by color electric flux tubes with  
 energy linearly dependent on the distance between them.

In Ref.\cite{Eto:2017rfr}, 
we have referred fractionally quantized vortices
winding in the first and second components as 
a up-type vortex (or u-vortex for short) 
and a down-type vortex (or d-vortex), respectively, 
borrowing the terminology of quarks in QCD.
Similarly, we have called
an antivortex of the first (second) component $\bar{\rm u}$- ($\bar{\rm d}$-) vortex.
In the presence of the Rabi coupling,
the elementally topological objects are composite defects of u-, d-vortices and
the solitons connecting them. 
There are two possibilities for the soliton to select
the vortices in a pair on its two endpoints: either a vortex and an antivortex in the same component such as u and $\bar{\rm u}$
(d and $\bar{\rm d}$), or vortices in different species such as u and d ($\bar{\rm u}$ and $\bar{\rm d}$).
We have called the former a mesonic vortex molecule and the latter a baryonic vortex molecule in analogy with QCD;
Let us regard quantized circulations  $n_S$  as the baryon number in QCD, 
and the winding  $n_R$ of the relative phase between the two components 
as the color charge. 
Then, u and d vortices carrying fractional baryon number like quarks 
carry a color charge and therefore cannot exist alone.
On the other hand, 
mesonic molecules $\bar{\rm u}$u and $\bar{{\rm d}}$d 
not carrying the baryon number like mesons 
and baryonic molecules ud carrying one baryon number 
both do not carry a color charge (color singlets) and therefore 
can exist stably. See Table \label{tab:winding} as 
a summary of possible states.

\begin{table}[h]
\begin{center}
\begin{tabular}{c||c|c|c|c|c}
 & $n_1$ & $n_2$  & $n_S$ {\tiny (baryon \#)} & $n_R$ {\tiny (color charge)} & $2n_R$ \\
\hline\hline
u & 1 & 0 & $1/2$ & $1/2$ & 1\\
d & 0 & 1 & $1/2$ & $- 1/2$ & $-1$\\
$\bar{\rm u}$ & $-1$ & 0 & $-1/2$ & $-1/2$ & $-1$\\
$\bar{\rm d}$ & 0 & $-1$ & $-1/2$ & $1/2$ & $1$ \\
\hline
$\bar{\rm u}$u & 0 & 0 & $0$ & 0  & $0$\\
$\bar{\rm d}$d & 0 & 0 & $0$ & 0  & $0$\\
$\bar{{\rm u}}$d & $-1$ & 1 & 0 & $-1$ & $-2$\\
$\bar{{\rm d}}$u & $1$ & $-1$ & 0 & 1 & $2$\\
\hline
ud & 1 & 1 & $1$ & 0  & $0$\\
$\bar{{\rm u}}\bar{{\rm d}}$ & $-1$ & $-1$ & $-1$ & 0 & $0$\\
\end{tabular}
\caption{Topological winding numbers $n_1$, $n_2$, $n_S$ and $n_R$ are shown.}
\label{tab:winding}
\end{center}
\end{table}

In this paper, in order to pursue further similarities between BECs and QCD, 
we focus on a few body dynamics of vortex molecules, more precisely their collisions,
in contrast to previous works focussing on 
dynamics of either single molecules 
\cite{Tylutki:2016mgy,Calderaro:2017,Eto:2017rfr}
or many molecules describing vortex lattices 
\cite{Cipriani:2013nya}
or 
Berezinskii-Kosterlitz-Thouless transition \cite{Kobayashi:2018ezm}.  
Our numerical studies are twofolds: the meson-meson scattering and the meson-baryon scattering.
First, we investigate the meson-meson scattering of the same spices ($\bar{\rm u}$u-$\bar{\rm u}$u). Firstly, we simulate 
the head-on collisions (zero impact parameter) by varying the incident angles.
We show that the constituent vortices swap the partners in collisions. 
The recombination can be understood as a collision of the SG and anti-SG solitons, and 
the swapping is nothing but the pair annihilation and creation 
of the confining SG solitons.
The simulation with the initial relative angle $\pi$ happens to show
the right angle scattering of the two mesons, which is very common among relativistic topological solitons.
We then develop a useful description by
using the Feynman diagrams.
We also study the meson-meson scatterings with non-zero impact parameters.

We then study the meson-meson scattering of the different spices 
($\bar{\text{u}}\text{u}$-$\bar{\text{d}}\text{d}$).  
We find that in  head-on collision with the 0 relative angle,  
the scattering of the two SG solitons occurs.
For the head-on collisions with the smaller relative angles,
the incoming u and d mesons are converted to the intermediate baryon and anti-baryon pair during the collision. 
The intermediate baryons rotate, and then they are reformed back into the mesons
at the second recombination. Forming the intermediate baryonic state results in 
the shift of the outgoing line from the ingoing one.
We also study the scatterings with non-zero impact parameters  in this case too.

Next we study the meson-baryon scatterings ($\bar{\rm u}$u and ud) and 
find that the meson and baryon swap their constituent u vortices,
and the new meson goes out while the new baryon stays at slightly shifted point from the original baryon point.
For scattering of 
a longer meson to the baryon,  
the recombination takes place also in this case,
so that a longer and kink bend baryon is formed at first stage,
and subsequently 
a long and bent molecule is unstable
and soon disintegrates into a set of shorter meson and baryon. 
As a result, the final state includes more molecules than the initial configuration, resembling what happens in real Hadron collider experiments.
We exaggeratedly call it a vortical hadron jet
in the VHC (vortical hadron collider) experiment. 

We then further discuss a connection between BEC and QCD 
comparing the Polyakov's dual photon model in $2+1$ dimensions to the low energy effective theory based on two-component BECs. 
We also point out that 
the so-called  Okubo-Zweig-Iizuka (OZI) rule \cite{Okubo:1963fa,Zweig:1981pd,Iizuka:1966fk}  rule in QCD, 
which is a phenomenological rule 
concerning with the final state of the collision 
found in 1960s \cite{Fritzsch:2018xbs},
seems to hold in vortex molecule collisions in BECs.

This paper is organized as follows.
In Sec.~\ref{sec:hadron}, we introduce our model and describe mesonic and baryonic vortex molecules.
In Secs.~\ref{sec:meson-meson} and \ref{sec:meson_meson_2}, we study the meson-meson scattering of the same spices ($\bar{\rm u}$u-$\bar{\rm u}$u) 
and of the different spices  ($\bar{\rm u}$u-$\bar{\rm d}$d), respectively. 
In Sec.~\ref{sec:meson_baryon}, we study the meson-baryon scatterings ($\bar{\rm u}$u and ud).
In Sec.~\ref{sec:connection_to_QCD}, we give comments on a connection between BEC and QCD.
Sec.~\ref{sec:summary} is devoted to a summary and discussion.


\section{Hadronic vortex molecules}
\label{sec:hadron}

Theoretically, 
dynamics of the condensates $\Psi_i$ can be described by
the coupled Gross--Pitaevskii (GP)
equations
\beq
\!\!\!\!\!\!\!\!
\left[  i\hbar \frac{\p}{\p t}
\!+\! \frac{\hbar^2}{2m}\nabla^2 \!- \! \left(g_i |\Psi_i|^2 \!+\! g_{12}|\Psi_{\hat i}|^2 \!-\!\mu_i \right)\!\right]\!\! \Psi_i
\!=\! -  \hbar \omega \Psi_{\hat i}\,,\qquad (i=1,2)\,,
\label{eq:gp1}
\eeq
where $\hat 1 = 2$, $\hat 2 = 1$, 
$g_{ij}$ represents the atom--atom coupling constants, $m$ is the mass of atom, and $\mu_i$ represents the chemical potential. 
The first and second condensates $\Psi_{1,2}$ are coherently coupled
through the Rabi (Josephson) terms with the Rabi frequency $\omega$. 
Experimentally, such a coherent coupling was achieved by the JILA group \cite{coherent}.
In the following, we assume
\beq
g_1 = g_2 \equiv g,\quad \mu_1 = \mu_2 \equiv \mu\,,
\label{eq:flavor_sym}
\eeq 
for simplicity, and focus on
a miscible BEC ($g > g_{12}$)
in which both condensates coexist with 
\beq
|\Psi_1| = |\Psi_2| = \sqrt{\frac{\mu + \hbar \omega}{g + g_{12}}} \equiv v\,.
\eeq

If the Rabi interaction is turned off $\omega = 0$,
the system has two $U(1)$ symmetries $U(1)_1 \times U(2)_2$ defined by
\beq
U(1)_1:\ (\Psi_1,\ \Psi_2) \to (e^{i\alpha_1} \Psi_1,\Psi_2)\,,\quad
U(1)_2:\ (\Psi_1,\ \Psi_2) \to (\Psi_1,e^{i\alpha_2}\Psi_2)\,.
\eeq
These symmetries can also be expressed as
\beq
[U(1)_S \times U(1)_R]/\mathbb{Z}_2:
(\Psi_1,\Psi_2) \to (e^{i\alpha}\Psi_1,e^{\pm i\alpha}\Psi_2)\,,
\eeq
where $+$ is for $U(1)_S$ and $-$ is for $U(1)_R$, and 
$\mathbb{Z}_2$ represents simultaneous rotations of 
$U(1)_S$ and $U(1)_R$ with angles $\alpha=\pi$ for both of them, 
which has been introduced to remove a redundancy.
When $\omega \neq 0$, the relative $U(1)_R$ is manifestly broken.
Namely, the number of each atom is not preserved but the total number of the first and second atoms is preserved.

In the following, we will numerically solve the GP equation (\ref{eq:gp1}). For that purpose
it is convenient first to rewrite it in terms of dimensionless variables
\beq
\tilde t = \frac{\mu}{\hbar}t,\quad
\tilde x_i = \frac{\hbar}{\sqrt{m\mu}}x_i,\quad
\tilde \omega = \frac{\hbar}{\mu}\omega,\quad
\tilde g_{12} = \frac{g_{12}}{g},\quad
\tilde \Psi_i = \sqrt{\frac{g}{\mu}}\Psi_i.
\eeq
Then, the GP equation can be rewritten as
\beq
\!\!\!\!\!\!\!\!
\left[  i \frac{\p}{\p \tilde t}
\!+\! \frac{1}{2}\tilde \nabla^2 \!- \! \left(|
\tilde \Psi_i|^2 \!+\! \tilde g_{12}|\tilde \Psi_{\hat i}|^2 \!-\!1 \right)\!\right]\!\! 
\tilde \Psi_i
\!=\! -   \tilde \omega \tilde \Psi_{\hat i}\,,\qquad (i=1,2)\,.
\label{eq:gp2}
\eeq
Thus, the essential parameters are only $\tilde \omega$ and $\tilde g_{12}$.
In what follows, we will assume $\tilde g_{12} \neq 1$.

\begin{figure*}[t]
\begin{center}
\includegraphics[width=0.4\hsize]{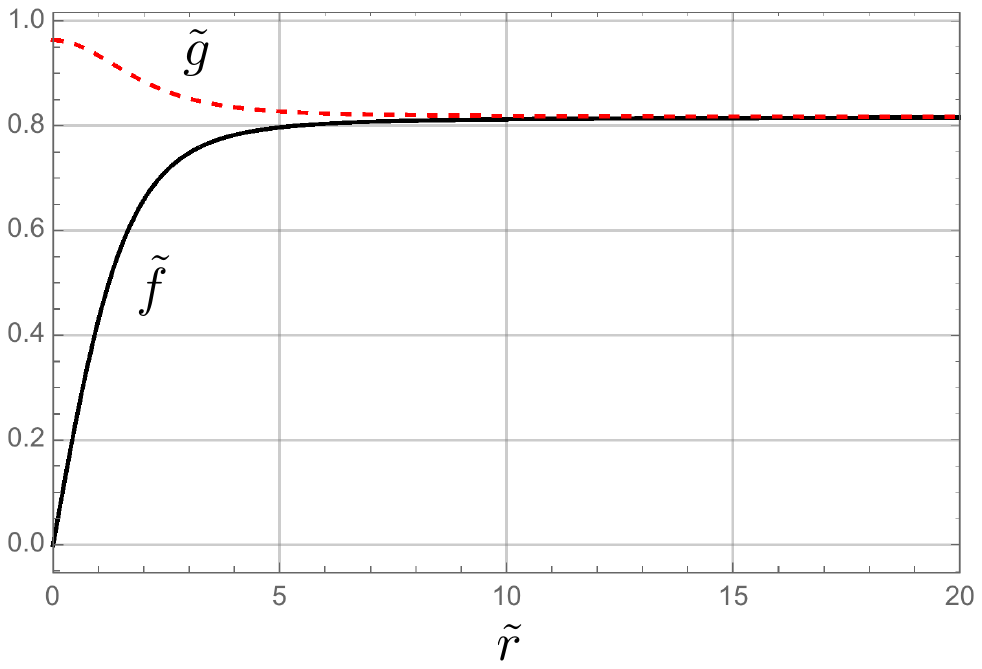}
\caption{The profile functions $\tilde f$ and $\tilde g$ of a u- or d-vortex for $\tilde g_{12} = 0.5$. The Rabi term is suppressed
($\tilde\omega = 0$).}
\label{fig:profile}
\end{center}
\end{figure*}

When $\tilde \omega = 0$, both the $U(1)_1 \times U(1)_2$ are spontaneously broken in the ground state. As a consequence,
there are two kinds of topologically stable vortices supported by topological winding number
$\pi_1(U(1)_1 \times U(1)_2) = \mathbb{Z}\times\mathbb{Z}$.
A vortex associated with the first $U(1)_1$ at the origin, which we will call the u-vortex, is given by
\beq
\text{u-vortex}:\quad 
\tilde \Psi_1 = \tilde f(\tilde r) e^{i\theta},\quad
\tilde \Psi_2 = \tilde g(\tilde r),
\label{eq:u}
\eeq
with $\tilde r = \sqrt{\tilde x_1^2 + \tilde x_2^2}$.
Similarly, a vortex associated with the second $U(1)_2$ at the origin, which we will call the d-vortex, is given by
\beq
\text{d-vortex}:\quad 
\tilde \Psi_1 = \tilde g(\tilde r),\quad
\tilde \Psi_2 = \tilde f(\tilde r) e^{i\theta}.
\label{eq:d}
\eeq
A u-vortex becomes an anti u-vortex (we will refer it to a $\bar{\rm u}$-vortex) by exchanging $\theta \to -\theta$.
Similarly, a d-vortex and a $\bar{\rm d}$-vortex are replaced by $\theta \to -\theta$.
The profile functions $\tilde f$ and $\tilde g$ satisfy the following second order ordinary differential equations
\beq
\left[
\frac{1}{2}\left(\frac{\p^2}{\p \tilde r^2} + \frac{1}{\tilde r}\frac{\p}{\p \tilde r}
- \frac{1}{\tilde r^2}\right) - \left(\tilde f^2 + \tilde g_{12} \tilde g^2 -1\right)
\right] \tilde f  = 0,\\
\left[
\frac{1}{2}\left(\frac{\p^2}{\p \tilde r^2} + \frac{1}{\tilde r}\frac{\p}{\p \tilde r}
\right) - \left(\tilde g^2 + \tilde g_{12} \tilde f^2 -1\right)
\right] \tilde g = 0.
\eeq
The appropriate boundary conditions are
\beq
\lim_{\tilde r \to 0} \tilde f = 0,\quad
\lim_{\tilde r \to \infty} \tilde f = \tilde v,\quad
\lim_{\tilde r \to 0} \frac{\p\tilde g}{\p\tilde r} = 0,\quad
\lim_{\tilde r \to \infty} \tilde g = \tilde v_0,
\eeq
where we have defined
\beq
\tilde v_0 = \frac{1}{\sqrt{1+\tilde g_{12}}}.
\eeq
Fig.~\ref{fig:profile} shows a typical numerical solution of 
$\tilde f$ and $\tilde g$ for $\tilde g_{12} = 0.5$ as an example.

\begin{figure*}[h]
\begin{center}
\includegraphics[width=0.85\hsize]{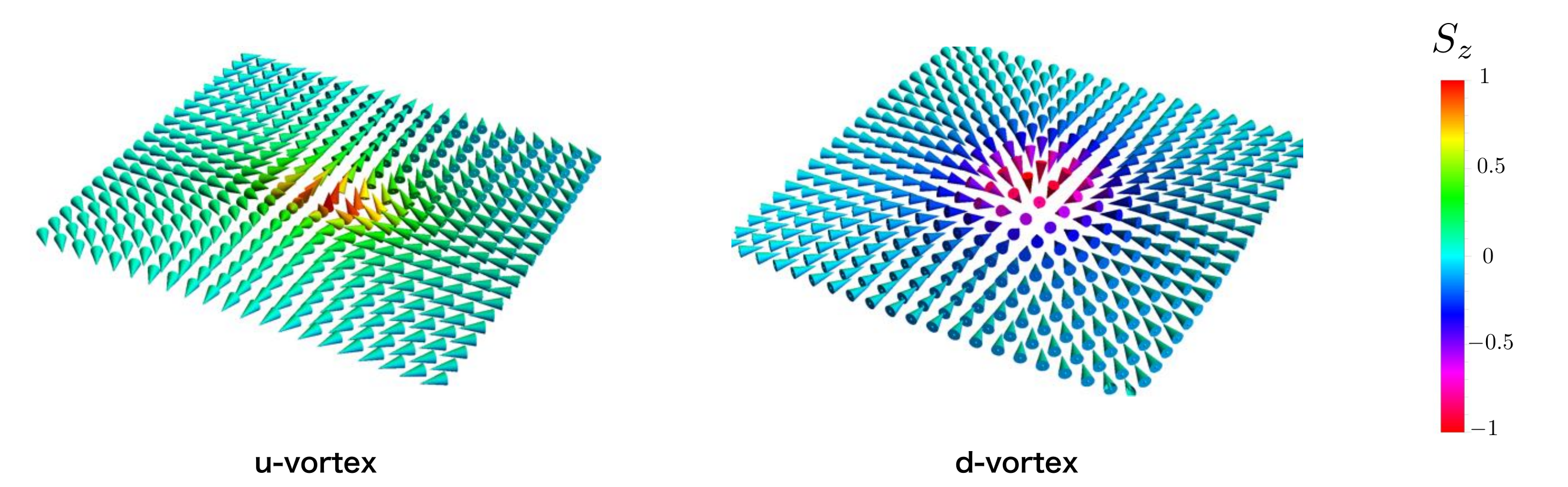}
\caption{Vector plots of the pseudo spin $\bm{S}(x^1,x^2)$
for the u- and d-vortices (constructed from the solution given in Fig.~\ref{fig:profile})
are plotted on the $x^1$-$x^2$ plane ($\tilde x^{1,2}\in[-5,5]$). The colors of arrows represent $S_z$.
The Rabi frequency is set to be $\tilde \omega =0$.}
\label{fig:pseudo_spin}
\end{center}
\end{figure*}

It is useful to introduce the pseudo-spin to distinguish 
u- and d-vortices;
\beq
\bm{S} = -\frac{{\vec \Psi}^\dagger \bm{\sigma} \vec\Psi }{{\vec\Psi}^\dagger\vec\Psi},\quad
\vec \Psi = \left(\Psi_1,\ \Psi_2\right)
\eeq
with the Pauli matrices  $\bm{\sigma}$.
$\bm{S}$ is a real three-vector satisfying $|\bm{S}| = 1$, and so it can be thought of as coordinates of an internal $S^2$ target space.
The pseudo spin $\bm{S}$ is transformed as a triplet $\bm{3}$ under the $SU(2)$ transformation $\vec \Psi \to U \Psi$ with
$U \in SU(2)$. Note that the $SU(2)$ is only manifest symmetry of a part of the GP equations, namely
the gradient terms of Eq.~(\ref{eq:gp1}). The rest terms of Eq.~(\ref{eq:gp1}) generally do not respect the $SU(2)$ symmetry
but the subgroup $U(1)_R \subset SU(2)$.
Only when $g = g_{12}$ ($\tilde g_{12} = 1$) together with the condition (\ref{eq:flavor_sym}) holds,
the system symmetry $U(1)_1 \times U(1)_2$ is enhanced to $[U(1)_S \times SU(2)]/\mathbb{Z}_2$.
Otherwise the $SU(2)$ is only approximate symmetry of generic GP equation  (\ref{eq:gp1}).
The u- and d-vortices have 
$\left(\tilde \Psi_1(0,0),\tilde \Psi_2(0,0)\right) = (0, \tilde g(0))$ 
 and $(\tilde g(0),0)$
at their cores,
so that their the pseudo spins are $\bm{S} = (0,0,1)$ (up) and $\bm{S} = (0,0,-1)$ (down), respectively. 
Fig.~\ref{fig:pseudo_spin} shows the pseudo
spins of the u- and d-vortices.

When $\tilde \omega = 0$, the vortex number can be measured by
\beq
n_i = \frac{1}{2\pi} \oint_{\rm C} d\theta_i = \frac{1}{2\pi} \int^{2\pi}_0 \frac{d\theta_i}{d\theta} d\theta,
\eeq
where $\theta_i$ is the phase of $\Psi_i$ ($\theta_i = \arg\Psi_i$), and  C is a closed curve in the $x^1$-$x^2$ plane.

When $\tilde \omega \neq 0$, neither u- nor d-vortex exists alone since $U(1)_R$ is manifestly broken.
The winding numbers $n_1$ and $n_2$ are no longer good topological numbers.
The Rabi term effectively works as a sine-Gordon (SG) type potential
\beq
V_{\rm Rabi} = - 2\hbar \omega v^2 \cos (\theta_1-\theta_2),
\label{eq:rabi_V}
\eeq
where we have set $\psi_i = v e^{i\theta_i}$.
Due to this, the u-vortex given in Eq.~(\ref{eq:u}) with $(\theta_1,\theta_2) = (\theta,0)$ is inevitably attached
by a SG soliton at $\theta = \pi$. By the same reason, the d-vortex given in Eq.~(\ref{eq:d}), $\bar{\rm u}$-vortex and
$\bar{\rm d}$-vortex are also
attached by a SG or an anti SG soliton.
Under the nonzero $\tilde \omega$, it turns out that the followings are more useful than $n_1$ and $n_2$
\beq
n_S &=& \frac{1}{2\pi}\oint_{\rm C} d\theta_S = \frac{1}{2\pi}\int^{2\pi}_0 
\frac{1}{2}\left(\frac{d\theta_1}{d\theta} + \frac{d\theta_2}{d\theta}\right)d\theta
= \frac{n_1+n_2}{2}\,,\\
n_R &=& \frac{1}{2\pi}\oint_{\rm C} d\theta_R = \frac{1}{2\pi}\int^{2\pi}_0 
\frac{1}{2}\left(\frac{d\theta_1}{d\theta} - \frac{d\theta_2}{d\theta}\right)d\theta
= \frac{n_1 - n_2}{2}\,,
\eeq
with $\theta_S = \left(\theta_1 + \theta_2\right)/2$ and $\theta_R = \left(\theta_1 - \theta_2\right)/2$.
Here, $n_S$ and $n_R$ are topological invariants taking values in half integers when $\tilde \omega = 0$. 
Once we turn on $\tilde \omega \neq 0$, $n_R$ is no longer a topological number. 
Though $n_R$ is not preserved, it has another physical meaning:
$2n_R$ corresponds to the total SG soliton number across 
the curve C. For a single u-vortex, we have $2n_R = 1$ for any C which 
encloses it. 
We summarize $n_1$, $n_2$, $n_S$ and $n_R$ in Table \ref{tab:winding}. From this, we see that
u- and $\bar{\rm d}$-vortices are attached by a SG soliton while $\bar{\rm u}$- and d-vortex 
are attached by an anti SG soliton.

\begin{figure*}[h]
\begin{center}
\includegraphics[width=0.75\hsize]{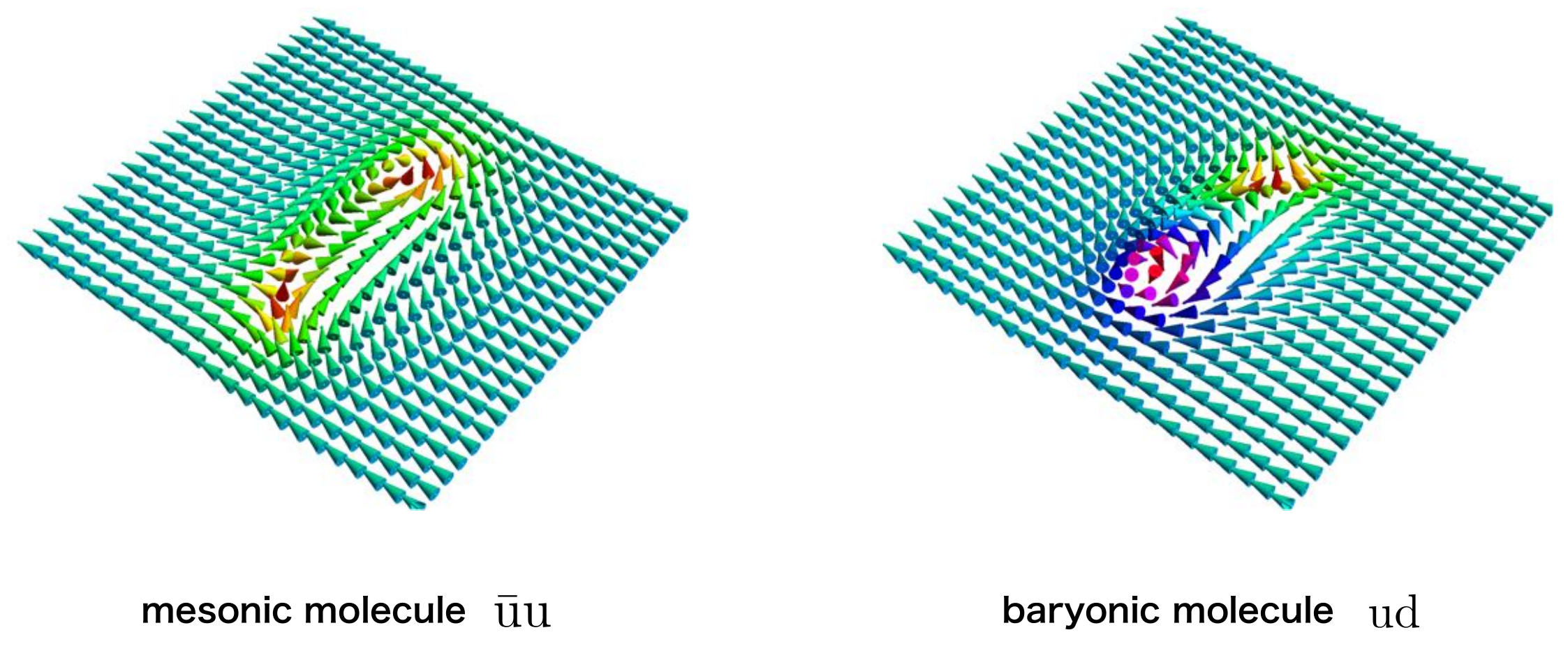}
\caption{Vector plots of the pseudo spin $\bm{S}(x^1,x^2)$
for the $\bar{\rm u}$u and ud are plotted on the $x^1$-$x^2$ plane. 
The plot region of the left panel is $\tilde x^1 \in [-10,10]$ and $\tilde x^2 \in [40,60]$,
and that for the right panel is $\tilde x^1 \in [-10,10]$ and $\tilde x^2 \in [-10,10]$. The color map is
the same as the one in Fig.~\ref{fig:pseudo_spin}.
The Rabi frequency is set to be $\tilde \omega =0.05$.}
\label{fig:pesudo_meson_baryon}
\end{center}
\end{figure*}

However, this is only a static picture. Since a semi-infinitely long soliton costs infinite energy,
it is dynamically unstable and disintegrate into shorter solitons. A finite soliton is terminated by two vortices,
namely it forms a vortex molecule with the SG soliton bounding two vortices.
We can figure out all possible types of molecules by seeing $n_R$. 
Since a vortex molecule is finite configuration, the corresponding $n_R$ for
a sufficiently large C enclosing it must be zero.
Therefore, there exist four kinds of molecules, 
$\bar{\rm u}$u, $\bar{\rm d}$d,
ud, and $\bar{\rm u}\bar{\rm d}$. 
We plot the pseudo spins of $\bar{\rm u}$u and ud molecules in Fig.~\ref{fig:pesudo_meson_baryon}.

This phenomenon resembles quark confinement in QCD. The quarks are elementary
particles in nature, but they are confined in the form of hadrons and we cannot take any quarks out from hadrons.
Quarks in QCD resemble our ``elementary" u- and d-vortices in the 2 component BECs.
One might envisage that the isospin of up and down quarks is a natural counterpart of the pseudo spin.
Furthermore, as is well known, 
a hadron consists of several quarks whose total color charge is zero (singlet of $SU(3)$ color group).
Now, it is natural to relate the $U(1)_R$ winding number $n_R$ 
of the ``elementary" vortices (u,d,$\bar{\rm u}$,$\bar{\rm d}$)
with the color charge of the quarks. Namely, a hadron in the two component BECs is singlet by means of $n_R = 0$.
Moreover, we can also relate the $U(1)_S$ winding number $n_S$ of the elementary vortices
(u- and d-vortices have 1/2, and $\bar{\rm u}$- and $\bar{\rm d}$- have $-1/2$ winding number)
with the baryon number of the quarks (a quark has 1/3 and an anti quark has $-$1/3 baryon number).
A hadron with the baryon number $+1$($-1$) are called a baryon (an anti-baryon), and that with no baryon number 
is called a meson in QCD. Borrowing the terminology from QCD, we may refer a ud vortex molecule to a baryon
since it has $n_S = 1$,
and the $\bar{\rm u}\bar{\rm d}$ vortex molecule to an anti-baryon with $n_S=-1$. 
Similarly, we call the $\bar{\rm u}$u
molecule as a u meson while the $\bar{\rm d}$d molecule as a d meson because they have $n_S=0$. 
Of course, though this analogy between the two component
BECs and QCD is limited, it is very useful. 
For example, there does not exist $\bar{\rm u}$d meson in the BEC system because its $U(1)_R$ winding number is not zero, see Table \ref{tab:winding}. We will give more discussions on a relation between QCD and BECs in Sec.~\ref{sec:connection_to_QCD}.

\begin{figure*}[h]
\begin{center}
\includegraphics[width=\hsize]{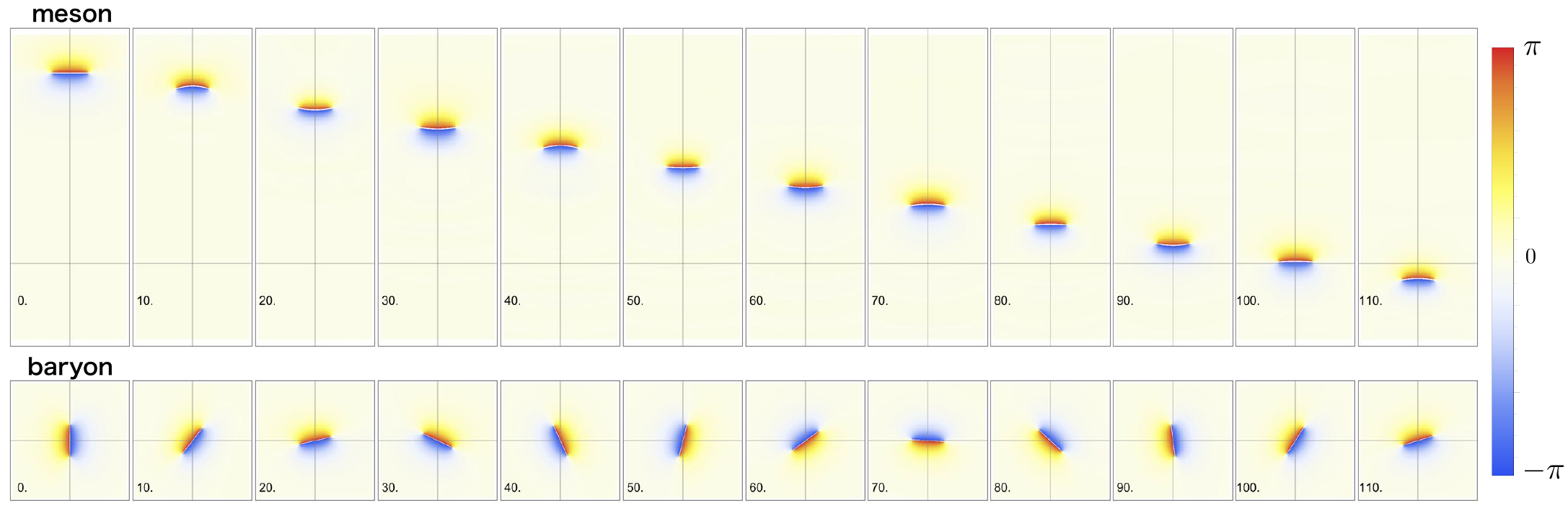}
\caption{The upper panel shows a motion of a u meson, and the lower panel shows a motion of a baryon. We choose $\tilde g_{12} = 0.5$
and $\tilde \omega = 0.05$. The color corresponds to $\arg(\Psi_1) - \arg(\Psi_2)$. The red corresponds to
$\arg(\Psi_1) - \arg(\Psi_2) = \pi$ while the blue corresponds to $\arg(\Psi_1) - \arg(\Psi_2) = -\pi$.
The upper panel ($\tilde x_1 \in [-15,15]$, $\tilde x_2 \in [-20,60]$): 
We put $\bar{\rm u}$-vortex at $(\tilde x_1, \tilde x_2) = (-5,50)$
and u-vortex at $(\tilde x_1, \tilde x_2) = (5,50)$ at $\tilde t = 0$ in the left-most figure.
The lower panel ($\tilde x_1 \in [-15,15]$, $\tilde x_2 \in [-15,15]$): we put u-vortex at
$(\tilde x_1,\tilde x_2) = (0,4)$ and d-vortex at $(\tilde x_1,\tilde x_2) = (0,-4)$ at $\tilde t=0$.
To see motions of the molecules, we also show snapshots with interval $\delta \tilde t = 10$.}
\label{fig:single_b_m}
\end{center}
\end{figure*}

Let us next mention the dynamics of the mesonic and baryonic vortex molecules. A meson goes straight with an almost 
constant velocity
toward the direction perpendicular to the molecule. This motion can be understood by a Magnus force between the vortex
and anti-vortex in the meson. 
For a longer meson the attractive force is dominated by the SG soliton and therefore
the moving speed is almost constant. 
On the other hand, the attractive force originated 
by the inter-vortex force dominates for a shorter meson. 
Therefore, the shorter meson moves faster. 
However, these observations are valid only for a meson
with reasonable length. If the meson is too short, it soon decays. If the meson is too long, it soon disintegrates \cite{Eto:2017rfr}.
We show a typical motion of a u meson in Fig.~\ref{fig:single_b_m}. 
We put a meson whose length is about 10 in terms of 
the dimensionless coordinate $\tilde x_i$. We put the meson at $(\tilde x_1,\tilde x_2) = (0,50)$ at $\tilde t = 0$
(the pseudo spin of the initial state is given in the left panel of Fig.~\ref{fig:pesudo_meson_baryon}).
It moves down and passes the origin around $\tilde t = 100$. During the run, the soliton periodically bends forward
and backward, and distance between u and $\bar{\rm u}$ periodically gets shorter and longer.

On the other hand, 
a baryon moves very differently. It does not run but rotates with an almost constant angular speed \cite{Tylutki:2016mgy,Eto:2017rfr}.
When the baryon is longer, the motion is dominated by attractive force due to the soliton. On the other hand,
when the baryon is shorter, the soliton tension and the inter-vortex force compete. When $g_{12} > 0$,
the inter-vortex force is repulsive \cite{Eto:2011wp}. The shorter the baryon is, the stronger the inter-vortex force is. 
On the contrary, when the baryon is too short, the soliton tension becomes negligible. Therefore, the rotating speed of
the baryon for $g_{12} > 0$ becomes gradually smaller and it vanishes at an equilibrium. For the baryon with $g_{12} < 0$, 
both the inter-vortex interaction and soliton tension give attractive forces \cite{Eto:2011wp} so that there does not exist such an equilibrium.
Similarly to the meson, too long baryon soon disintegrates. However, the baryon never vanishes since the $U(1)_S$ winding
number is topological.
Fig.~\ref{fig:single_b_m} shows a typical rotating baryon. We put the baryon of the length about 8 in the dimensionless
unit at the origin at $\tilde t = 0$
(the pseudo spin of the initial state is given in the right panel of Fig.~\ref{fig:pesudo_meson_baryon}). 
It rotates clockwise and return to the original angle around $\tilde t = 90$.
As meson, it slightly vibrates during the rotation.  
Baryons shorter than equilibrium distance rotate counterclockwise.

Before closing this section, let us mention significant difference between relativistic and non-relativistic dynamics
of vortices. Both straight moving of the mesonic molecule and rotation of the baryonic molecule are peculiar to
non-relativistic system. Due to these characteristic motions, molecules are quasi stable. On the contrary, in relativistic
systems two vortices separate (close on) when the inter vortex force is repulsive (attractive), 
so that no stable molecules in general exist.
In what follows, we will make use of these non-relativistic property, and numerically simulate scatterings of mesonic and baryonic
molecules in two-component BECs as a vortical hadron collider.\footnote{The movies of our numerical simulations studied below
are available as the Supplemental Material.}

Here, one comment is in order. 
Vortex dynamics in two-component BEC in the absence of the Rabi interaction 
was studied in Ref. \cite{Kobayashi:2013wra,Kasamatsu:2015cia} in which case fractional vortices are 
liberated. Even in such a case, dynamics is quite nontrivial but it is out of the scope of the present paper.

\section{
Meson-meson scattering: the case of $\bar{\rm u}{\rm u}$-$\bar{\rm u}{\rm u}$ }
\label{sec:meson-meson}

\subsection{$\bar{\text{u}}\text{u}$-$\bar{\text{u}}\text{u}$ head-on collision 
}
\label{sec:uu_headon1}

\begin{figure*}[h]
\begin{center}
\includegraphics[width=0.95\hsize]{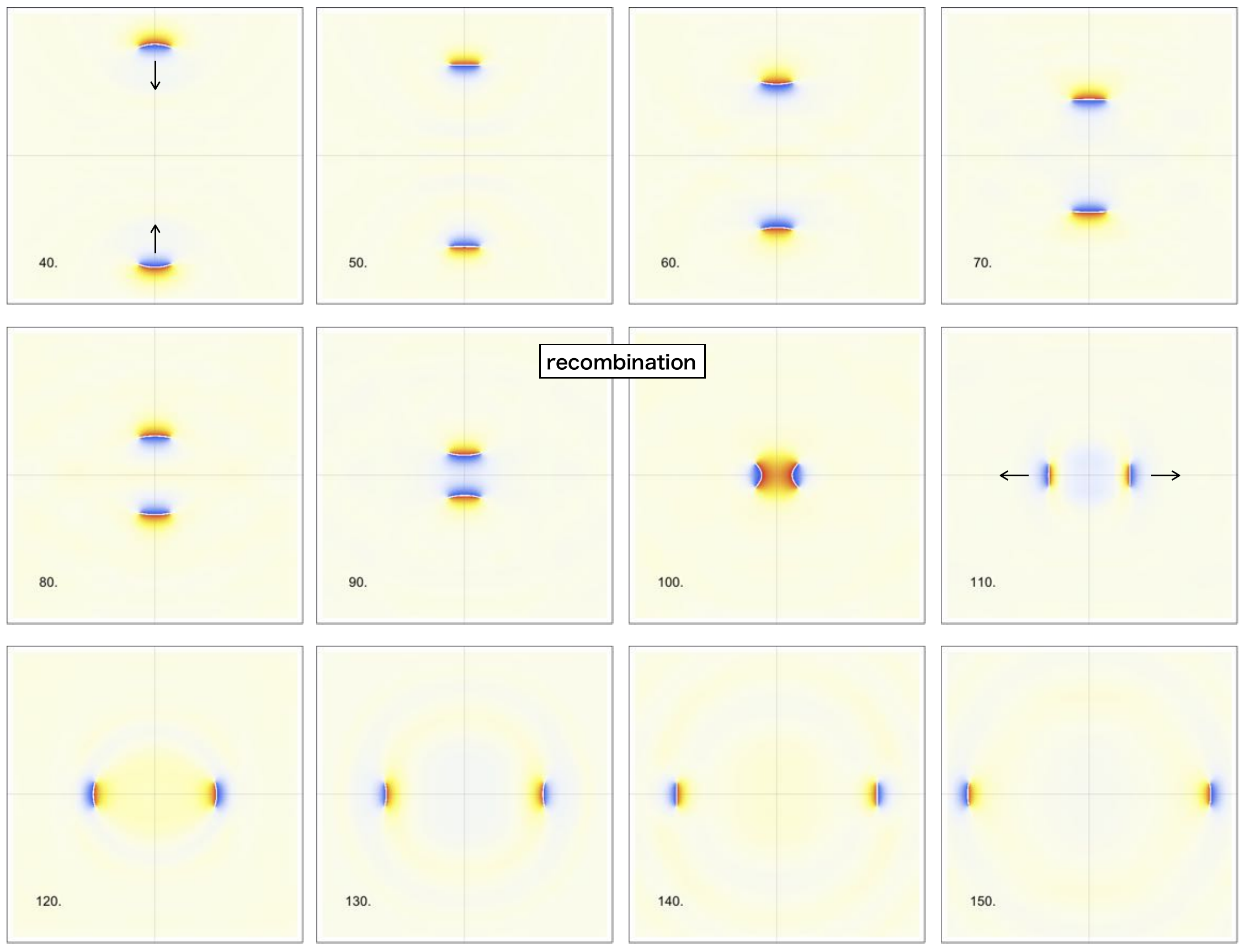}
\caption{Two u mesons scattering
: Color density plots of the relative phase $\arg(\Psi_1) - \arg(\Psi_2)$. 
We initially $(\tilde t = 0)$ set the same u meson ($\bar{\rm u}_1{\rm u}_1$) at $(\tilde x_1,\tilde x_2) = (0,50)$ 
as the one given in the left-most panel of the upper line of Fig.~\ref{fig:single_b_m},
and the same u meson ($\bar{\rm u}_2{\rm u}_2$) but rotated by 180 degree at $(\tilde x_1,\tilde x_2) = (0,-50)$.
They straightly run with almost constant speed and collide around the origin. During the collision, the sine-Gordon (SG) solitons 
partially reconnect and the recombination takes place. As a consequence, the u mesons in the head-on collision scatter through a right angle.
We show the snapshots from $\tilde t =40$ to $150$ with an interval $\delta \tilde t = 10$, and the plot region is
$\tilde x_{1,2} \in [-40,40]$.}
\label{fig:ss_DP_set06_ex01_umum}
\end{center}
\end{figure*}

\begin{figure*}
\begin{center}
\includegraphics[width=0.95\hsize]{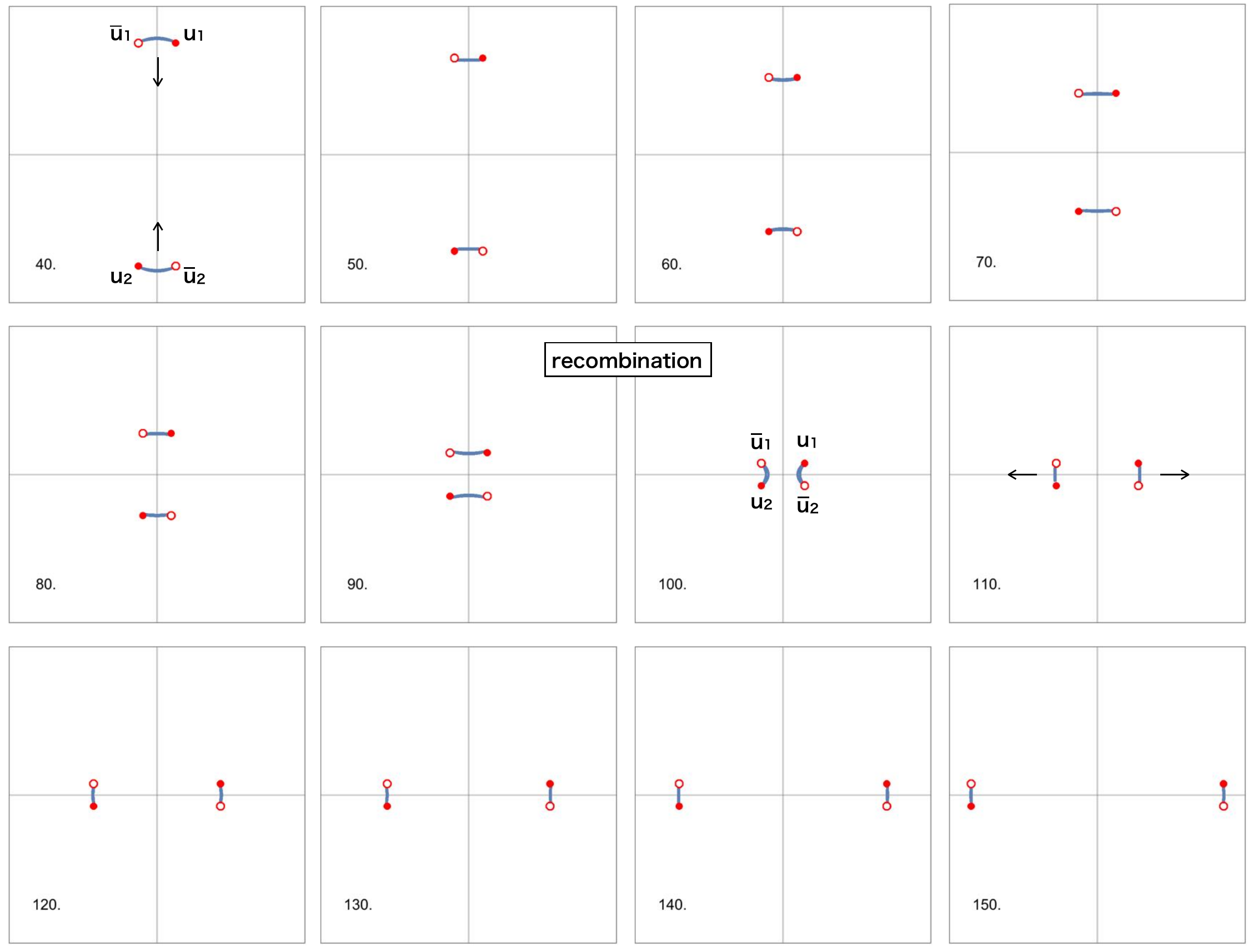}
\caption{A simplified plot of the u mesons scattering shown in Fig.~\ref{fig:ss_DP_set06_ex01_umum}.
We put painted red disks (unpainted red circle) at the points corresponding to the u ($\bar{\rm u}$) vortex centers.
Gray regions bridging the u and $\bar{\rm u}$ vortices show the SG solitons. The gray regions are those
where the relative phases take the values within $3 \le |\arg(\Psi_1) - \arg(\Psi_2)| \le \pi$, 
which are numerically obtained from Fig.~\ref{fig:ss_DP_set06_ex01_umum}.
It is easier to figure out the vortex and anti vortex in this representation than Fig.~\ref{fig:ss_DP_set06_ex01_umum}.}
\label{fig:ss_KP_set06_ex01_umum}
\end{center}
\end{figure*}

Let us begin with the most elementary process, namely head-on collision of two mesons of the same species.
Since we have concentrated on the symmetric model under replacement $\Psi_1$ and $\Psi_2$, we can choose the
u mesons without loss of generality.
We prepare an initial configuration as follows. First, we create a configuration $(\Psi_1^{(1)},\Psi_2^{(1)})$
for the u meson ($\bar{\rm u}_1{\rm u}_1$) corresponding to the one given in the left-most panel of the first row in Fig.~\ref{fig:single_b_m}.
Similarly, we prepare another configuration $(\Psi_1^{(2)},\Psi_2^{(2)})$ for the u meson ($\bar{\rm u}_2{\rm u}_2$)
by rotating the $\bar{\rm u}_1{\rm u}_1$ by 180 degree around the origin. Then, we superpose these 
two configurations \`a la the Abrikosov as
\beq
\Psi^{\rm (ini)}_i = \frac{1}{v}\Psi^{(1)}_i\Psi^{(2)}_i,\quad (i=1,2).
\eeq
In this way, we have the initial configuration which has two incoming mesons $\bar{\rm u}_1{\rm u}_1$ at 
$(\tilde x_1,\tilde x_2) = (0,50)$ [precisely speaking, we put a u vortex at $(\tilde x_1,\tilde x_2) = (5,50)$ and
a $\bar{\rm u}$ vortex at $(\tilde x_1,\tilde x_2) = (-5,50)$]
and the $\bar{\rm u}_2{\rm u}_2$ and $(\tilde x_1,\tilde x_2) = (0,-50)$
[we put a u vortex at $(\tilde x_1,\tilde x_2) = (-5,-50)$ and
a $\bar{\rm u}$ vortex at $(\tilde x_1,\tilde x_2) = (5,-50)$].
With the initial configuration at hand, next we numerically integrate the Gross-Pitaevskii equations. The result
is shown in Figs.~\ref{fig:ss_DP_set06_ex01_umum} and \ref{fig:ss_KP_set06_ex01_umum}.

Up to slightly before the moment of the collision ($\simeq 100$), each meson goes straight toward the origin as if
the other meson does not exist. However, an interesting vortical reaction occurs during the collision. It is a recombination 
of the SG solitons binding the constituent vortices $\bar{\rm u}$ and ${\rm u}$. Before the collision
$\bar{\rm u}_1{\rm u}_1$ and $\bar{\rm u}_2{\rm u}_2$ are well separated, and the SG solitons bridge $\bar{\rm u}_1$ and
${\rm u}_1$, and also $\bar{\rm u}_2$ and ${\rm u}_2$, respectively.
As can be seen in Fig.~\ref{fig:ss_KP_set06_ex01_umum}, $\bar{\rm u}_1$ and ${\rm u}_2$ (${\rm u}_1$ and $\bar{\rm u}_2$)
collide head-on, so that two distances between $\bar{\rm u}_1$ and ${\rm u}_1$, and $\bar{\rm u}_1$ and ${\rm u}_2$ become
comparable about the moment of the collision. Then, the SG solitons reconnect different pair of $\bar{\rm u}$ and ${\rm u}$
from the initial pairs. The new SG solitons are vertical, so that the new outgoing mesons fly along the $x^1$-axis.
Namely, the u mesons in the head-on collision scatter through a right angle as a consequence of the recombination.
We may describe this process as follows 
\beq
\bar{\rm u}_1{\rm u}_1 \quad + \quad \bar{\rm u}_2{\rm u}_2 
\quad \to \quad
\bar{\rm u}_1{\rm u}_2 \quad + \quad \bar{\rm u}_2{\rm u}_1\,.
\label{eq:mm_01}
\eeq
Note that the subscriptions (1 and 2) are introduced just for our convenience; The u$_1$ and u$_2$ vortices are
the same vortices, so they are indistinguishable.
In analogy with chemical and nuclear reactions, we call processes concerning vortex molecules such as Eq.~(\ref{eq:mm_01}) as ``vortical  reactions.''

Note also that the right angle scattering of two topological solitons are common in relativistic field theories.
For instance, magnetic monopoles, vortices and so on do so. 
However, the right angle scattering is usually observed in the collisions of
two solitons with the same topological charges. 
The right angle scattering here is very different
since mesons have zero topological charge and it occurs as a consequence of the recombination.

\begin{figure*}[h]
\begin{center}
\includegraphics[width=0.75\hsize]{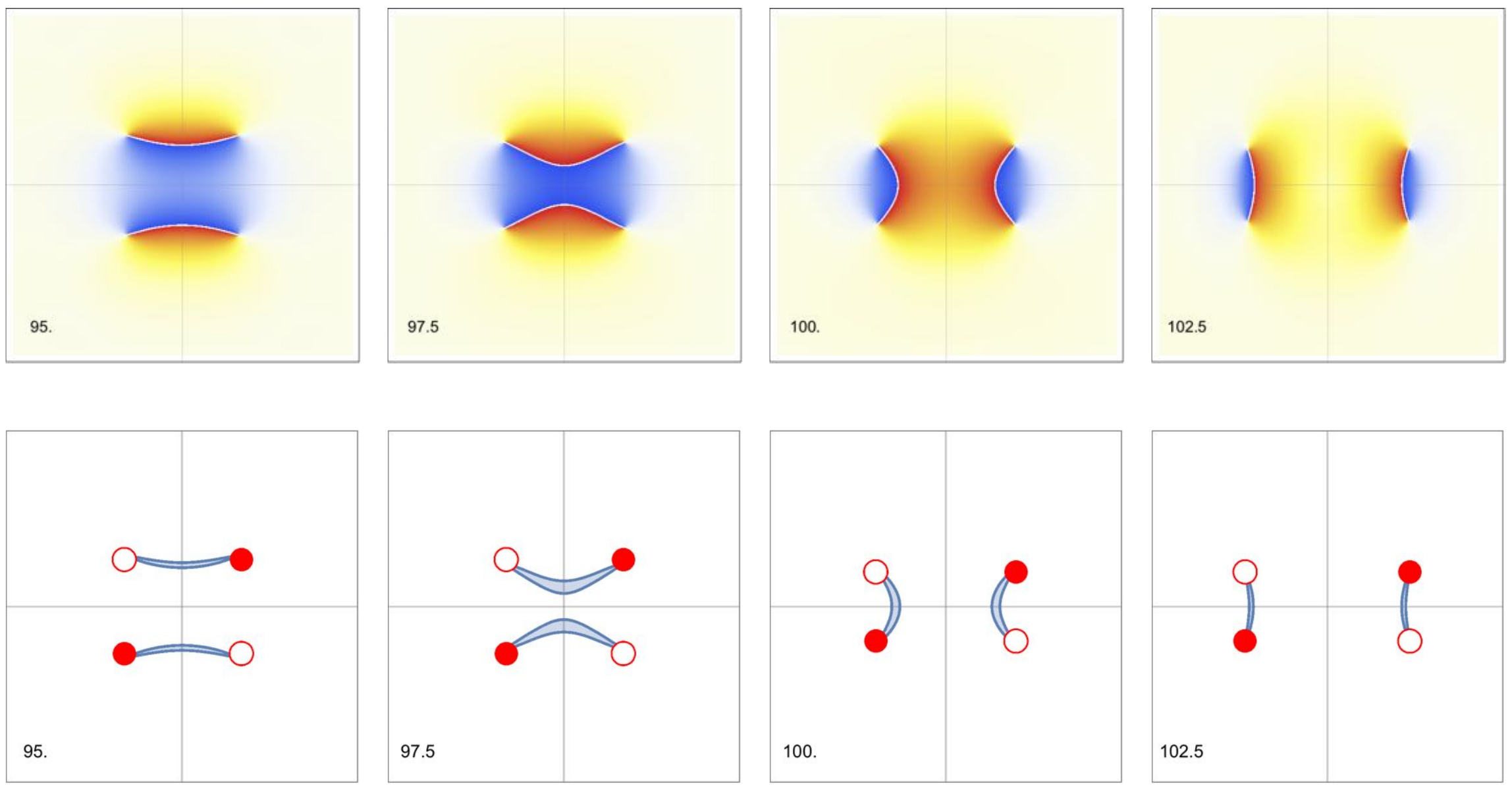}
\caption{
Snapshots in close-up for $\tilde x_i \in [-15,15]$ at $\tilde t = 95, 97.5, 100, 102.5$ for the u meson scattering given in Figs.~\ref{fig:ss_DP_set06_ex01_umum} and \ref{fig:ss_KP_set06_ex01_umum}.
The horizontal SG soliton incoming from the top and the anti-soliton from the bottom gradually bend as they are close by, and 
they partially annihilate at the tips and reconnect to form the vertical SG solitons.}
\label{fig:ss_DPKP_set06_ex01_umum}
\end{center}
\end{figure*}

\begin{figure*}
\begin{center}
\includegraphics[width=0.95\hsize]{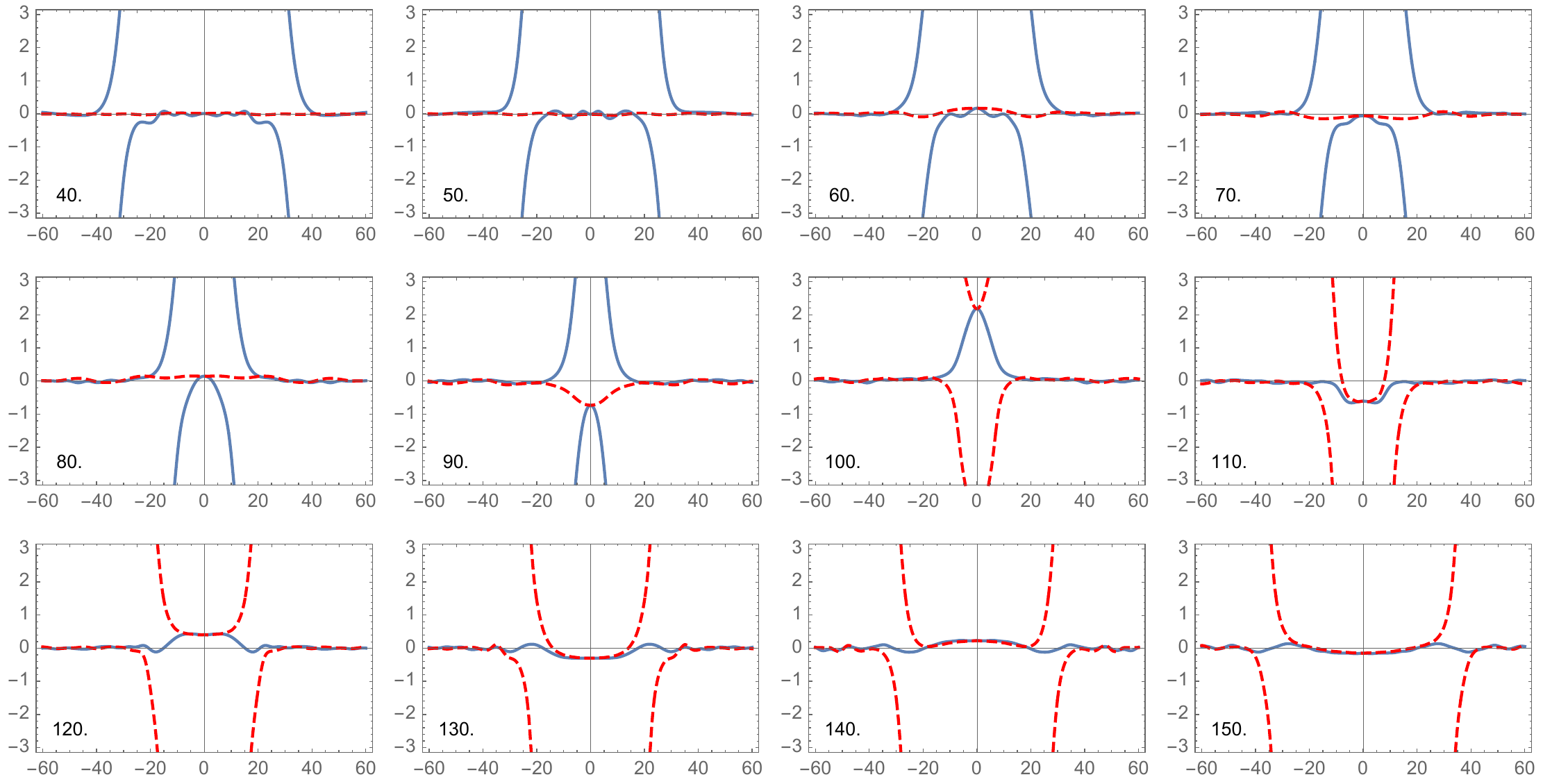}
\caption{The phase plots corresponding to the u meson collision in Fig.~\ref{fig:ss_DP_set06_ex01_umum}. The solid blue curves show the relative phase $\arg(\Psi_1) - \arg(\Psi_2)$ along the $\tilde x_2$-axis
of Fig.~\ref{fig:ss_DP_set06_ex01_umum}. The SG soliton coming from the right-hand side and the anti-SG soliton coming
from the left-hand side collide and annihilate. Similarly, the broken red curves show 
$\arg(\Psi_1) - \arg(\Psi_2)$ along the $\tilde x_1$-axis of Fig.~\ref{fig:ss_DP_set06_ex01_umum}.
The SG and anti-SG solitons are pairwisely created around the moment of collision.}
\label{fig:SG_set06_ex01_umum}
\end{center}
\end{figure*}

Let us observe the recombination phenomena more carefully.
Fig.~\ref{fig:ss_DPKP_set06_ex01_umum} shows snapshots in close-up at $\tilde t = 95, 97.5, 100, 102.5$. 
We should pay attention to the orientation of the SG solitons. Looking at the left-most panel at $\tilde t = 95$ 
of Fig.~\ref{fig:ss_DPKP_set06_ex01_umum}
from the top to the bottom along the $\tilde x^2$-(vertical) axis, the color firstly changes from red to blue representing the upper SG soliton,
and then it changes back from blue to red representing the lower SG soliton. The deepest red region corresponds to $\theta_1-\theta_2 = \pi$
while the deepest bluish region corresponds to $\theta_1-\theta_2 = - \pi$. Thus, when we focus on the SG solitons, 
the head-on collision of the two u mesons is nothing but a scattering of the SG and anti-SG solitons.
As can be seen in the panel at $\tilde t = 97.5$ of Fig.~\ref{fig:ss_DPKP_set06_ex01_umum}, 
the SG soliton and anti-SG soliton greatly bend especially around their centers due to an attractive force
so that they collide before the u and $\bar{\rm u}$ vortices at the edges of the SG solitons do. 
Then, their tips annihilate to each other, and they proceed to complete the recombination process, see
transformation from $\tilde t = 97.5$ to $\tilde t = 100$ shown in Fig.~\ref{fig:ss_DPKP_set06_ex01_umum}.
The annihilation of the SG soliton and anti solitons can  be clearly seen in Fig.~\ref{fig:SG_set06_ex01_umum}
where we plot the relative phase $\theta_1-\theta_2$ on the $\tilde x^2$-axis by the blue curve. The SG soliton
corresponds to the jump $\pi \to -\pi$ from left to right of the horizontal axis (the $\tilde x^2$-axis) while
the anti-SG soliton corresponds to the opposite jump from $-\pi \to \pi$. They collide and annihilate 
about $\tilde t = 100 \sim 110$. To be complete, let us also look at the relative phase along the $\tilde x^1$-axis.
It is also plotted in Fig.~\ref{fig:SG_set06_ex01_umum} by the red broken curve. The horizontal axis corresponds
to the $\tilde x^1$-axis for the red broken curves. Before the collision, no SG solitons exist along the $\tilde x^1$-axis. However, as the SG and anti SG solitons along the $\tilde x^2$-axis annihilate, a new pair of the SG and
anti-SG solitons on the $\tilde x^1$-axis is created. Hence, we find that the recombination phenomenon is taken over
by the pair annihilation and creation of the SG solitons.

\clearpage

\subsection{$\bar{\text{u}}\text{u}$-$\bar{\text{u}}\text{u}$ 
 scattering at $\pi/8$ angle}
\label{sec:uu_headon2}

Let us next study u meson scattering similar to that in Sec.~\ref{sec:uu_headon1}, but in this time 
the upper and lower mesons are rotated by $-\pi/8$ and $\pi/8$ from those in Fig.~\ref{fig:ss_DP_set06_ex01_umum}. 
\begin{figure*}[h]
\begin{center}
\includegraphics[width=0.95\hsize]{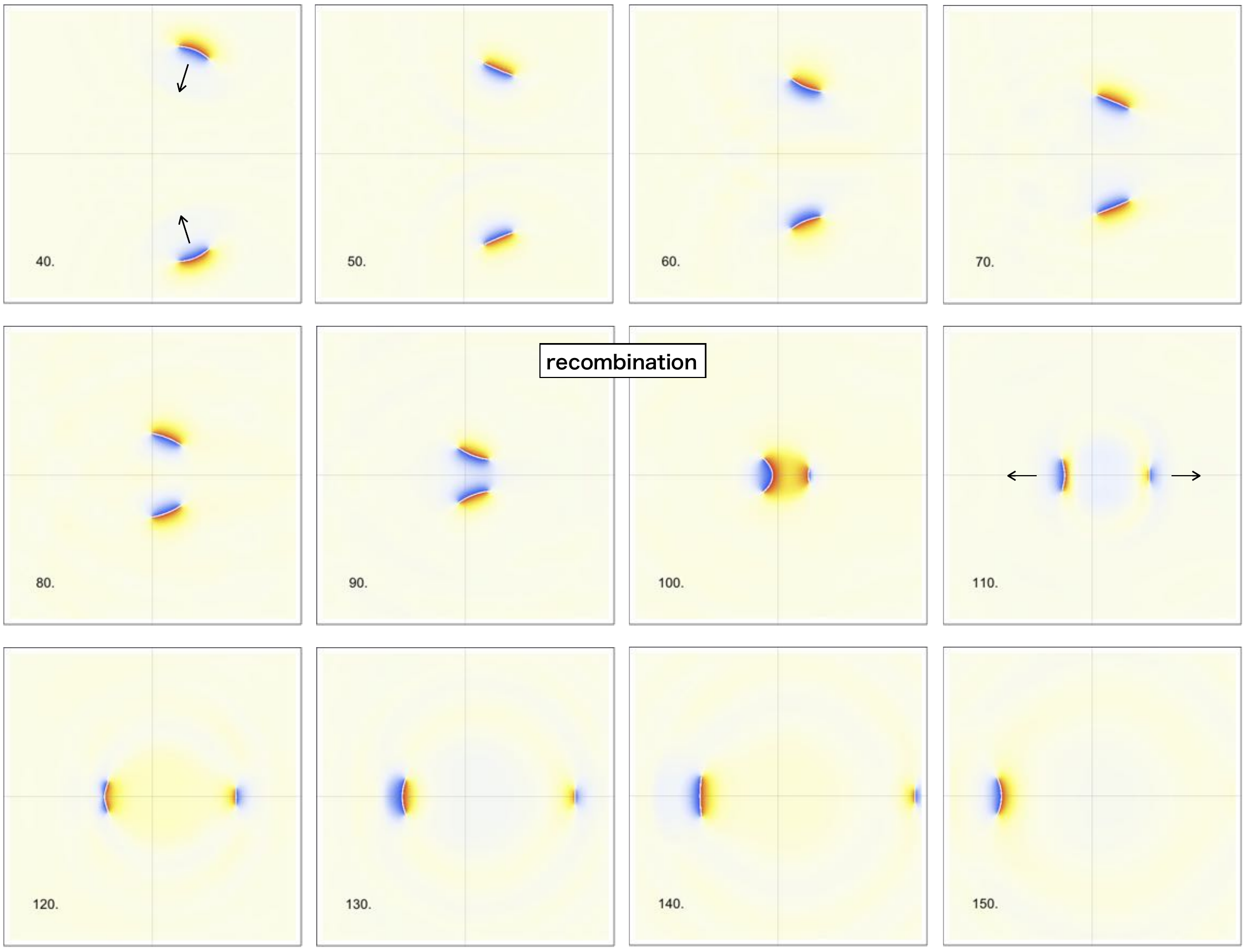}
\caption{Two slightly tilted u mesons scattering
: Color density plots of the relative phase $\arg(\Psi_1) - \arg(\Psi_2)$. 
We initially $(\tilde t = 0)$ set the u meson ($\bar{\rm u}_1{\rm u}_1$) at 
$(\tilde x_1,\tilde x_2) = (50 \sin\pi/8,50 \cos\pi/8)$,
and the other u meson ($\bar{\rm u}_2{\rm u}_2$) at $(\tilde x_1,\tilde x_2) = (50\sin\pi/8,-50 \cos\pi/8)$.
We only show the snapshots from $\tilde t =40$ to $150$ with interval $\delta \tilde t = 10$, and the plot region is
$\tilde x_{1,2} \in [-40,40]$.}
\label{fig:s_DP_set04_ex01_umum}
\end{center}
\end{figure*}
One can see how the scattering is going on in Figs.~\ref{fig:s_DP_set04_ex01_umum} and \ref{fig:ss_KP_set04_ex01_umum}.
Indeed, it goes qualitatively in the same way as the previous case.
The mesons go straight with almost constant speed until they are close by, then a recombination takes place during
the collision. Due to the tilts of incoming mesons, newly formed mesons are in different sizes 
and scatter off toward the left and right directions as the previous case.
The asymmetry can also be seen in the speeds of the out-going mesons.
The meson $\bar{\rm u}_1{\rm u}_2$ moves faster than the meson $\bar{\rm u}_2{\rm u}_1$
as shown in Fig.~\ref{fig:ss_KP_set04_ex01_umum}. This is because the former is shorter than
the latter. Fig.~\ref{fig:ss_DPKP_set04_ex01_umum} shows snap shots in close-up at
$\tilde t = 92.5, 95, 97.5, 100$, in which we again observe that a partial annihilation of the SG and
anti-SG solitons leads to the recombination. Compared to the previous case,
bend of the SG solitons are milder, because the bending points are not the center and are shifted towards the edges
${\rm u}_1$ and $\bar{\rm u}_2$.
\begin{figure*}
\begin{center}
\includegraphics[width=0.95\hsize]{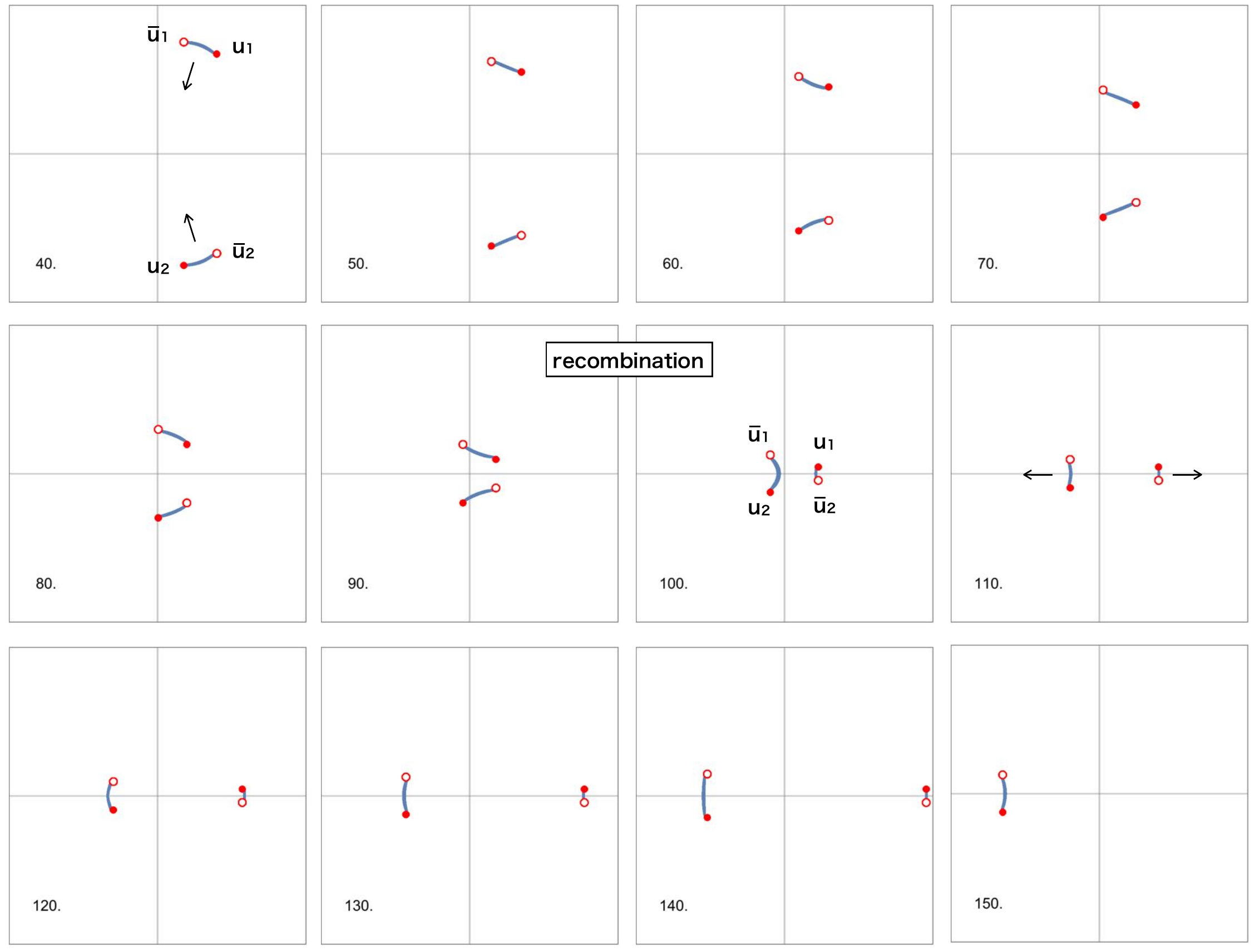}
\caption{A simplified plot of the u mesons scattering shown in Fig.~\ref{fig:s_DP_set04_ex01_umum}.
For details, see the caption of Fig.~\ref{fig:ss_KP_set06_ex01_umum}.}
\label{fig:ss_KP_set04_ex01_umum}
\end{center}
\end{figure*}
\begin{figure*}
\begin{center}
\includegraphics[width=0.75\hsize]{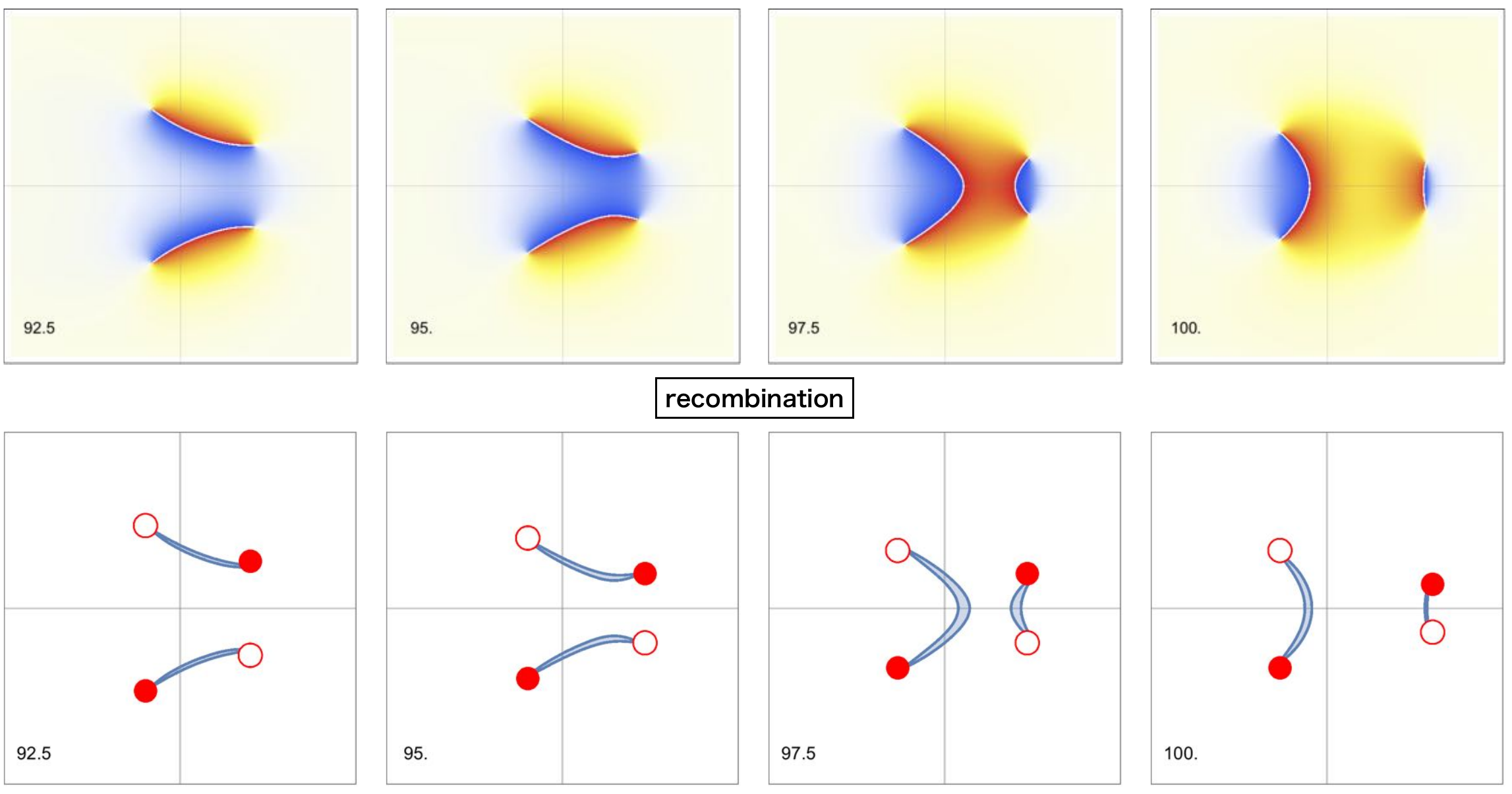}
\caption{Snapshots in close-up for $\tilde x_i \in [-15,15]$ at $\tilde t = 92.5, 95, 97.5, 100$ 
for the u meson scattering given in Figs.~\ref{fig:s_DP_set04_ex01_umum} and \ref{fig:ss_KP_set04_ex01_umum}.
The points where the incoming SG solitons steeply bend are not the centers.
The incoming SG solitons partially annihilate around the steepest bending point 
and reconnect to form vertical SG solitons.}
\label{fig:ss_DPKP_set04_ex01_umum}
\end{center}
\end{figure*}

\clearpage

\subsection{$\bar{\text{u}}\text{u}$-$\bar{\text{u}}\text{u}$ 
scattering at $\pi/4$ angle}

\begin{figure*}[h]
\begin{center}
\includegraphics[width=\hsize]{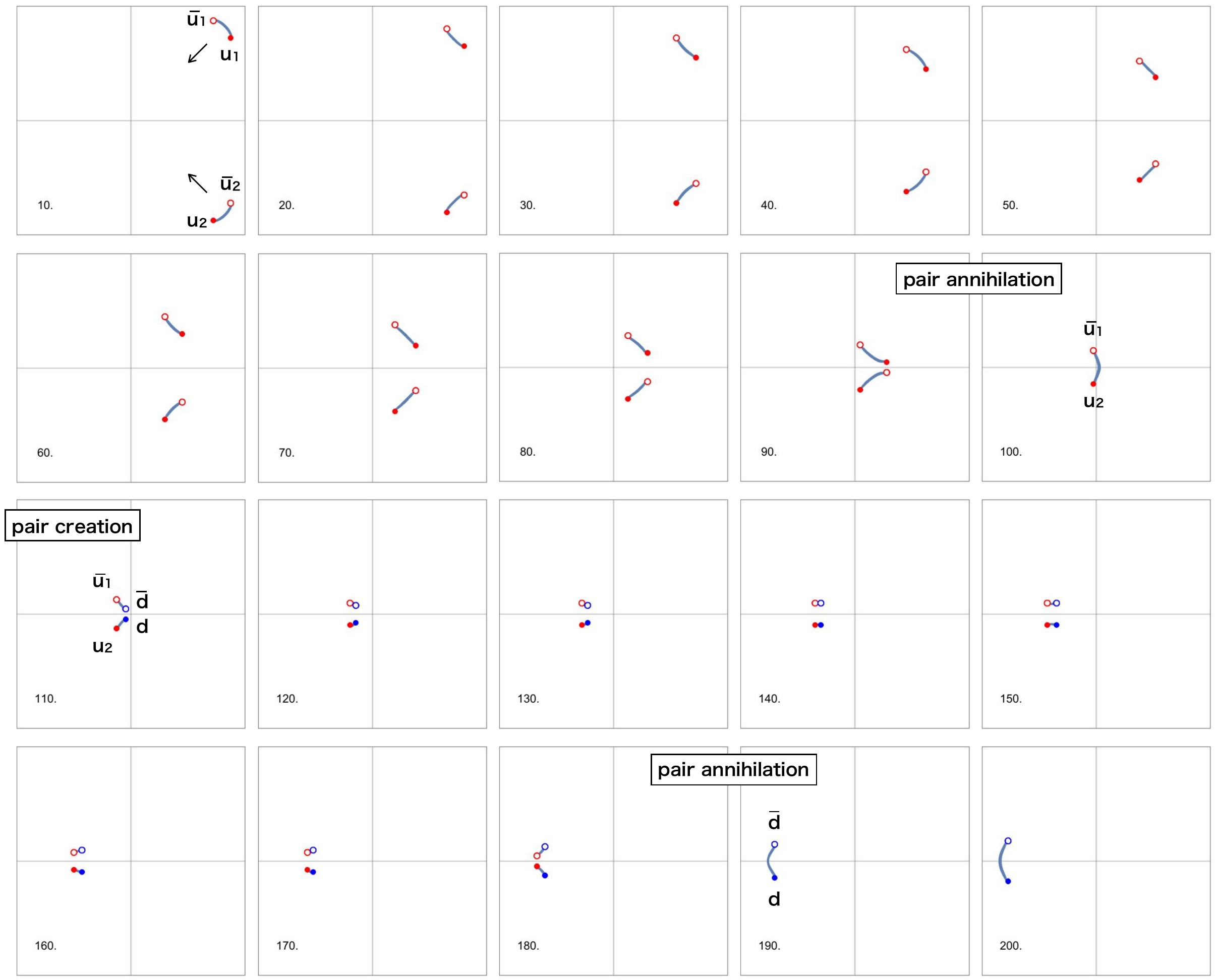}
\caption{Two tilted u meson scattering with $\pi/4$ angle:
We initially $(\tilde t = 0)$ set one u meson ($\bar{\rm u}_1{\rm u}_1$) at 
$(\tilde x_1,\tilde x_2) = (50 \sin\pi/4,50 \cos\pi/4)$,
and the other u meson ($\bar{\rm u}_2{\rm u}_2$) at $(\tilde x_1,\tilde x_2) = (50\sin\pi/4,-50 \cos\pi/4)$.
We only show the snapshots from $\tilde t =10$ to $200$ with interval $\delta \tilde t = 10$, and the plot region is
$\tilde x_{1,2} \in [-40,40]$.}
\label{fig:ss_KP_set03_ex01_umum}
\end{center}
\end{figure*}

\begin{figure*}
\begin{center}
\includegraphics[width=0.75\hsize]{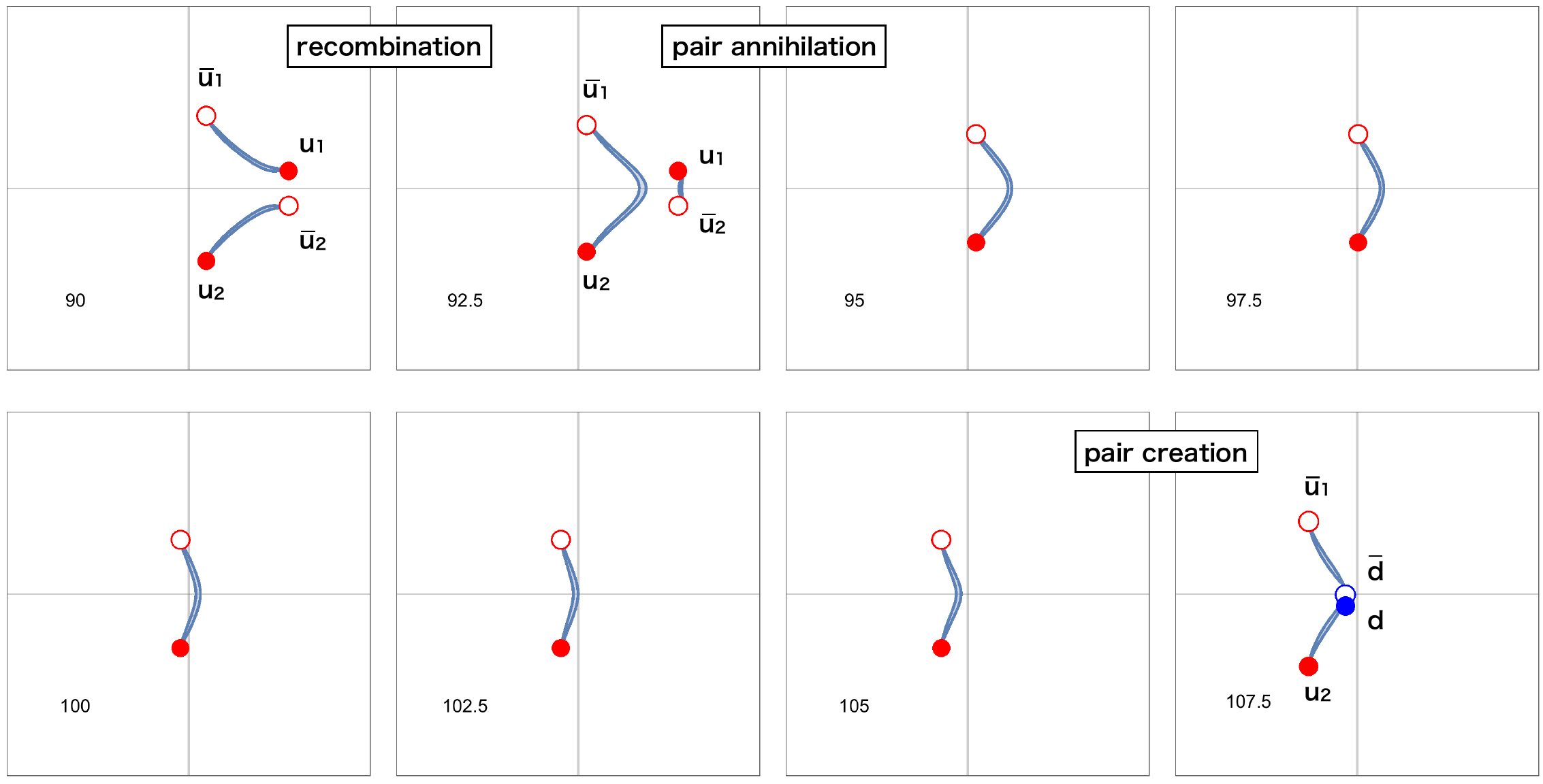}
\caption{Snapshots in close-up for $\tilde x_i \in [-15,15]$ at $\tilde t = 90, 92.5, 95, 97.5, 100, 102.5, 105, 107.5$ 
for the u meson scattering given in Figs.~\ref{fig:ss_KP_set03_ex01_umum}. The recombination, the pair annihilation 
of $\bar{\rm u}{\rm u}$, and the pair creation of $\bar{\rm d}{\rm d}$ occurs in order.}
\label{fig:DPKP_set03_ex01_umum}
\end{center}
\end{figure*}

Let us attempt to simulate one more u meson scattering experiment by rotating further the incoming mesons
by $\pi/4$. As expected, the motion of the mesons before the collision is almost unchanged from the previous two cases,
see Fig.~\ref{fig:ss_KP_set03_ex01_umum}.  
On the contrary, the states after the collision
are distinctive. Firstly, we only observe one meson $\bar{\rm u}_1{\rm u}_2$ after the collision.
Looking at the moment of the collision in more details, the very short $\bar{\rm u}_2{\rm u}_1$ is created but
it is soon annihilated, see Fig.~\ref{fig:DPKP_set03_ex01_umum}.
This occurs because the relative angle of the incoming mesons is too large.
If we further rotate the initial mesons, there is no enough period for the SG solitons to bend to be annihilated. Then,
the constituent vortices, ${\rm u}_1$ and $\bar{\rm u}_2$, are annihilated and 
the two SG solitons join to form a long $\bar{\rm u}_1{\rm u}_2$ meson.

After the collision, the long meson $\bar{\rm u}_1{\rm u}_2$ runs toward the left but such a long meson is
unstable. As can be seen in Fig.~\ref{fig:DPKP_set03_ex01_umum}, it soon breaks up into two pieces, the baryon
(${\rm u}_2{\rm d}$) and the anti-baryon ($\bar{\rm u}_1\bar{\rm d}$),
by creating d and $\bar{\rm d}$ 
vortices at the center of the long SG soliton.
It is notable that this process is peculiar to the 2 component BECs and it never happens in scalar BEC systems.
The baryon ${\rm u}_2{\rm d}$ spins clockwise whereas the anti-baryon spins counterclockwise,
see the panels with $\tilde t = 120 \sim 180$ of Fig.~\ref{fig:ss_KP_set03_ex01_umum}.
At the same time, the pair of baryon and anti-baryon behave as a pair of an integer vortex and an anti-integer vortex.
Thus, the baryon and anti-baryon move parallel toward the left direction.
After a while, the baryon and anti-baryon join to form a meson $\bar{\rm d}{\rm d}$ with $\bar{\rm u}_1$ and 
${\rm u}_2$ being annihilated.

Thus the corresponding reaction process can be summarized as
\beq
\bar{\rm u}_1{\rm u}_1
\quad + \quad
\bar{\rm u}_2{\rm u}_2
\quad \to \quad
\bar{\rm u}_1{\rm u}_2
\quad \to \quad
{\rm u}_2{\rm d} 
\quad + \quad
\bar{\rm u}_1\bar{\rm d}
\quad \to \quad
\bar{\rm d}{\rm d}.
\label{eq:mm_02}
\eeq
Note that the baryon number is preserved to be 0 throughout the reaction.

\subsection{Interaction vertices and Feynman diagrams}

\begin{figure*}[h]
\begin{center}
\includegraphics[width=0.85\hsize]{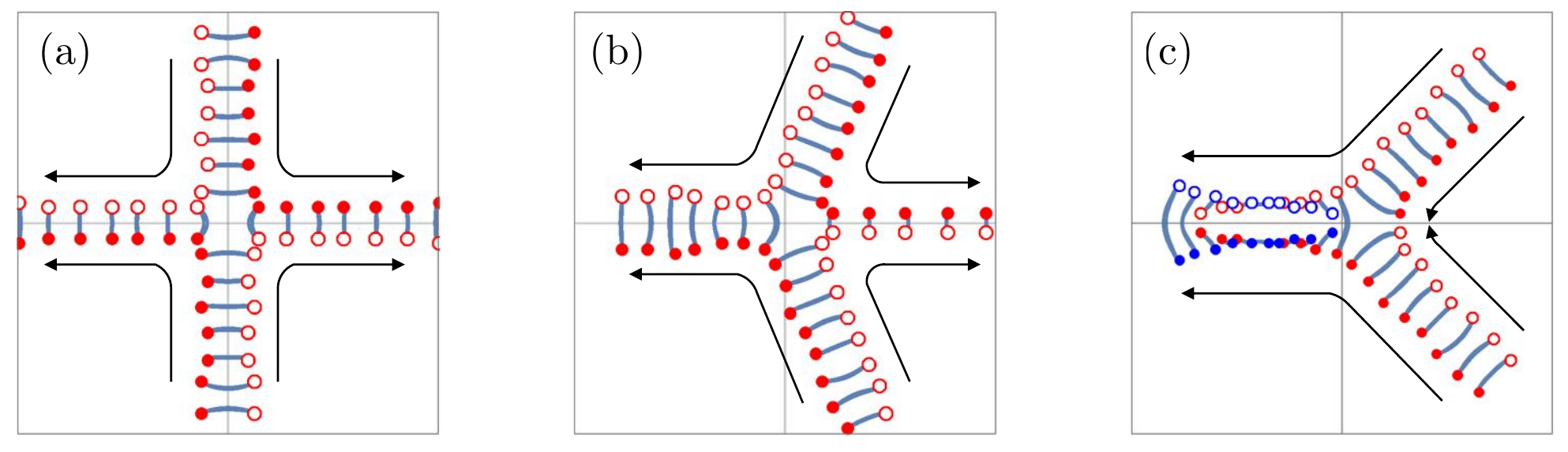}
\caption{Summary on u meson and u meson collider experiments:
(a) corresponds to Fig.~\ref{fig:ss_KP_set06_ex01_umum}, (b) to Fig.~\ref{fig:ss_KP_set04_ex01_umum},
and (c) to Fig.~\ref{fig:ss_KP_set03_ex01_umum}.}
\label{fig:mm_scatterings}
\end{center}
\end{figure*}

Let us summarize the two u meson head-on scattering given above. We found that
the scattering sensitively depends on the collision angle which is direct evidence for a meson
to have a substructure. We have shown the three examples the head-on collisions
with the relative angle $\pi$ in Fig.~\ref{fig:ss_KP_set06_ex01_umum}, 
$3\pi/4$ in Fig.~\ref{fig:ss_KP_set04_ex01_umum}, and
$\pi/2$ in  Fig.~\ref{fig:ss_KP_set03_ex01_umum}. An overview of the scatterings is shown in Fig.~\ref{fig:mm_scatterings}.

\begin{figure*}[h]
\begin{center}
\includegraphics[width=0.65\hsize]{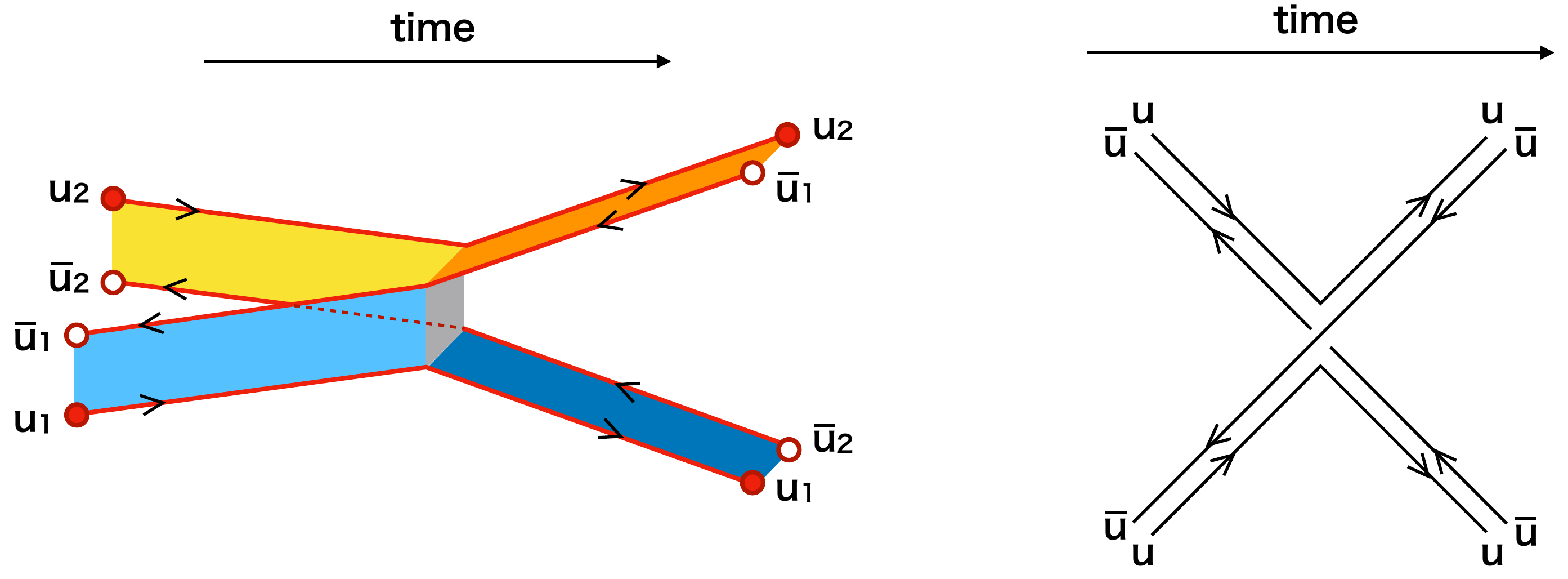}
\caption{Meson-meson-meson-meson vertex:
The left panel is a 3D diagram describes (a) of Fig.~\ref{fig:mm_scatterings}. The right panel is a Feynman diagram
which is a simplified 2D expression of the left one.}
\label{fig:4meson_vertex}
\end{center}
\end{figure*}

Here, we describe these results by a particle physics point of view.
We first deal with the first process given in (a) of Fig.~\ref{fig:mm_scatterings} whose
vortical reaction is given in Eq.~(\ref{eq:mm_01}). The two incoming u mesons scatter into the two outgoing u mesons.
We interpret this process as an elementary interaction among mesons.
Namely, this yields a meson-meson-meson-meson vertex. 
In order to emphasize this view point, let us give
a 3D spacetime (2 spatial and 1 temporal)
diagram given in the left figure of Fig.~\ref{fig:4meson_vertex}. It corresponds to the scattering experiment
of (a) of Fig.~\ref{fig:mm_scatterings}, 
consisting of the world lines of the constituent vortices together with the world sheet of the SG solitons.
The 3D diagram can be further simplified as the right figure of Fig.~\ref{fig:4meson_vertex}. It is a 2D diagram and
resembling the so-called twig diagram (a sort of Feynman diagram which includes only quark lines) 
known in QCD \cite{Zweig:1981pd}. Resembling standard Feynman diagrams in quantum field theories,
an anti-vortex is represented by a line with an arrow opposite to 
the time evolution. 
Thus, we find a four meson vertex as an elemental interaction
among mesons. 

The second process corresponding (b) of Fig.~\ref{fig:mm_scatterings} is essentially same as (a).
So the scattering (b) is also described by the same 4 meson vertex in Fig.~\ref{fig:4meson_vertex}.

\begin{figure*}[h]
\begin{center}
\includegraphics[width=0.4\hsize]{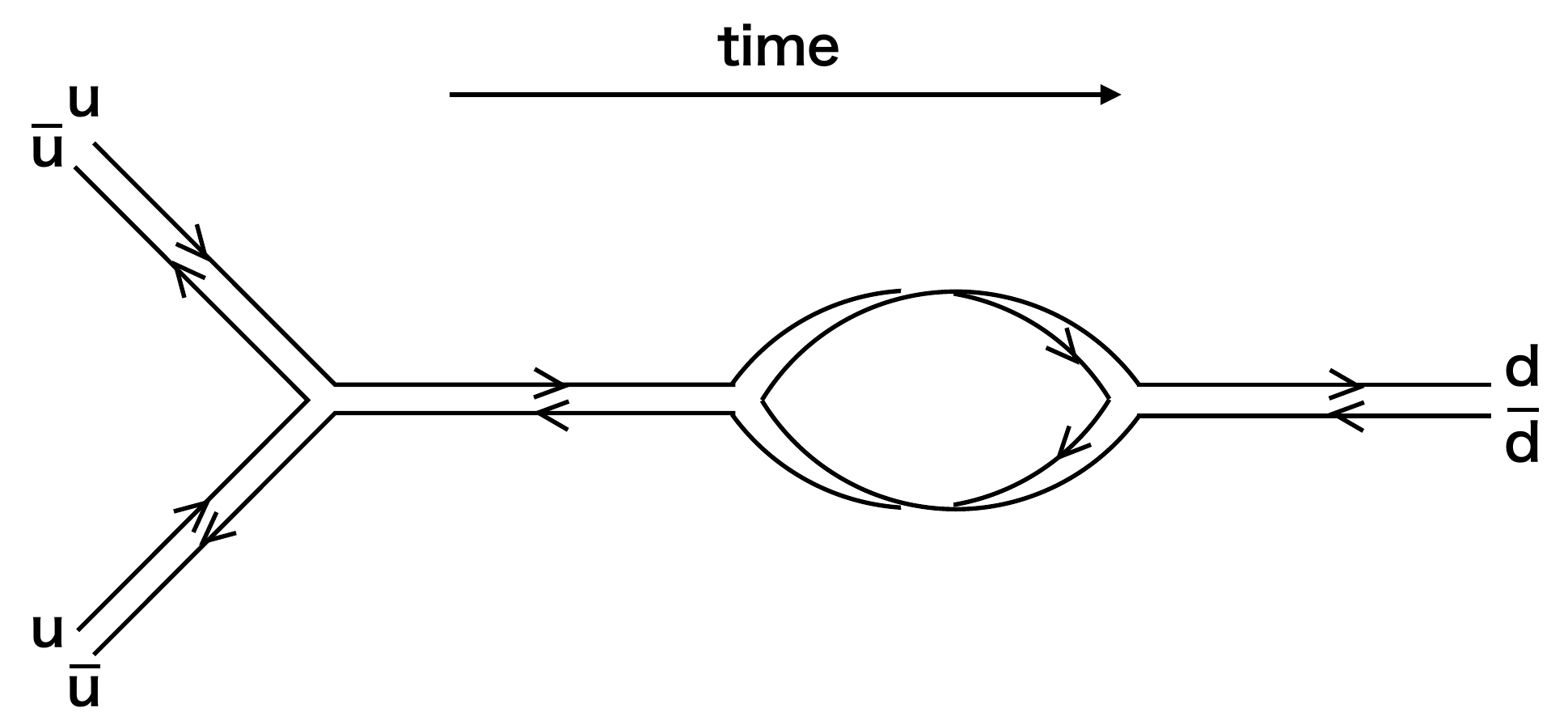}
\caption{The 1 loop diagram for $\bar{\rm u}{\rm u} + \bar{\rm u}{\rm u} \to \bar{\rm d}{\rm d}$ corresponding to
(c) of Fig.~\ref{fig:mm_scatterings}.}
\label{fig:mm_1_loop}
\end{center}
\end{figure*}

\begin{figure*}[h]
\begin{center}
\includegraphics[width=0.65\hsize]{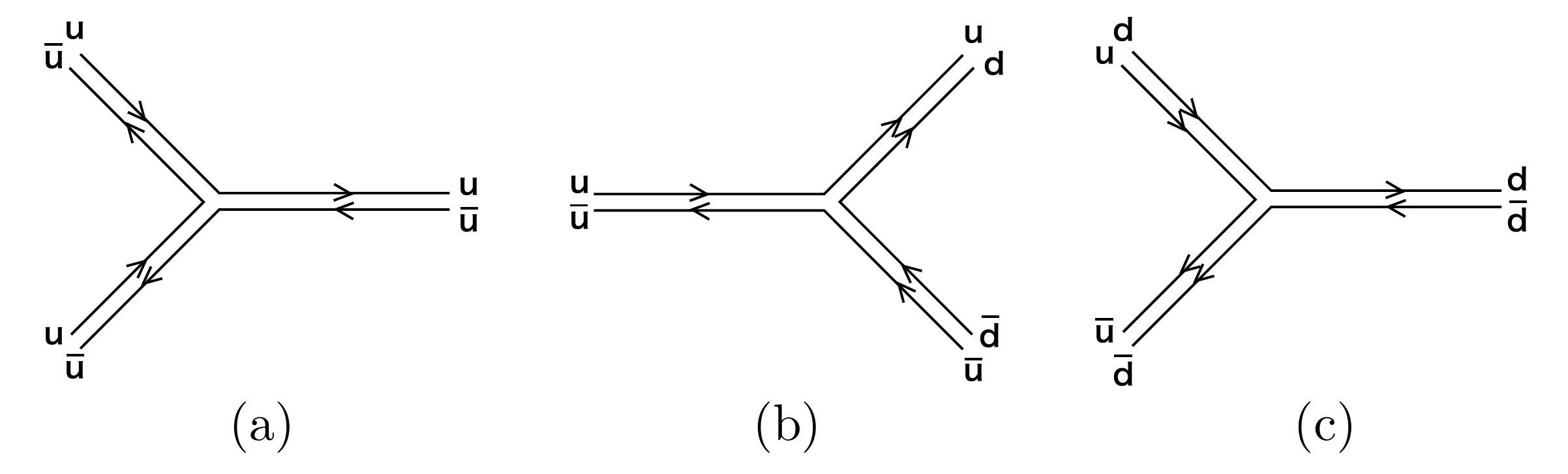}
\caption{The three 3-vertices included in the 1 loop diagram in Fig.~\ref{fig:mm_1_loop}. The time direction is from left to right.}
\label{fig:3vertex}
\end{center}
\end{figure*}

Next, let us make a diagram for the third process (c) of Fig.~\ref{fig:mm_scatterings}.
This scattering is more complicated. 
Indeed, 
the corresponding reaction formula given in Eq.~(\ref{eq:mm_02}) consists of three proceeding steps.
The Feynman diagram for this scattering is given in Fig.~\ref{fig:mm_1_loop} including a loop and three different 3-vertices.
 The left most vertex of Fig.~\ref{fig:mm_1_loop}
is a meson-meson-meson where all mesons are the u type, as shown as (a) of Fig.~\ref{fig:3vertex}. 
The middle and right most vertices are
meson-baryon-baryon vertices, summarized as (b) and (c) of Fig.~\ref{fig:3vertex}.

\begin{figure*}[h]
\begin{center}
\includegraphics[width=0.9\hsize]{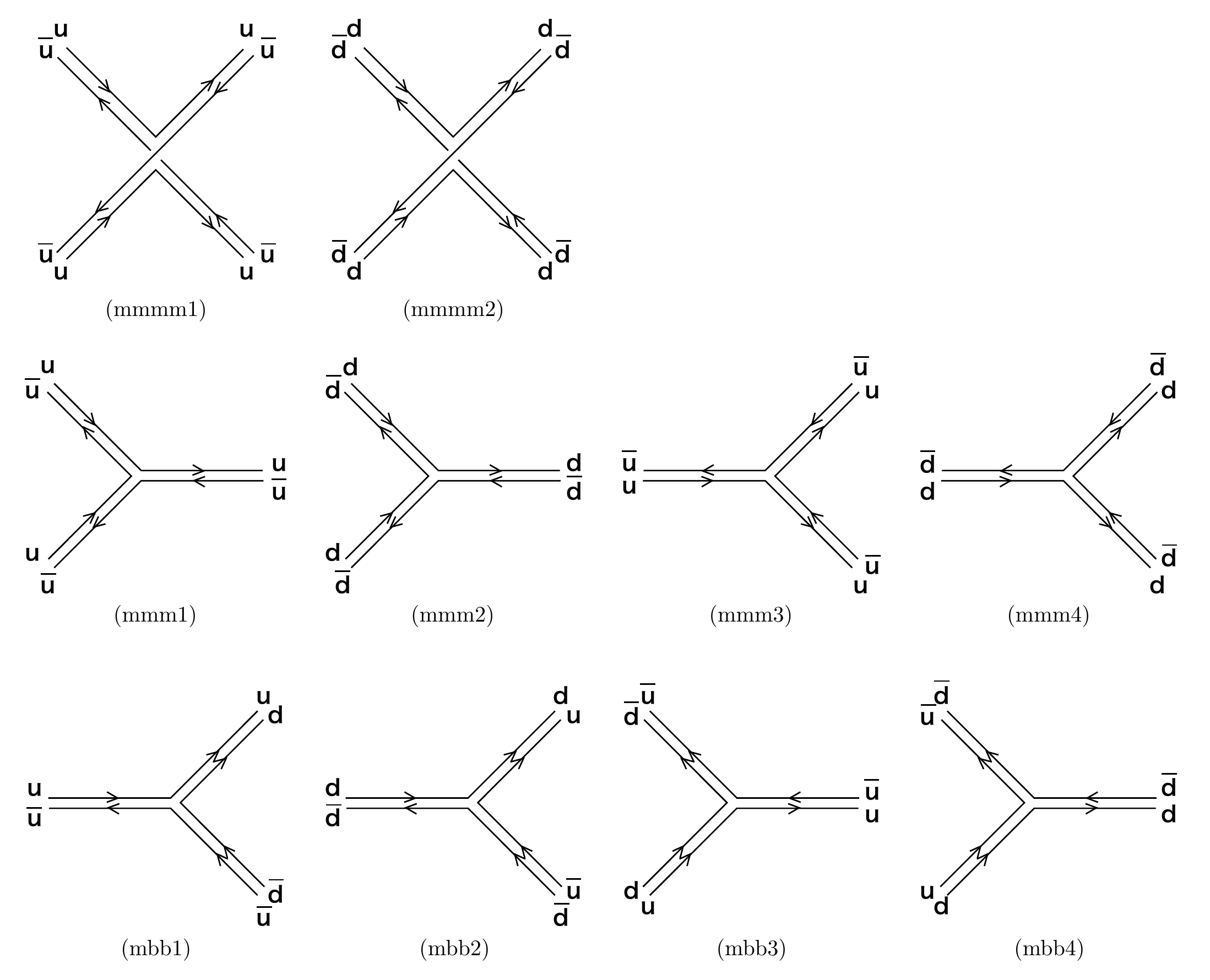}
\caption{Feynman diagrams found in two u (d) mesons scattering experiments.
The first row shows the meson-meson-meson-meson vertices. The second row shows
the meson-meson-meson vertices, and the third row shows
the meson-baryon-baryon vertices. }
\label{fig:all_vertex}
\end{center}
\end{figure*}

Once we get these elementary vertices, it is easy for us to expect what kind of scatterings are possible 
without performing numerical simulations. Therefore, next our task is to track down all possible vertices.
To this end, symmetry is helpful. 
Our system has the symmetry $F:\Psi_1 \leftrightarrow \Psi_2$ due to the special choice 
of the parameters given in Eq.~(\ref{eq:flavor_sym}). This ensures the ``flavor" symmetry 
among the vortices as
$F: \left(X_{\rm u}(t),Y_{\rm u}(t)\right)\leftrightarrow \left(X_{\rm d}(t),Y_{\rm d}(t)\right)$,
where $\left(X_{\rm u,d}(t),Y_{\rm u,d}(t)\right)$ stands for the position of a u or d vortex at time $t$.
Furthermore, the GP equations are invariant under the time reversal symmetry 
$T:\Psi_i(x_1,x_2,t) \leftrightarrow \Psi_i^*(x_1,x_2,-t)$,
and the parity transformation $P:\Psi_i(x_1,x_2,t) \leftrightarrow \Psi_i(x_1,-x_2,t)$.
Note that the parity transformation in even spatial dimensions is identical to a reflection symmetry on an axis.
With respect to the vortices, the former transforms a u vortex to a $\bar{\rm u}$ vortex as
$T: \left(X_{\rm u}(t),Y_{\rm u}(t)\right)\leftrightarrow \left(X_{\bar{\rm u}}(-t),Y_{\bar{\rm u}}(-t)\right)$.
The same holds for d and $\bar{\rm d}$ vortices. Similarly, the latter also transforms a u vortex to $\bar{\rm u}$
since it exchanges $\theta = {\rm arg}\,(x_1 + i x_2) \leftrightarrow -\theta = {\rm arg}\,(x_1-ix_2)$.
Hence, the parity transformation is
$P: \left(X_{\rm u}(t),Y_{\rm u}(t)\right)\leftrightarrow \left(X_{\bar{\rm u}}(t),-Y_{\bar{\rm u}}(t)\right)$,
and similar for d and $\bar{\rm d}$ vortices. By using the $F$, $T$, and $P$ transformations, 
we can exhaust 
all possible diagrams as summarized in Fig.~\ref{fig:all_vertex}.
For example, the  diagrams  labelled by (mmmm1) and (mmmm2) 
are exchanged by the $F$ transformation, while those labelled by 
(mmm1) and (mmm3) are related by the $T$ transformation. All the diagrams in Fig.~\ref{fig:all_vertex}
are invariant under the $P$ transformation.

\subsection{$\bar{\rm u}{\rm u}$-$\bar{\rm u}{\rm u}$ collisions with impact parameters}
\label{sec:impact_uu}

\begin{figure*}[h]
\begin{center}
\includegraphics[width=0.95\hsize]{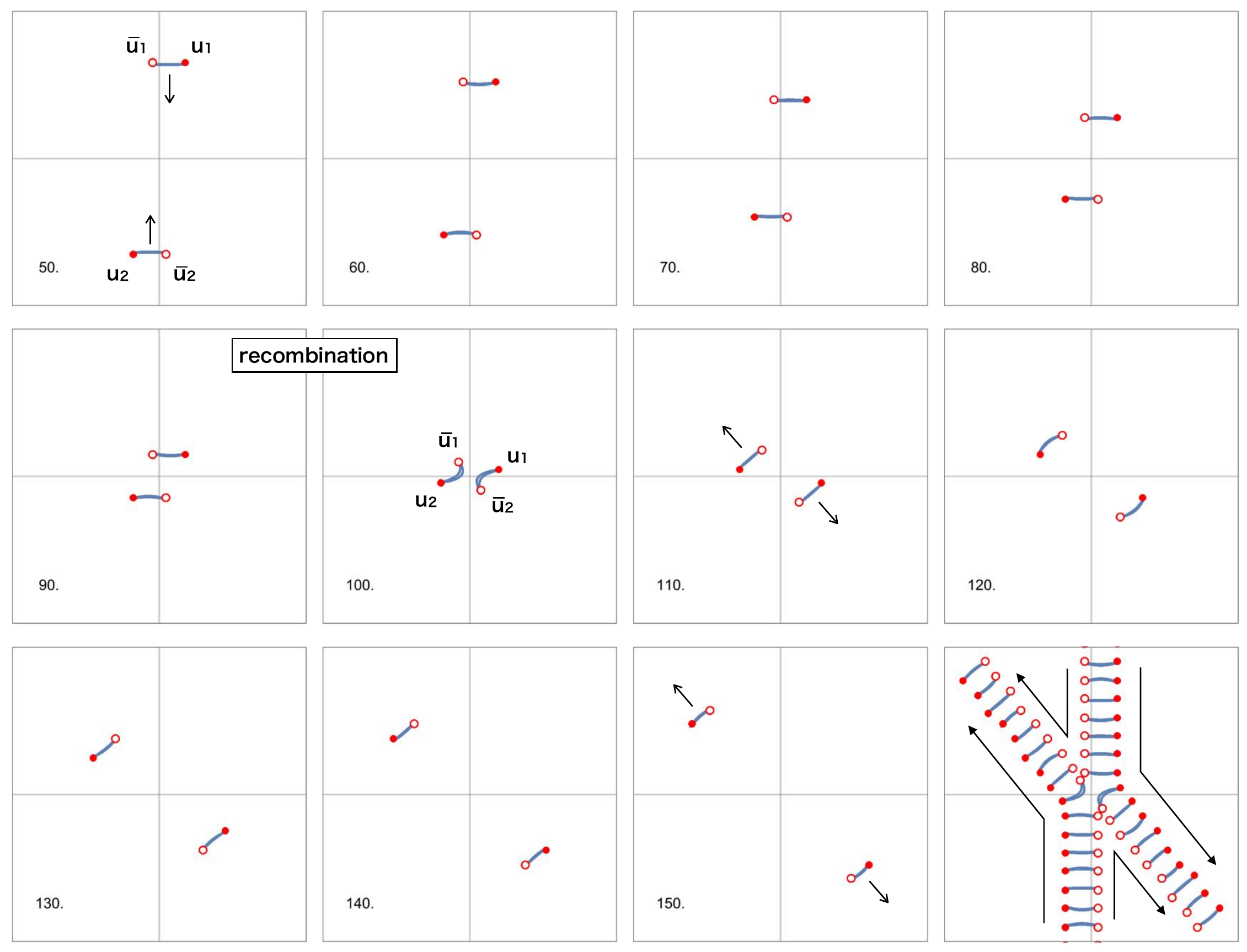}
\caption{Meson-meson scattering ($\bar{\rm u}{\rm u}$-$\bar{\rm u}{\rm u}$) with an impact parameter. Initially,
the two mesons are placed at $(\tilde x_1,\tilde x_2) = (\tilde b,50)$, and $(-\tilde b,50)$ with the impact
parameter $\tilde b = 2.5$. We show the snap shots with $\tilde t = 50 \sim 150$ with interval $\delta\tilde t = 10$.
The panel at the right-bottom corner is a sequence photograph. 
The plot region is $\tilde x_{1,2} \in [-40,40]$.}
\label{fig:ss_KP_set06_a_ex01_umum}
\end{center}
\end{figure*}

The final process for the $\bar{\rm u}{\rm u}$-$\bar{\rm u}{\rm u}$ collisions are scatterings
with impact parameters. 
The initial mesons are placed at $(\tilde x_1,\tilde x_2) = (\pm\tilde b,50)$.
First, we horizontally shift the initial mesons of Fig.~\ref{fig:ss_KP_set06_ex01_umum} by $\tilde b = 2.5$ 
as shown in Fig.~\ref{fig:ss_KP_set06_a_ex01_umum}. The scattering goes almost similarly to 
the head-on collision shown in Fig.~\ref{fig:ss_KP_set06_ex01_umum} except for a scattering angle. 
As can be seen in the panel of $\tilde t = 100$ of Fig.~\ref{fig:ss_KP_set06_a_ex01_umum}, the new mesons
created after the recombination are not vertical but diagonal.

\begin{figure*}[h]
\begin{center}
\includegraphics[width=0.95\hsize]{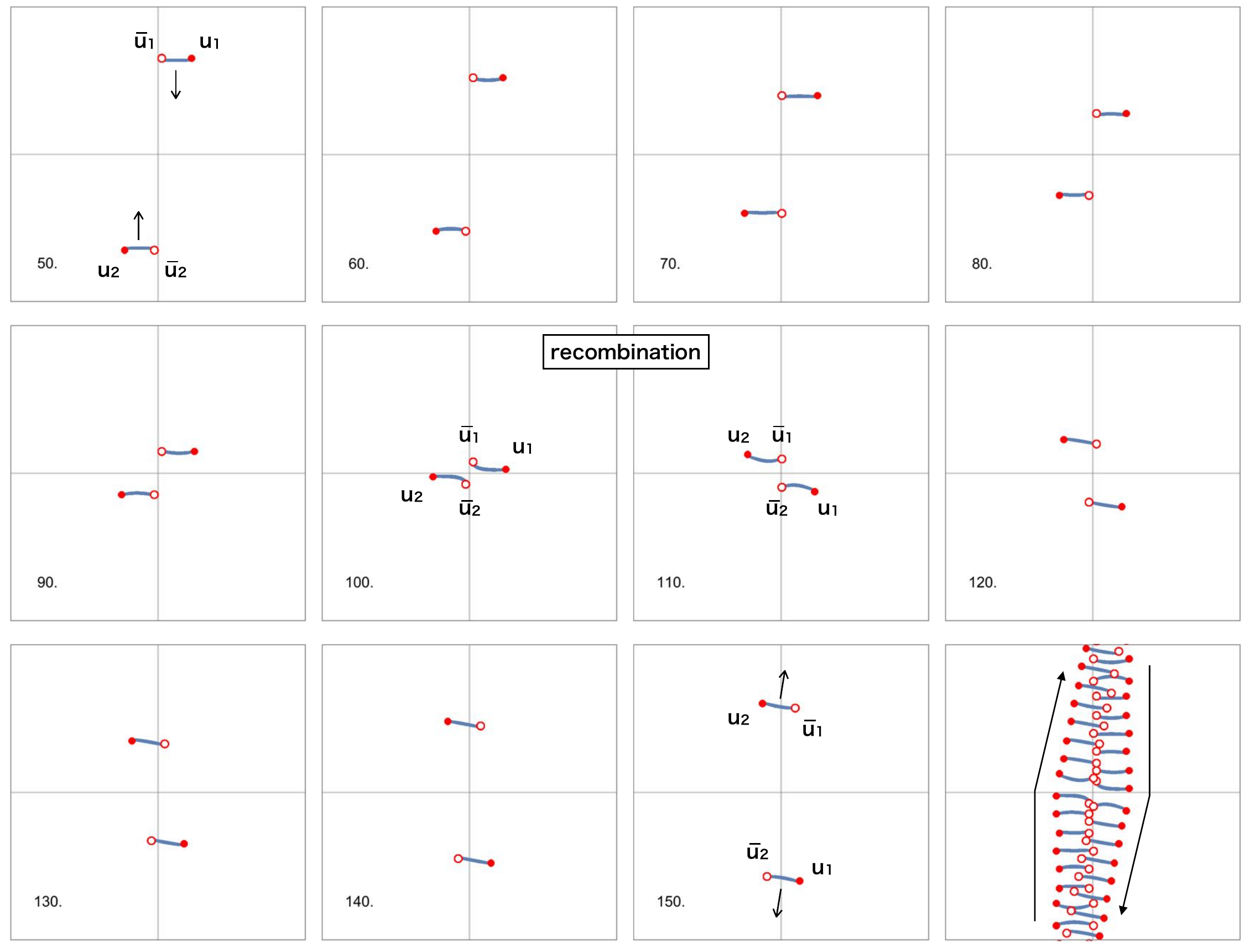}
\caption{Meson-meson scattering ($\bar{\rm u}{\rm u}$-$\bar{\rm u}{\rm u}$) with an impact parameter. Initially,
the two mesons are placed at $(\tilde x_1,\tilde x_2) = (\pm \tilde b,50)$ with the impact
parameter $\tilde b = 5$. We show the snapshots with $\tilde t = 50 \sim 150$ with interval $\delta\tilde t = 10$.
The panel at the right-bottom corner is a sequence photograph.
The plot region is $\tilde x_{1,2} \in [-40,40]$.}
\label{fig:ss_KP_set06_b_ex01_umum}
\end{center}
\end{figure*}

As is naturally expected, the scattering angle gets smaller as the impact parameter becomes larger.
Fig.~\ref{fig:ss_KP_set06_b_ex01_umum} shows a scattering with impact parameter $\tilde b = 5$. 
Namely, the two mesons are not initially
overlapped horizontally. Nevertheless, the recombination occurs during the collision, and the mesons collide with
a negative scattering angle.

\begin{figure*}[h]
\begin{center}
\includegraphics[width=0.95\hsize]{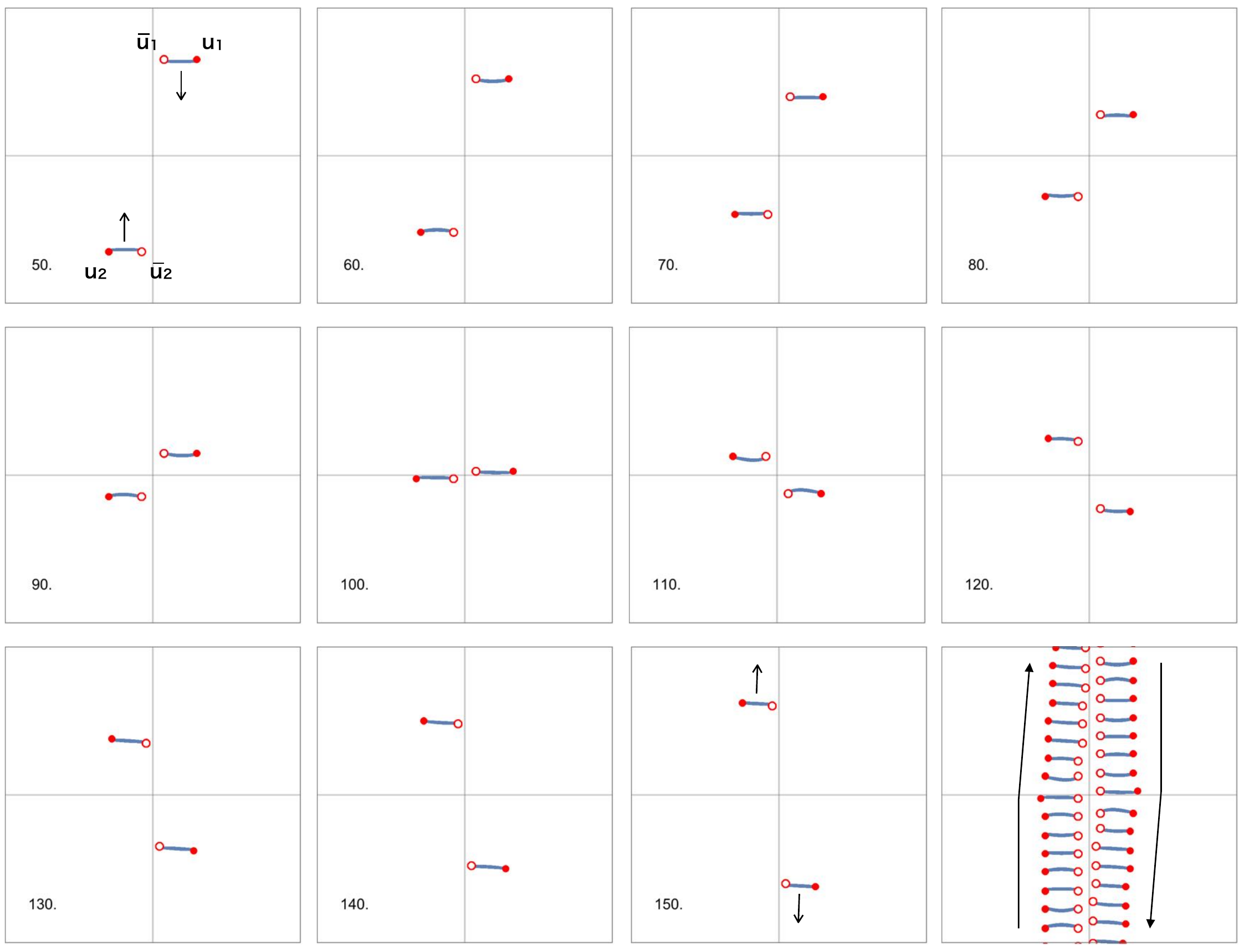}
\caption{Meson-meson scattering ($\bar{\rm u}{\rm u}$-$\bar{\rm u}{\rm u}$) with an impact parameter. Initially,
the two mesons are placed at $(\tilde x_1,\tilde x_2) = (\pm \tilde b,50)$ with the impact
parameter $\tilde b = 7.5$. We show the snapshots with $\tilde t = 50 \sim 150$ with interval $\delta\tilde t = 10$.
The panel at the right-bottom corner is a sequence photograph.
The plot region is $\tilde x_{1,2} \in [-40,40]$.}
\label{fig:ss_KP_set06_c_ex01_umum}
\end{center}
\end{figure*}

\begin{figure*}[h]
\begin{center}
\includegraphics[width=0.35\hsize]{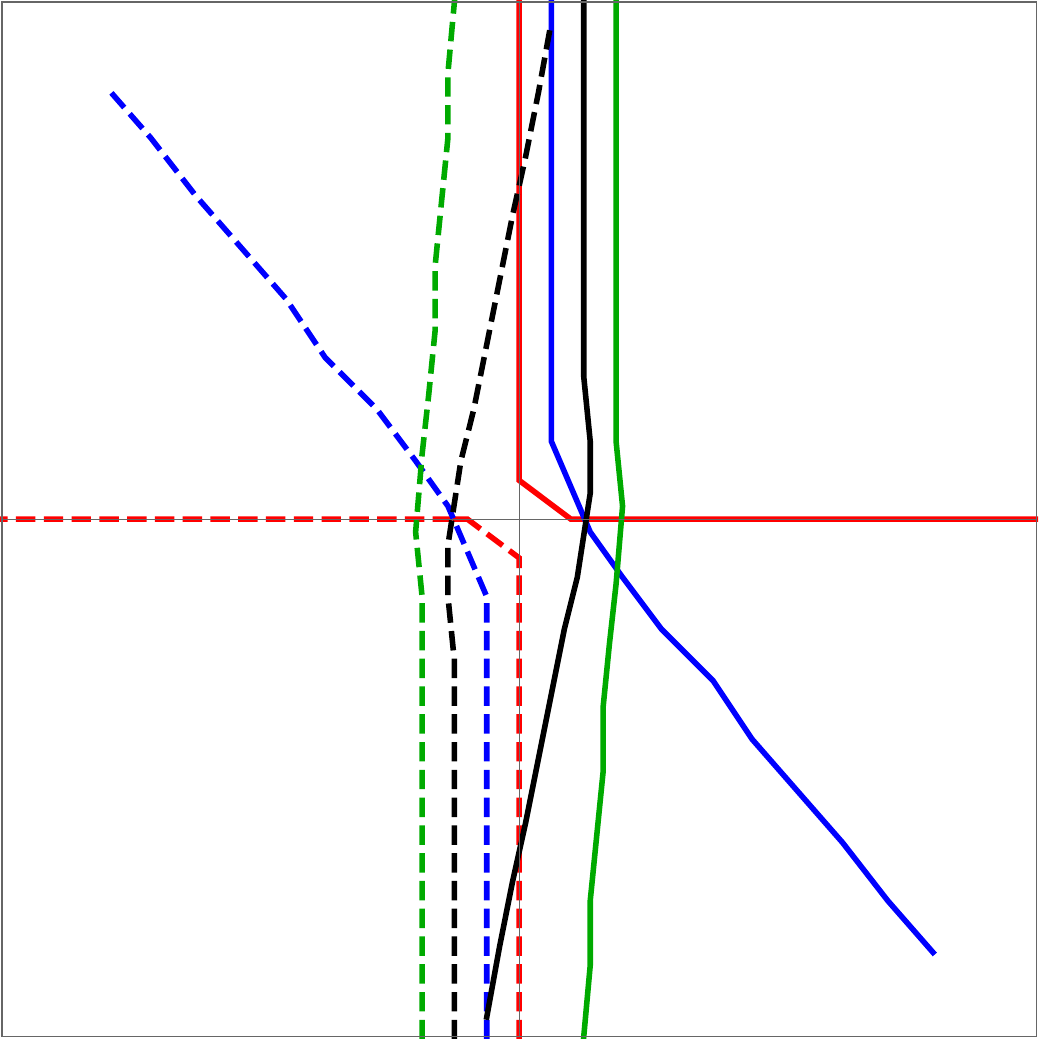}
\caption{The orbits of centers of the mesons during the meson-meson scatterings with the impact parameters
$\tilde b = 0, 2.5, 5$, and $7.5$ corresponding to Figs.~\ref{fig:ss_KP_set06_ex01_umum},
\ref{fig:ss_KP_set06_a_ex01_umum}, \ref{fig:ss_KP_set06_b_ex01_umum}, and \ref{fig:ss_KP_set06_c_ex01_umum}.
The plotted region is $\tilde x_i \in [-40,40]$.}
\label{fig:mm_impact_parameter}
\end{center}
\end{figure*}

If we put the initial mesons with a larger impact parameter, they pass through almost without interactions.
Fig.~\ref{fig:ss_KP_set06_c_ex01_umum} shows the scattering with the impact parameter $\tilde b = 7.5$.
The mesons go almost straight with a negative but small scattering angle. This suggests that the asymptotic meson-meson
interaction is attractive. We summarize the scatterings with impact parameters $\tilde b = 0, 2.5, 5$ and $7.5$
in Fig.~\ref{fig:mm_impact_parameter}.

\clearpage

\section{Meson-meson scattering: the case of $\bar{\text{u}}\text{u}$-$\bar{\text{d}}\text{d}$}
\label{sec:meson_meson_2}

\subsection{$\bar{\text{u}}\text{u}$-$\bar{\text{d}}\text{d}$ head-on collision at 0 angle}

\begin{figure*}[h]
\begin{center}
\includegraphics[width=0.95\hsize]{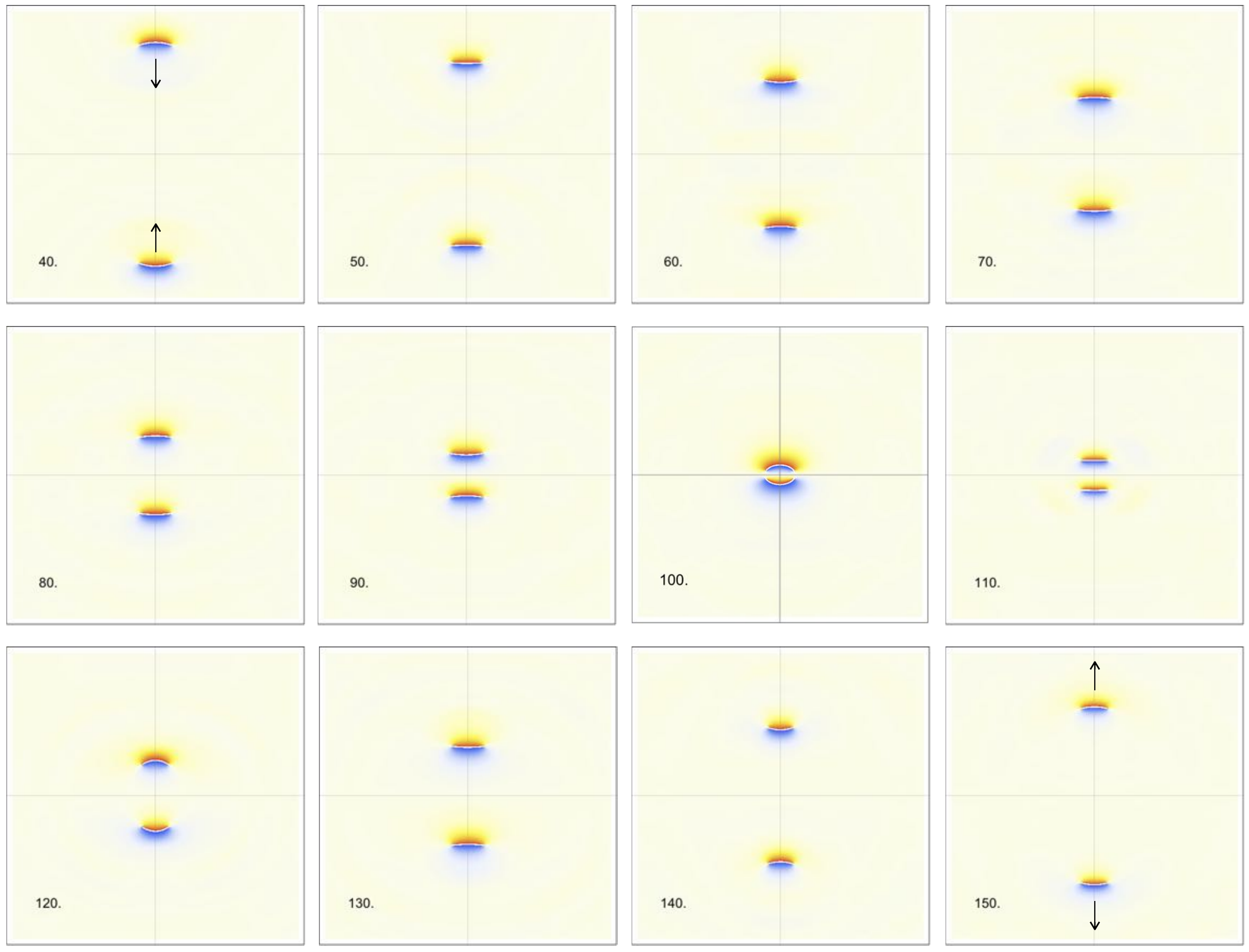}
\caption{The u and d meson scattering: 
Color density plots of the relative phase $\arg(\Psi_1) - \arg(\Psi_2)$. 
We initially $(\tilde t = 0)$ set the u meson at $(\tilde x_1,\tilde x_2) = (0,50)$,
and the d meson at $(\tilde x_1,\tilde x_2) = (0,-50)$.
They straightly run with almost constant speeds and pass through each other without an interaction.
We show the snapshots from $\tilde t =40$ to $150$ with interval $\delta \tilde t = 10$, and the plot region is
$\tilde x_{1,2} \in [-40,40]$.}
\label{fig:ss_DP_set06_ex01_umdm}
\end{center}
\end{figure*}

\begin{figure*}
\begin{center}
\includegraphics[width=0.95\hsize]{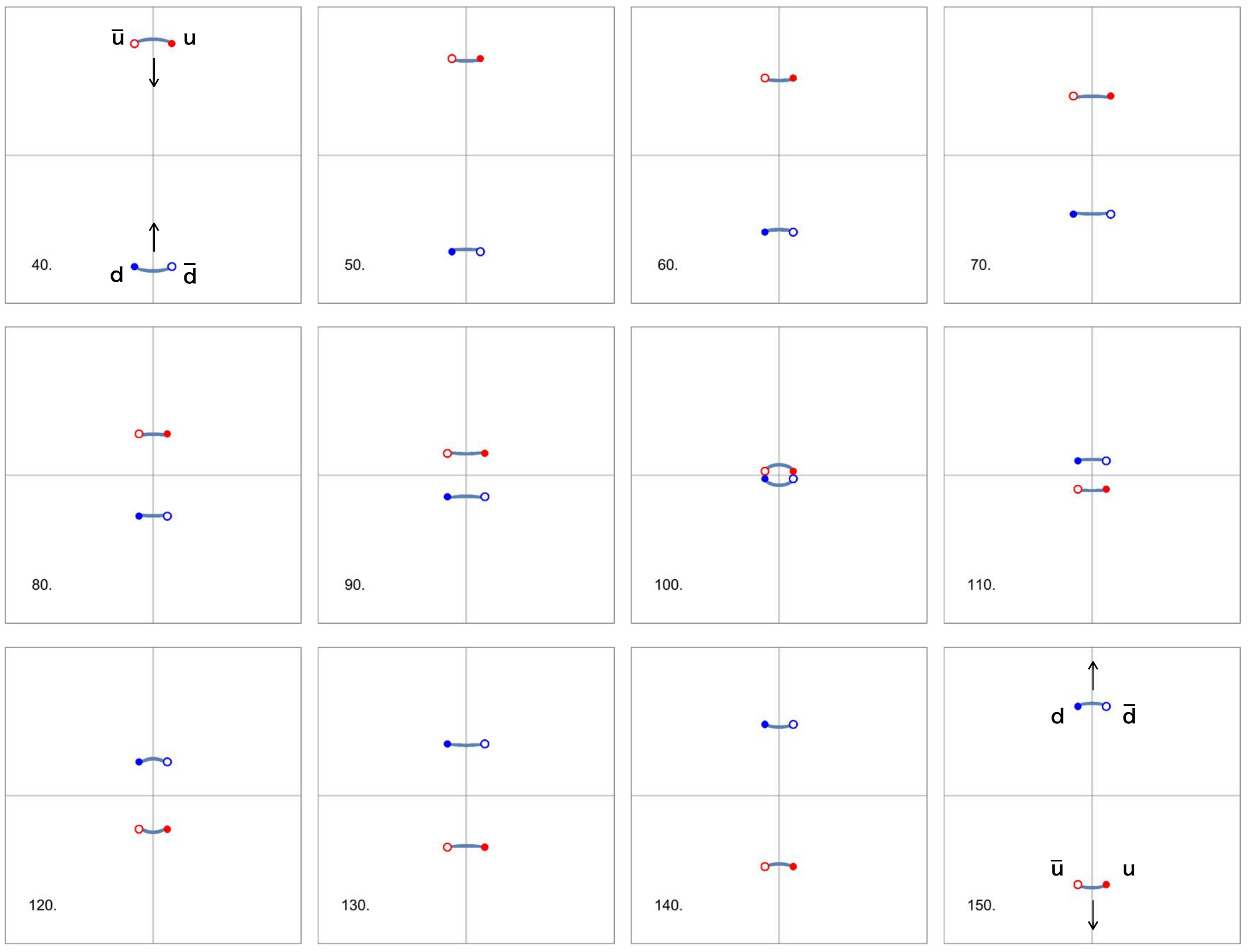}
\caption{A simplified plot of the u meson and d meson scattering shown in Fig.~\ref{fig:ss_DP_set06_ex01_umdm}.
We put painted red disks (unpainted red circle) at the points corresponding to the u ($\bar{\rm u}$) vortex centers.
Similarly, we put painted blue disks (unpainted blue circle) at the points corresponding to the d ($\bar{\rm d}$) vortex centers.
Gray regions bridging the u and $\bar{\rm u}$ vortices (d and $\bar{\rm d}$) show the SG solitons. 
The gray regions are those where the relative phases take the values within $3 \le |\arg(\Psi_1) - \arg(\Psi_2)| \le \pi$.}
\label{fig:ss_KP_set06_ex01_umdm}
\end{center}
\end{figure*}

Let us next study scatterings of a u meson and a d meson. We first collide the two mesons head-on.
The initial configuration is prepared by superposing a u meson at $(\tilde x_1,\tilde x_2) = (0,50)$
and the d meson at $(\tilde x_1,\tilde x_2) = (0,-50)$ 
[precisely speaking, we put a u vortex at $(\tilde x_1,\tilde x_2) = (5,50)$ and
a $\bar{\rm u}$ vortex at $(\tilde x_1,\tilde x_2) = (-5,50)$,
and a d vortex at $(\tilde x_1,\tilde x_2) = (-5,-50)$ and
a $\bar{\rm d}$ vortex at $(\tilde x_1,\tilde x_2) = (5,-50)$].
The outlook of the scattering is shown in Fig.~\ref{fig:ss_DP_set06_ex01_umdm}, which one might
think is not so interesting
compared to the $\bar{\rm u}{\rm u}$-$\bar{\rm u}{\rm u}$ head-on collision 
given in Fig.~\ref{fig:ss_DP_set06_ex01_umum}.
Indeed, the two mesons seem to pass through without 
an interaction, see Figs.~\ref{fig:ss_DP_set06_ex01_umdm} and \ref{fig:ss_KP_set06_ex01_umdm}.
This observation is true if we look at only the vortices. However, it is not true for the SG solitons.
As can be seen from Fig.~\ref{fig:ss_DP_set06_ex01_umdm}, this scattering can be understood as a collision of
one SG soliton (belonging to the u meson) and the other SG soliton (belonging to the d meson). 
This can be seen by noting that 
the color orders along the $\tilde x_2$-axis are the same for the u and d mesons.
This situation is in contract to  the meson-meson scattering
in Fig.~\ref{fig:ss_DP_set06_ex01_umum} in which 
the same spices involve the SG and anti-SG solitons.

\begin{figure*}
\begin{center}
\includegraphics[width=0.75\hsize]{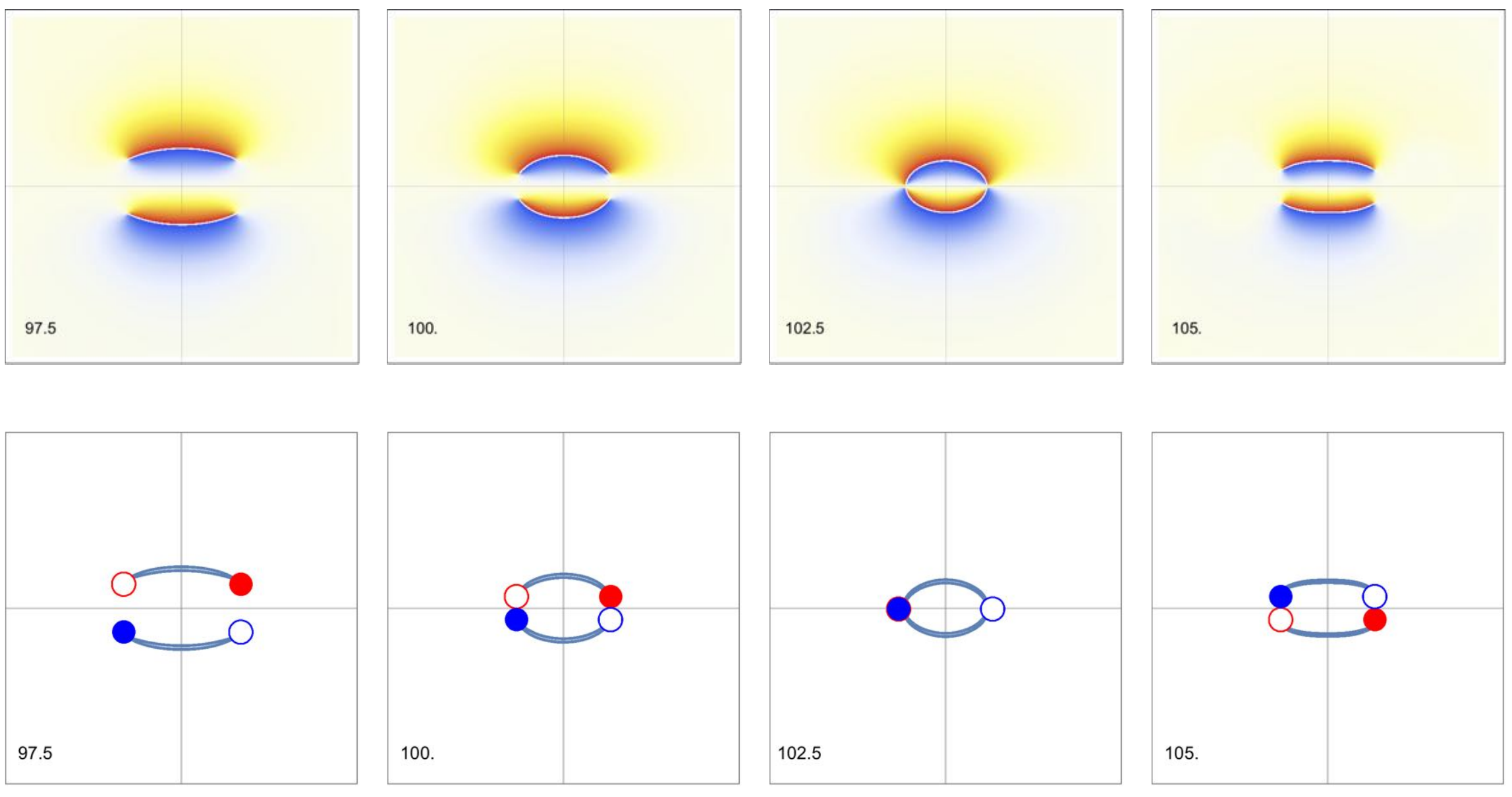}
\caption{Snapshots in close-up for $\tilde x_i \in [-15,15]$ at $\tilde t = 97.5, 100, 102.5, 105$ 
for the u meson and d meson scattering given in Figs.~\ref{fig:ss_DP_set06_ex01_umdm} and \ref{fig:ss_KP_set06_ex01_umdm}.
The SG solitons from the top and bottom gradually bend as they are close by and backscatter, whereas
the vortices at the ends of the mesons pass through each other. 
Namely, the mesons interchange the SG solitons during the collision.}
\label{fig:ss_DPKP_set06_ex01_umdm}
\end{center}
\end{figure*}

\begin{figure*}
\begin{center}
\includegraphics[width=0.95\hsize]{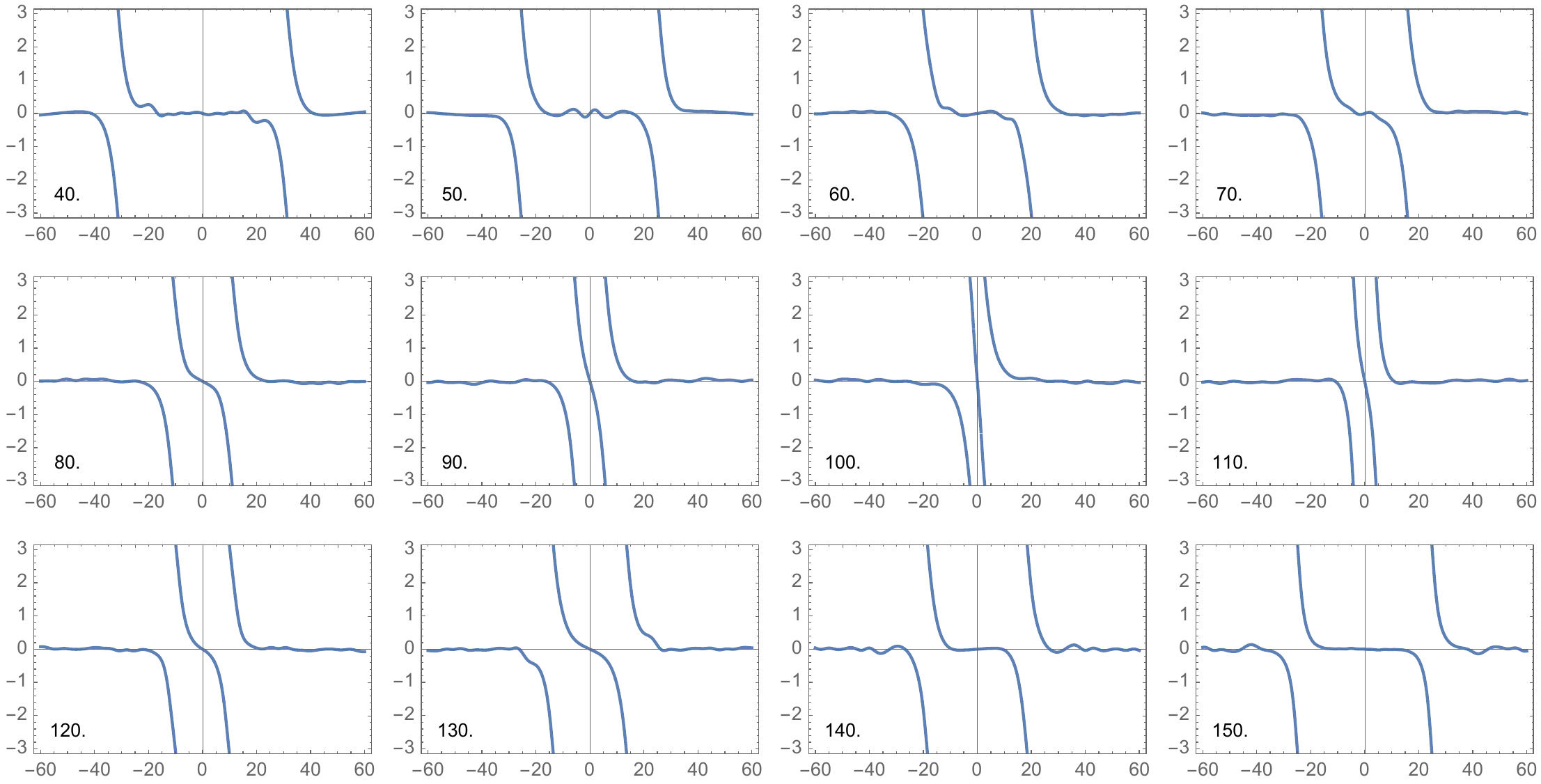}
\caption{The solid blue curve shows the relative phase $\arg(\Psi_1) - \arg(\Psi_2)$ along the $\tilde x_2$-axis
of Fig.~\ref{fig:ss_DP_set06_ex01_umdm}. The SG solitons coming from the right and left-hand sides collide and scatter 
backward. Throughout the scattering, the SG winding number $(=2)$ is preserved.}
\label{fig:SG_set06_ex01_umdm}
\end{center}
\end{figure*}

To see a nontrivial phenomenon in this scattering, let us carefully observe the period of passing.
Fig.~\ref{fig:ss_DPKP_set06_ex01_umdm} shows snapshots in close-up at $\tilde t = 97.5, 100, 102.5, 105$. 
While the u and $\bar{\rm d}$ (d and $\bar{\rm u}$) vortices pass through each other, the SG solitons largely bend toward the opposite
direction compared to Fig.~\ref{fig:ss_DPKP_set06_ex01_umum} due to repulsive interaction between the SG solitons.
Thus, distance between two SG solitons at the center of mesons never vanish,
and the SG solitons backscatter. 
Fig.~\ref{fig:SG_set06_ex01_umdm} shows the relative phase 
$\arg(\Psi_1) - \arg(\Psi_2)$ on the $\tilde x_2$ axis.
Throughout the scattering, there always exist two SG solitons (jumps from $-\pi\to\pi$). Since they are topologically
protected, they cannot be annihilated. This is the reason why their distance does not vanish and they backscatter.

Thus, we found that the SG solitons backscatter whereas the vortices at the ends of mesons pass through.
Along with this observation, we conclude that the mesons interchange the SG solitons before and after the
head-on collision.

\clearpage

\subsection{$\bar{\text{u}}\text{u}$-$\bar{\text{d}}\text{d}$ 
scattering at $\pi/8$ angle}

The collision of the u and d mesons studied in the previous section is not so interesting in
the sense that it does not include recombination phenomena which commonly occurs for the u and u meson
scatterings. 
To see if the recombination is peculiar to the u and u meson scattering only or it is general for
most scatterings, let us investigate the collision of the u and d mesons with angles here and in
the next subsection.

\begin{figure*}[h]
\begin{center}
\includegraphics[width=0.95\hsize]{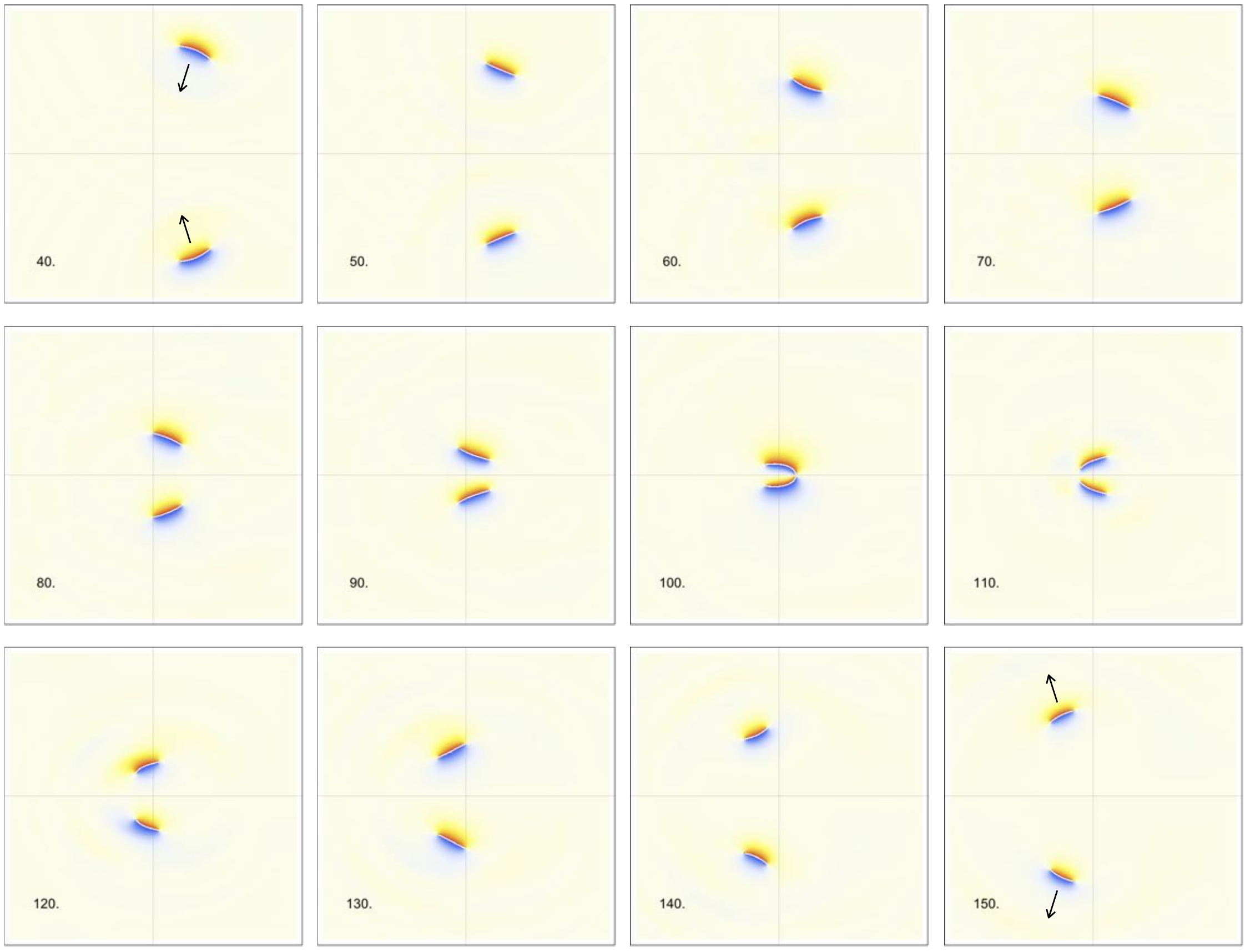}
\caption{Scattering of the slightly tilted u and d mesons with angle $\pi/8$.
We initially $(\tilde t = 0)$ set the u meson at 
$(\tilde x_1,\tilde x_2) = (50 \sin\pi/8,50 \cos\pi/8)$,
and the d meson at $(\tilde x_1,\tilde x_2) = (50\sin\pi/8,-50 \cos\pi/8)$.
We only show the snapshots from $\tilde t =40$ to $150$ with interval $\delta \tilde t = 10$, and the plot region is
$\tilde x_{1,2} \in [-40,40]$.}
\label{fig:ss_DP_set04_ex01_umdm}
\end{center}
\end{figure*}

\begin{figure*}
\begin{center}
\includegraphics[width=0.95\hsize]{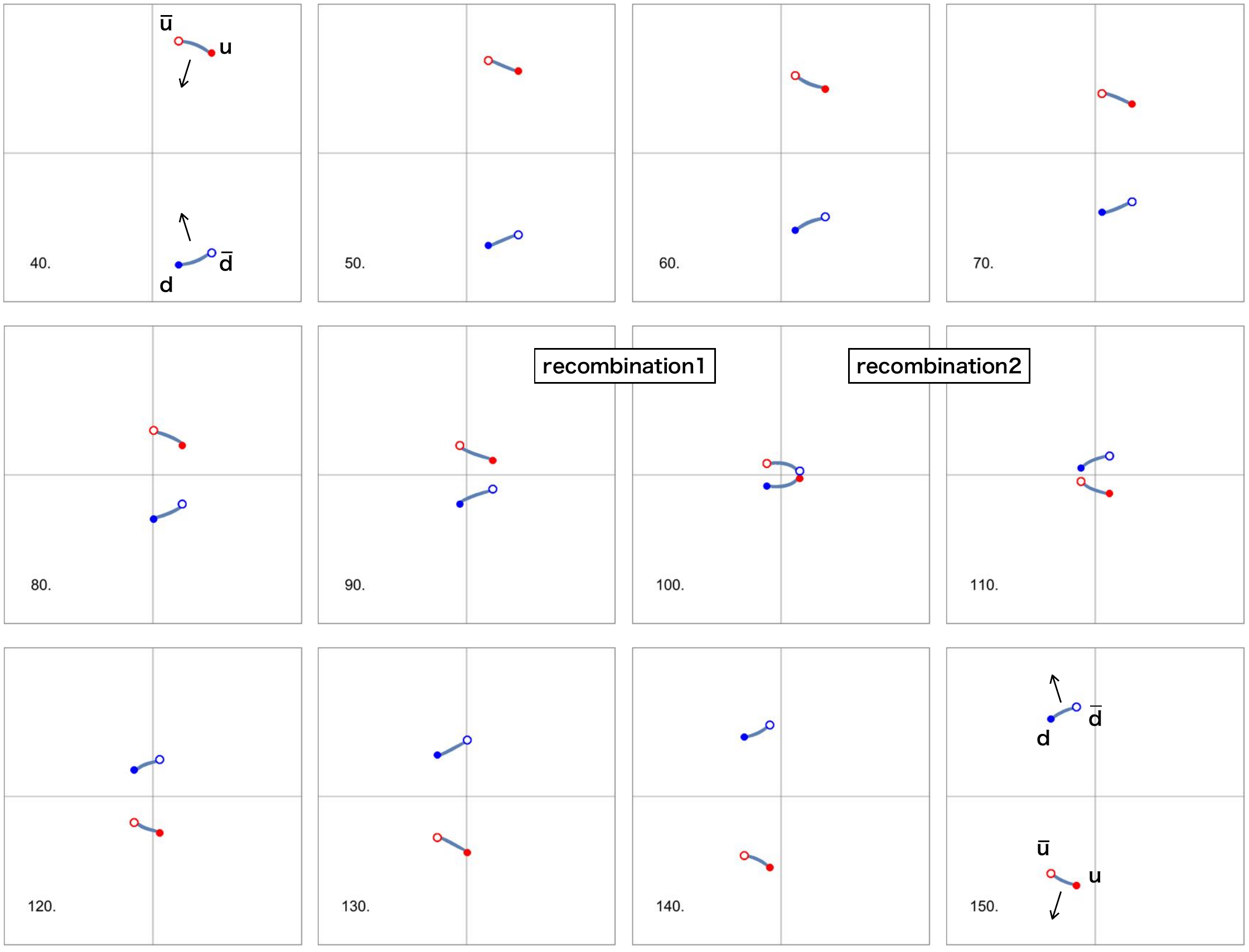}
\caption{A simplified plot of the u mesons scattering shown in Fig.~\ref{fig:ss_DP_set04_ex01_umdm}.}
\label{fig:ss_KP_set04_ex01_umdm}
\end{center}
\end{figure*}

Fig.~\ref{fig:ss_DP_set04_ex01_umdm} shows the outlook of the head-on collision of the u and d mesons with
angle $\pi/8$. Compared to Fig.~\ref{fig:s_DP_set04_ex01_umum}, the lower $\bar{\rm u}{\rm u}$ meson is replaced 
by $\bar{\rm d}{\rm d}$. 
At a first glance, the mesons seem to mere pass through each other, 
just as the previous scattering in Fig.~\ref{fig:ss_DP_set06_ex01_umdm}.
However, it is not such trivial. Since the mesons are tilted, there is a certain period that 
the u-vortex at the lower edge of the u meson and
the $\bar{\rm d}$-vortex at the upper edge of the d meson exchange their positions, while the remaining constituent vortices
are left to be unchanged, see the snapshot at $\tilde t = 100$ of Fig.~\ref{fig:ss_KP_set04_ex01_umdm}.
Soon after, the second passing of the remaining $\bar{\rm u}$ and d vortices follows, see 
the snapshot at $\tilde t = 110$ of Fig.~\ref{fig:ss_KP_set04_ex01_umdm}.
The interchange of constituent vortices is accompanied by the recombination of the SG solitons. The period between
the first and the second recombinations, the molecules are not u and d mesons but the baryon ud and anti-baryon
$\bar{\rm u}\bar{\rm d}$. Indeed, the SG solitons bridge the u and d vortices ($\bar{\rm u}$ and $\bar{\rm d}$ vortices)
in the snapshot at $\tilde t = 100$ of Fig.~\ref{fig:ss_KP_set04_ex01_umdm}.

\begin{figure*}
\begin{center}
\includegraphics[width=0.9\hsize]{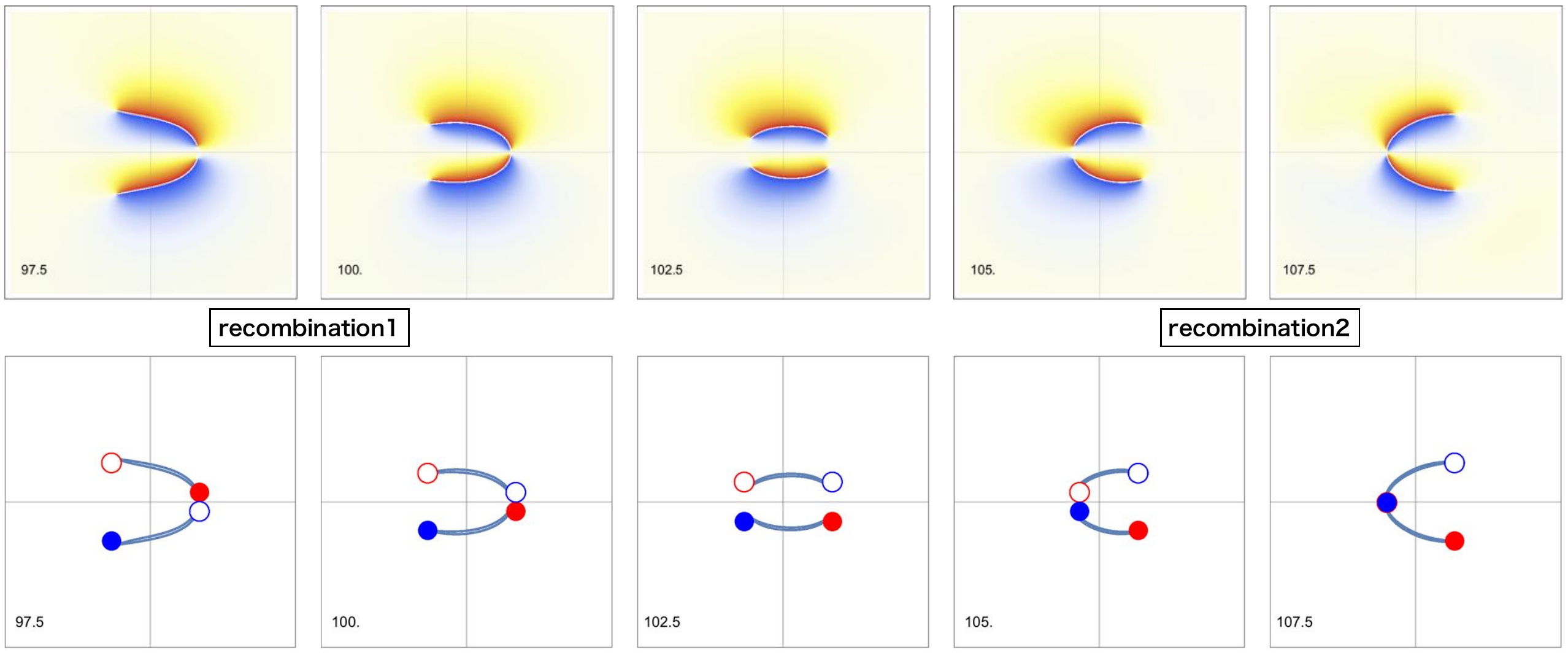}
\caption{Snapshots in close-up for $\tilde x_i \in [-15,15]$ at $\tilde t = 97.5, 100, 102.5, 105, 107.5$ 
for the u meson and d meson scattering given in Figs.~\ref{fig:ss_DP_set04_ex01_umdm} and \ref{fig:ss_KP_set04_ex01_umdm}.
The mesons are converted into the baryon and anti-baryon pair for a short period.}
\label{fig:ss_DPKP_set04_ex01_umdm}
\end{center}
\end{figure*}

Thus, the simulation here implies that the recombination can  happen 
universally not only for the mesons of the same spices but also for the different spices. We have not observed the recombination
in Fig.~\ref{fig:ss_DP_set06_ex01_umdm}, just because the two mesons are finely tuned to be exactly parallel.

Let us cut out the period with created baryons 
and see their motion in more details.
Fig.~\ref{fig:ss_DPKP_set04_ex01_umdm} shows 
snapshots in close-up for $\tilde x_i \in [-15,15]$ at $\tilde t = 97.5, 100, 102.5, 105$, and $107.5$.
The baryon (anti-baryon) formed after the first recombination rotates counterclockwise (clockwise) until
they are reformulated back into the mesons. During the period with baryons, they almost do not run. As a consequence,
the meson orbits before and after the collision slightly shift due to delay of forming baryons.

\subsection{$\bar{\text{u}}\text{u}$-$\bar{\text{d}}\text{d}$ 
scattering at $\pi/4$ angle}

\begin{figure*}[h]
\begin{center}
\includegraphics[width=0.95\hsize]{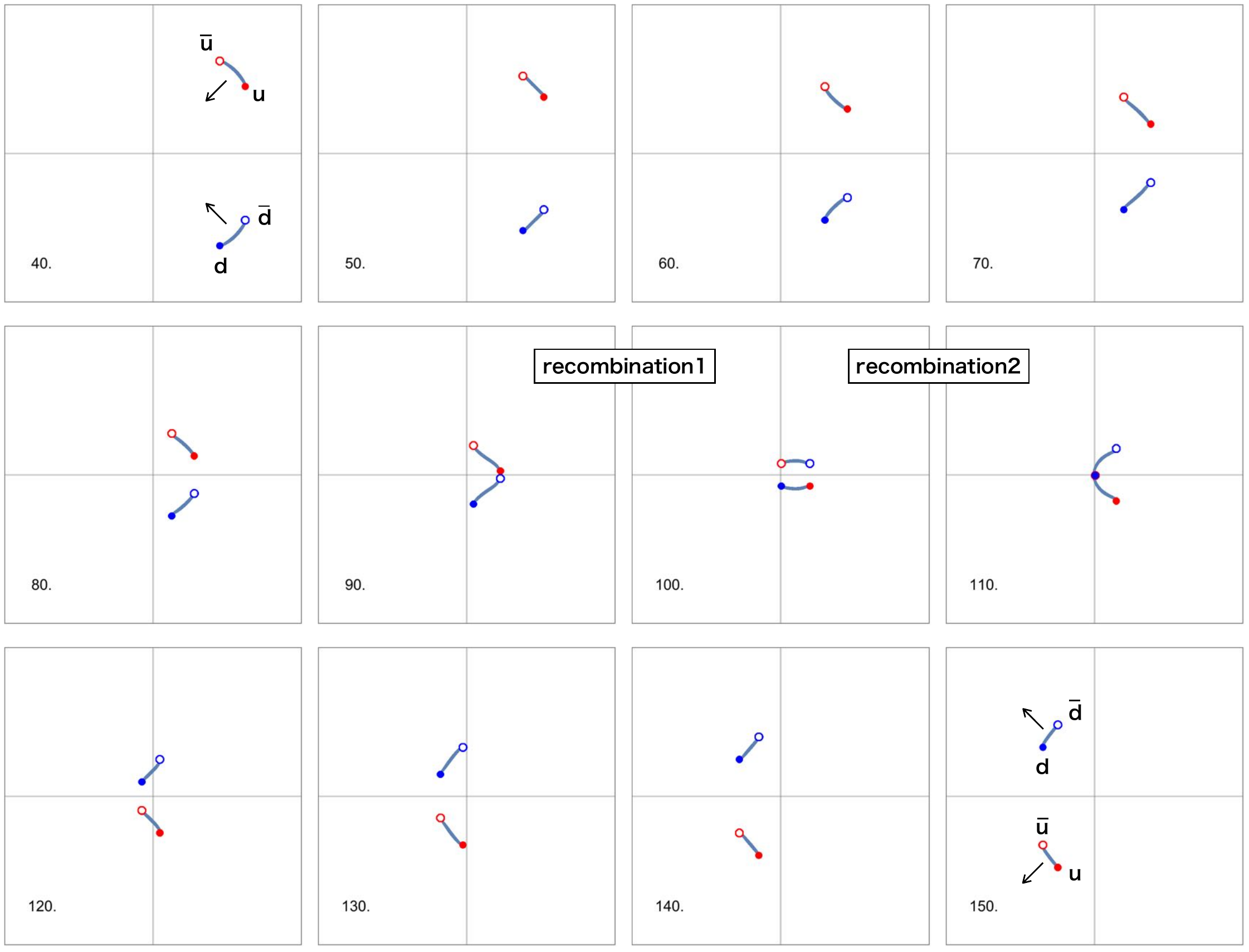}
\caption{Scattering of the tilted u and d mesons with angle $\pi/4$.
We initially $(\tilde t = 0)$ set the u meson at 
$(\tilde x_1,\tilde x_2) = (50 \sin\pi/4,50 \cos\pi/4)$,
and the d meson at $(\tilde x_1,\tilde x_2) = (50\sin\pi/4,-50 \cos\pi/4)$.
We only show the snapshots from $\tilde t =40$ to $150$ with interval $\delta \tilde t = 10$, and the plot region is
$\tilde x_{1,2} \in [-40,40]$.}
\label{fig:ss_KP_set03_ex01_umdm}
\end{center}
\end{figure*}

The observations in the previous subsection can be more sharply seen in the scattering of
the u and d mesons with larger angle; we take $\pi/4$ here as an example.
Fig.~\ref{fig:ss_KP_set03_ex01_umdm} shows an outlook of the scattering process.
Qualitatively, it goes very similarly to that with the angle $\pi/8$ given in Fig.~\ref{fig:ss_KP_set04_ex01_umdm}.
Namely, the incoming u and d mesons are converted to the intermediate baryon and anti-baryon pair at the
first recombination. The baryons rotate at their position without run for a while. Then they are reformed back into the mesons
at the second recombination and run straight toward the boundary. 
\begin{figure*}
\begin{center}
\includegraphics[width=0.75\hsize]{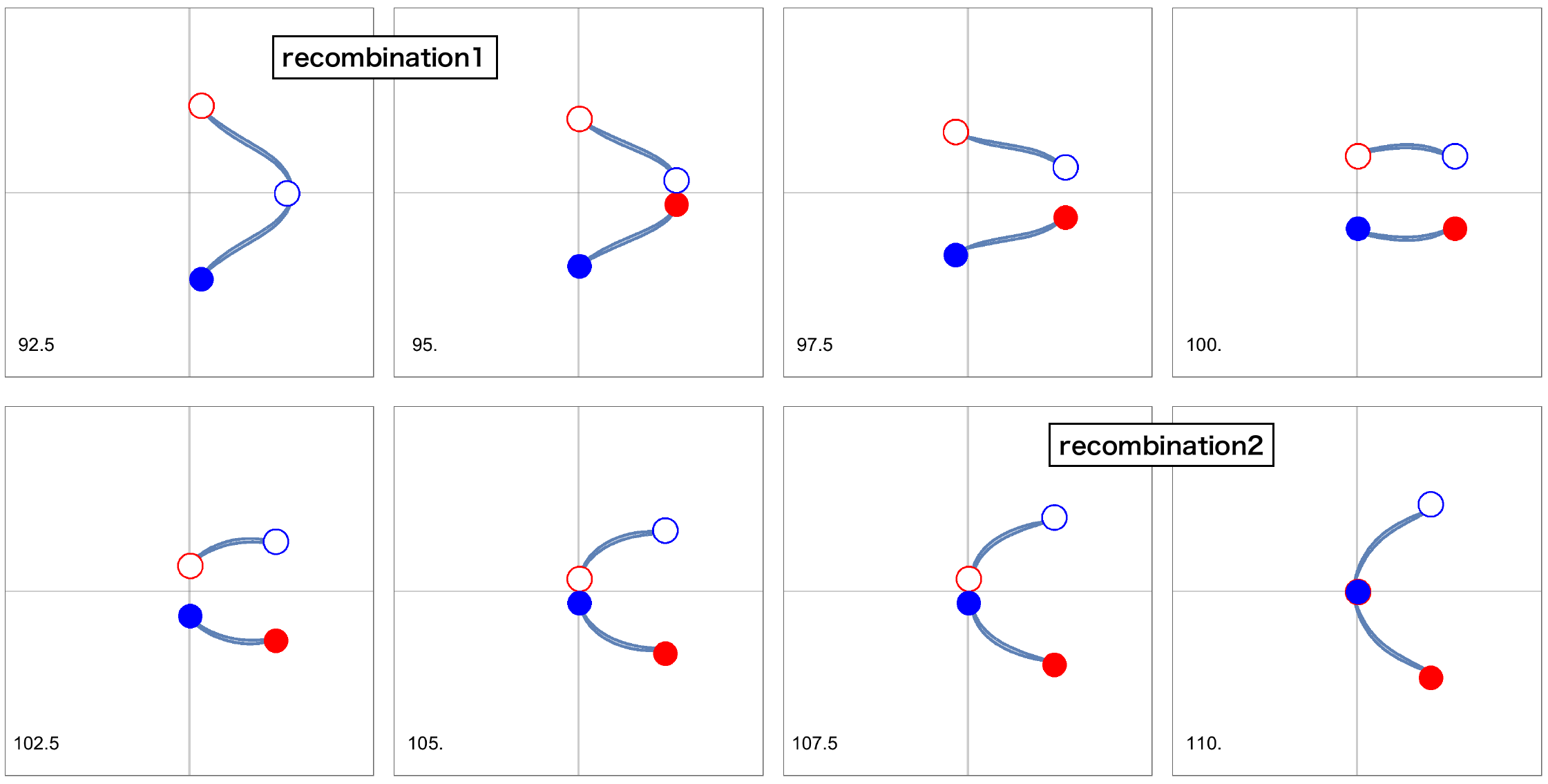}
\caption{Snapshots in close-up for $\tilde x_i \in [-15,15]$ at $\tilde t = 92.5, 95, 97.5, 100, 102.5, 105, 107.5, 110$ 
for the u meson and d meson scattering given in Fig.~\ref{fig:ss_KP_set03_ex01_umdm}.}
\label{fig:DPKP_set03_ex01_umdm}
\end{center}
\end{figure*}
The baryonic period in this scattering is longer than the previous one because of the steeper angle of the incoming mesons.
Comparing Figs.~\ref{fig:ss_DPKP_set04_ex01_umdm} and \ref{fig:DPKP_set03_ex01_umdm}, the baryonic period is
$\tilde t \simeq 97.5$-$105$ for the former while $\tilde t \simeq 95$-$107.5$ for the latter.  
A longer lifetime
of the baryonic period makes the rotation of the baryons
clearer to be seen.

\begin{figure*}[h]
\begin{center}
\includegraphics[width=0.55\hsize]{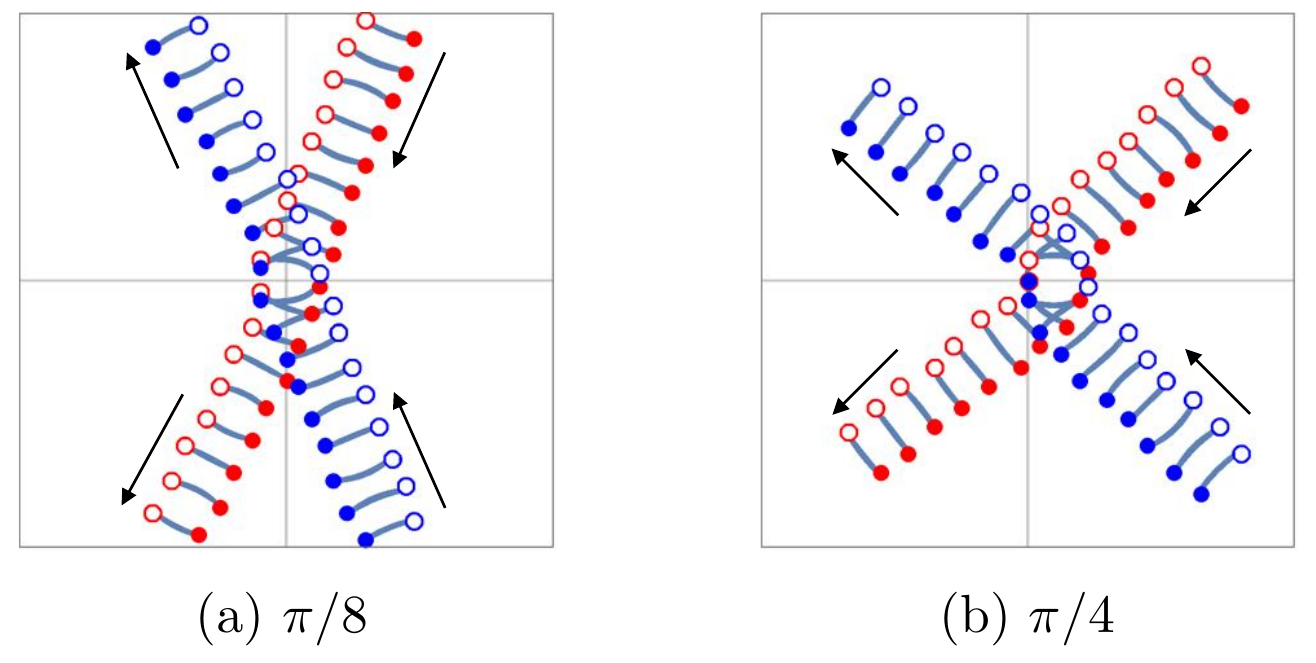}
\caption{Sequence photographs for the u and d meson 
scatterings with $\pi/8$ and $\pi/4$ angles given in Figs.~\ref{fig:ss_KP_set04_ex01_umdm}
and \ref{fig:ss_KP_set03_ex01_umdm}.}
\label{fig:delay}
\end{center}
\end{figure*}

As mentioned above, the conversion of mesons to baryon and anti-baryon
results in delay of the straight orbits of the incoming and outgoing mesons.
In order to see the delay clearer, we show sequence photographs for the u and d meson 
scatterings with $\pi/8$ and $\pi/4$ angles in Fig.~\ref{fig:delay}. One can see the orbits of
the constituent vortices steeply bend at the timing of recombinations.
Since vertical motions of the vortices in Fig.~\ref{fig:delay} are simple shift with almost constant speed, we can approximately
regard the vertical axis as the time. Then the horizontal motions would remind us a phase shift
which commonly appears in soliton scatterings 
or particle scattering. The phase shift in our system is due to the formation of
unstable intermediate baryonic states with finite life time. 
It is interesting that 
a phase shift common for scattering is observed here, 
although the mechanism is peculiar to our system and has never been seen elsewhere. 

\subsection{Interaction vertices and Feynman diagrams}

\begin{figure*}[h]
\begin{center}
\includegraphics[width=0.7\hsize]{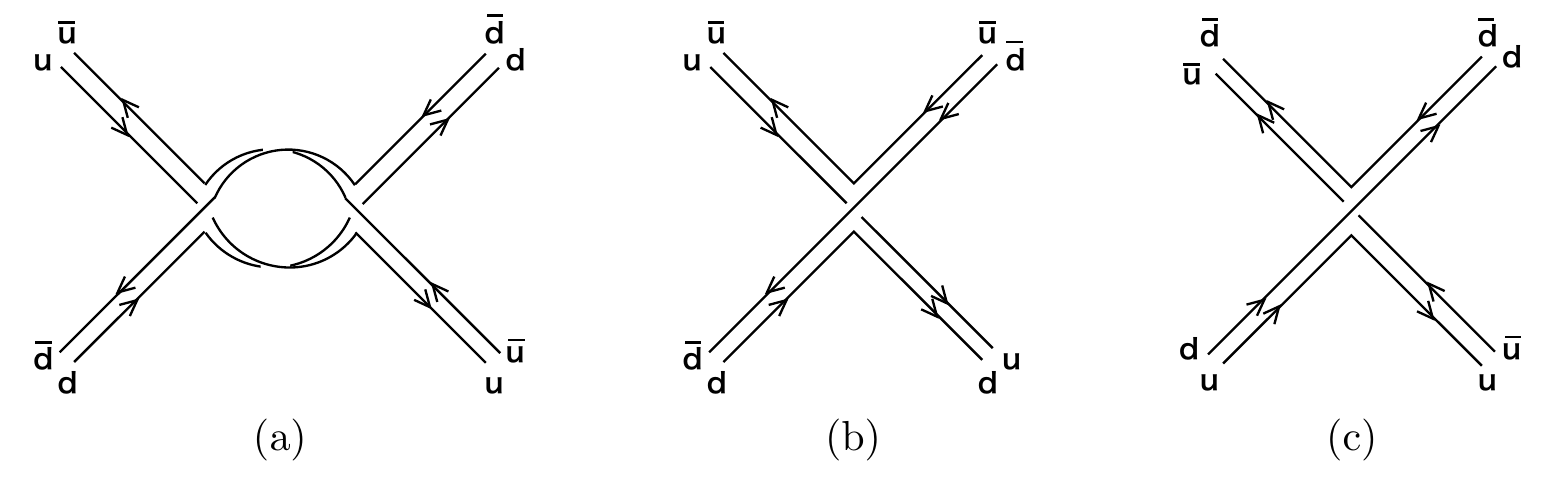}
\caption{(a) the Feynman diagram corresponding to the real processes in Fig.~\ref{fig:ss_KP_set03_ex01_umdm}. (b) and (c)
shows the meson-meson-baryon-baryon vertices which are building blocks of the twig diagram (a).}
\label{fig:1loop_umdm}
\end{center}
\end{figure*}

As in the u and u mesons scattering, we can describe the u and d mesons scattering by the Feynman diagram.
The diagram for Fig.~\ref{fig:delay} is a 1-loop diagram given in Fig.~\ref{fig:1loop_umdm} (a).
It includes meson-meson-baryon-baryon vertices which are shown in Fig.~\ref{fig:1loop_umdm} (b) and (c).
They are invariant under the $F$ and $P$ transformations while
they are exchanged by the $T$ transformation. Hence, no more diagrams are generated by
the symmetries. 

\begin{figure*}[h]
\begin{center}
\includegraphics[width=0.85\hsize]{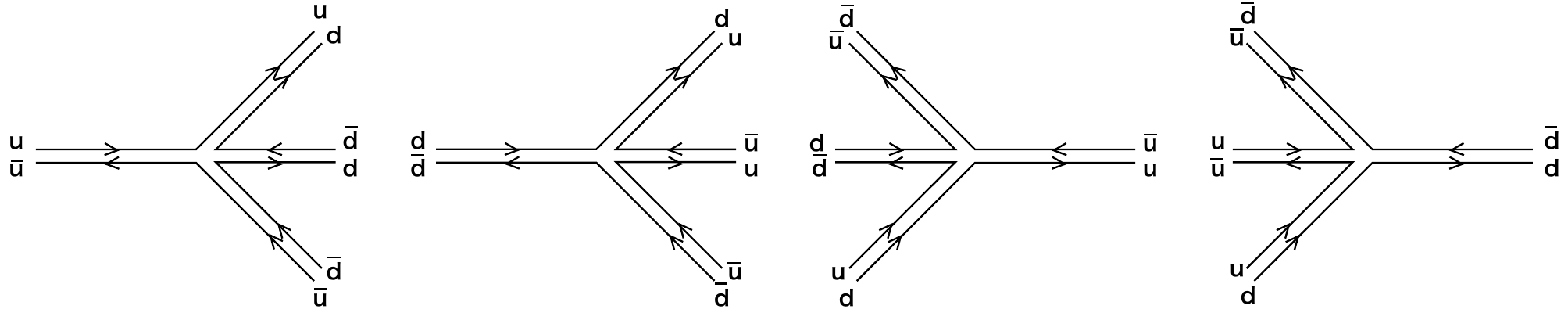}
\caption{The meson-meson-baryon-baryon vertices correspond to and relate to a long meson disintegration into
a baryon, an anti baryon, and a meson.}
\label{fig:meson_disintegration}
\end{center}
\end{figure*}

\begin{figure*}[h]
\begin{center}
\includegraphics[width=0.85\hsize]{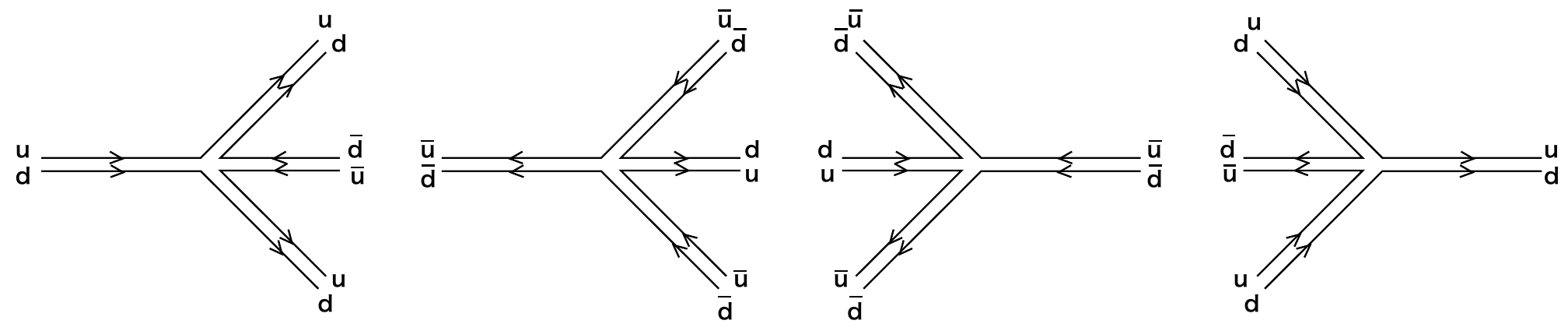}
\caption{The baryon-baryon-baryon-baryon vertices correspond to and relate to a long baryon disintegration into
two baryons and one anti-baryon.}
\label{fig:baryon_disintegration}
\end{center}
\end{figure*}

\begin{figure*}[h]
\begin{center}
\includegraphics[width=0.55\hsize]{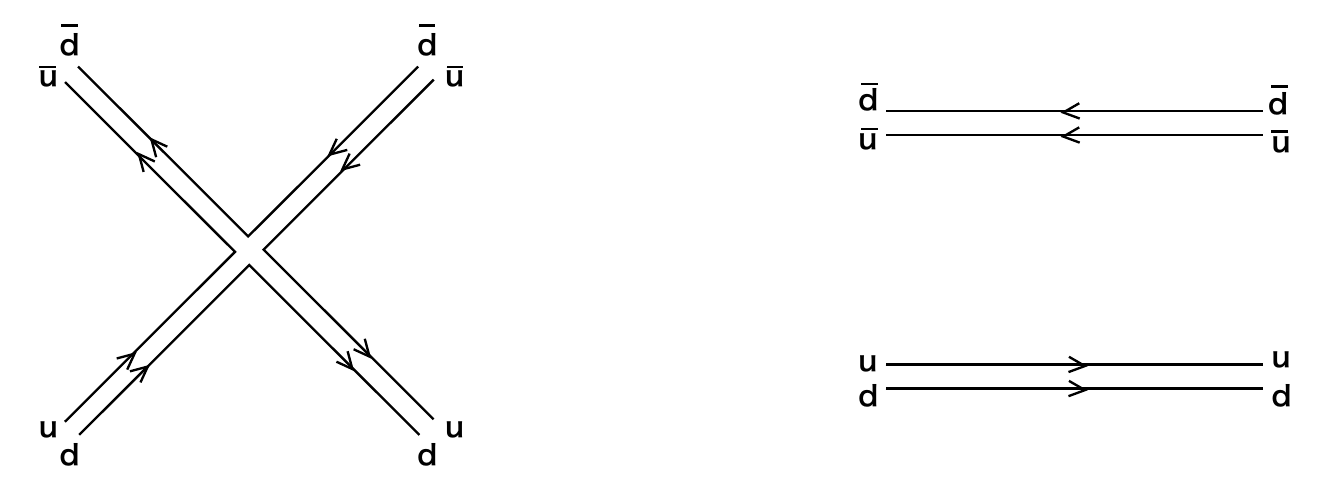}
\caption{The baryon-baryon-baryon-baryon vertices correspond to and relate to a long baryon disintegration into
two baryons and  an anti baryon.}
\label{fig:bbbb}
\end{center}
\end{figure*}

Related to these meson-meson-baryon-baryon vertices,
we observed in our previous work Ref.~\cite{Eto:2017rfr} that a long meson disintegrates into
three short molecules as
\beq
\bar{\rm u}{\rm u} \quad\to\quad
{\rm u}{\rm d} \quad+\quad
\bar{\rm d}{\rm d} \quad + \quad
\bar{\rm d}\bar{\rm u}.
\eeq
The Feynman diagrams corresponding to this and other related diagrams via the $F$, $T$, and $P$ symmetries
are given in Fig.~\ref{fig:meson_disintegration}. Interestingly, these diagrams can be generated by
horizontally flipping an external leg of the diagrams (b) and (c) of Fig.~\ref{fig:1loop_umdm}.
In a usual relativistic quantum field theory, the presence of the vertex (b) of Fig.~\ref{fig:1loop_umdm}
immediately means that the vertex of the left-most panel of Fig.~\ref{fig:meson_disintegration} also exists
as either a real or virtual process. However, our theory here is not relativistic, and moreover we are
dealing with only real processes which are solution of the GP equations. Therefore, for our system,
the vertex (b) of Fig.~\ref{fig:1loop_umdm}
does not automatically ensure the vertex of the left-most panel of Fig.~\ref{fig:meson_disintegration}.
We should emphasize that we put the four diagrams of Fig.~\ref{fig:meson_disintegration} in our list of the real processes
because we found them in real dynamics.
To make contrast clearer, let us mention another process 
which we previously encountered in Refs.~\cite{Tylutki:2016mgy,Eto:2017rfr}: 
\beq
{\rm u}{\rm d} \quad\to\quad
{\rm u}{\rm d} \quad+\quad
\bar{\rm d}\bar{\rm u} \quad + \quad
{\rm u}{\rm d}.
\eeq
The corresponding Feynman diagram and the other diagrams obtained via the $F$, $T$, and $P$ transformations 
are given in Fig.~\ref{fig:baryon_disintegration}.
By flipping the out-going external ud leg of the left-most diagram in Fig.~\ref{fig:baryon_disintegration}, we can generate  
the diagram shown in the left panel of Fig.~\ref{fig:bbbb}.
However, a real process corresponding to this diagram does not seem to happen. This is because 
a baryon ud and an anti baryon $\bar{\rm u}\bar{\rm d}$ 
are a pair of the integer vortex and anti-vortex, so that they move parallel with their distance being kept constant,
as can be seen in the right panel of Fig.~\ref{fig:bbbb}.
Hence, we have learned that we cannot freely flip any external legs of a vertex from left (right) to right (left).

\subsection{$\bar{\rm u}{\rm u}$-$\bar{\rm d}{\rm d}$ collisions with impact parameters}

\begin{figure*}[h]
\begin{center}
\includegraphics[width=0.95\hsize]{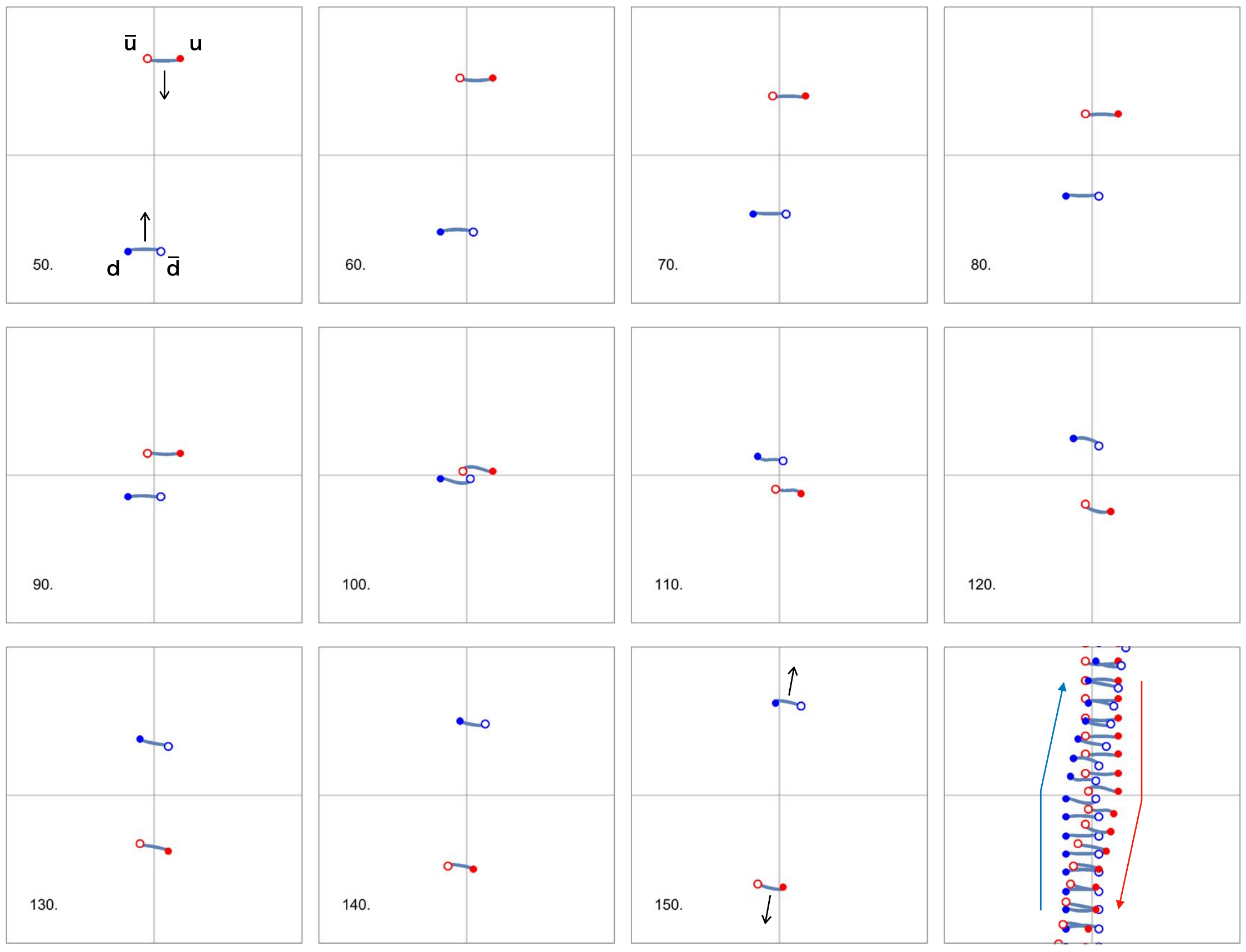}
\caption{Initial configuration is the same as that in Fig.~\ref{fig:ss_KP_set06_a_ex01_umum} except for the
fact that the lower meson is not $\bar{\rm u}{\rm u}$ but $\bar{\rm d}{\rm d}$.
The plot region is $\tilde x^i \in [-40,40]$ and $\tilde b = 2.5$.}
\label{fig:ss_KP_set06_a_ex01_umdm}
\end{center}
\end{figure*}

\begin{figure*}[h]
\begin{center}
\includegraphics[width=0.95\hsize]{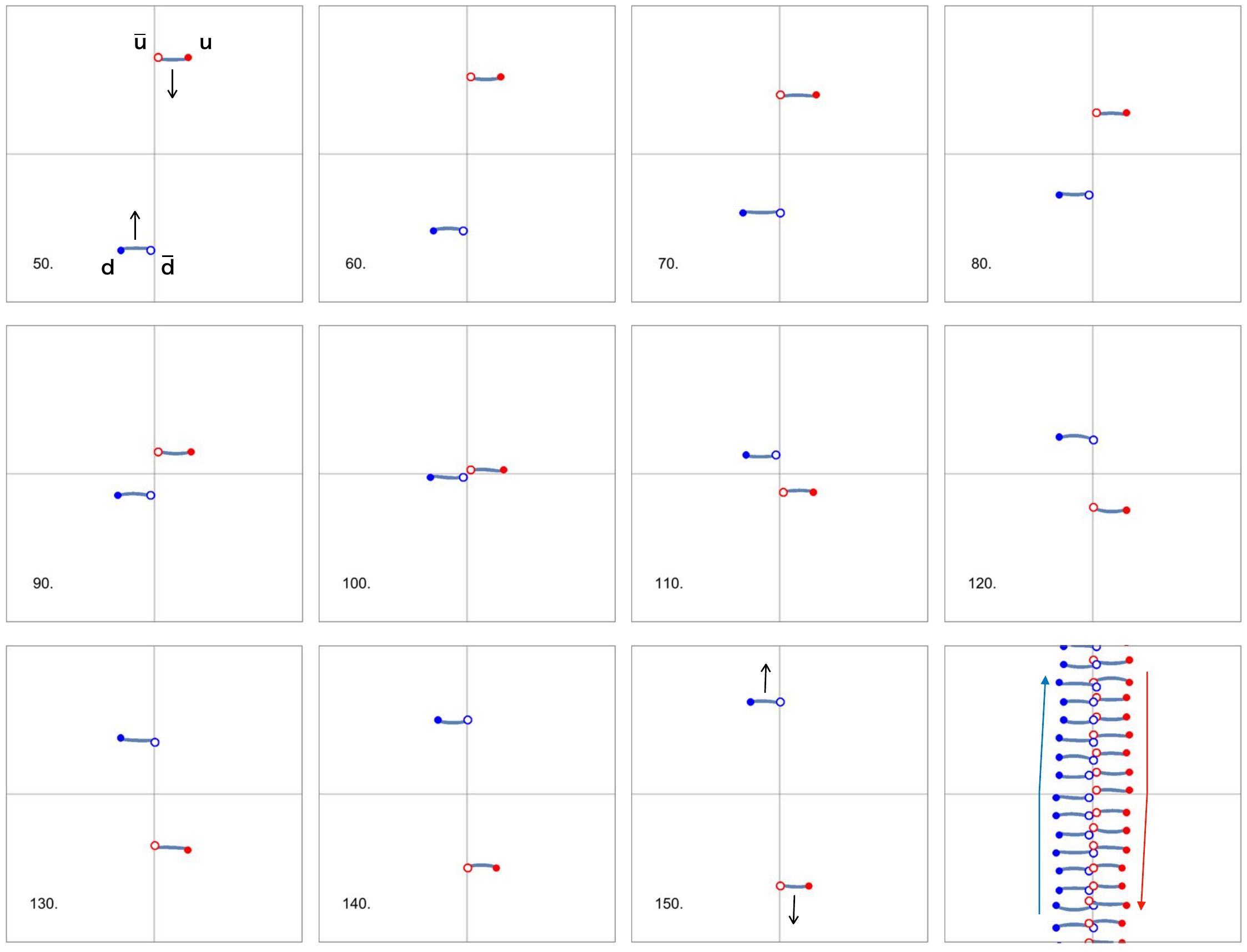}
\caption{Initial configuration is the same as that in Fig.~\ref{fig:ss_KP_set06_b_ex01_umum} except for the
fact that the lower meson is not $\bar{\rm u}{\rm u}$ but $\bar{\rm d}{\rm d}$.
The plot region is $\tilde x^i \in [-40,40]$ and $\tilde b = 5$.}
\label{fig:ss_KP_set06_b_ex01_umdm}
\end{center}
\end{figure*}

\begin{figure*}[h]
\begin{center}
\includegraphics[width=0.95\hsize]{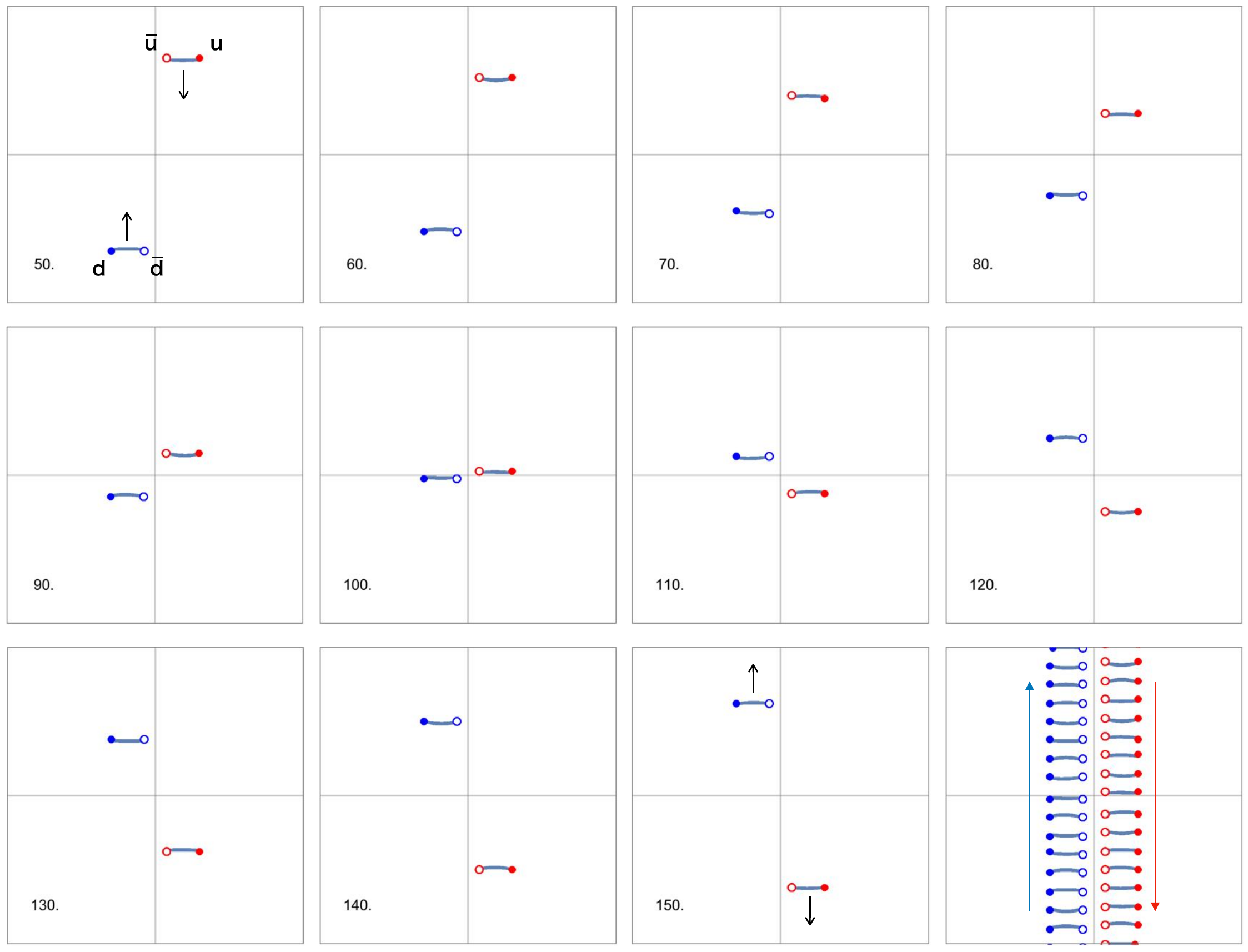}
\caption{Initial configuration is the same as that in Fig.~\ref{fig:ss_KP_set06_c_ex01_umum} except for the
fact that the lower meson is not $\bar{\rm u}{\rm u}$ but $\bar{\rm d}{\rm d}$.
The plot region is $\tilde x^i \in [-40,40]$ and $\tilde b = 7.5$.}
\label{fig:ss_KP_set06_c_ex01_umdm}
\end{center}
\end{figure*}

To make our study self-contained, let us repeat the same simulations for $\bar{\rm u}{\rm u}$ and 
$\bar{\rm d}{\rm d}$, as those for $\bar{\rm u}{\rm u}$ and $\bar{\rm u}{\rm u}$ in Sec.~\ref{sec:impact_uu}.
Namely, we study the u and d mesons collisions with the impact parameter $\tilde b$.
As counterparts of Figs.~\ref{fig:ss_KP_set06_a_ex01_umum}, \ref{fig:ss_KP_set06_b_ex01_umum}, and 
\ref{fig:ss_KP_set06_c_ex01_umum}, we do the same simulations by replacing the lower $\bar{\rm u}{\rm u}$ meson
with the $\bar{\rm d}{\rm d}$. The results are shown in Figs.~\ref{fig:ss_KP_set06_a_ex01_umdm},
\ref{fig:ss_KP_set06_b_ex01_umdm}, and \ref{fig:ss_KP_set06_c_ex01_umdm}.
Comparing Figs.~\ref{fig:ss_KP_set06_c_ex01_umum} and \ref{fig:ss_KP_set06_c_ex01_umdm}, it tuns out that
the asymptotic interaction between the u and d mesons are much smaller than the one between the u and u mesons.
Therefore, we see that the contact interaction dominates for the former.

\clearpage

\section{Baryon-Meson scattering}
\label{sec:meson_baryon}

Our final vortical collider simulations are scatterings of a meson and a baryon.
Since baryons do not run but rotate, we put a baryon at the origin as a target. Then,
we put an incoming meson sufficiently far from the target baryon, and they collide quite
similar to usual collider experiments. As expected, details of the scatterings sensitively 
depend on the timings of collision, namely geometric information such as relative positions
and angles of the baryon and meson. It is impossible to simulate all cases, and so we introduce
two typical scatterings which illustrate general features. 
In addition, we exhibit one
more example which has a special phenomenon, a jet, which remind us hadron collider experiments.

\subsection{Typical collisions}

\begin{figure*}[h]
\begin{center}
\includegraphics[width=0.85\hsize]{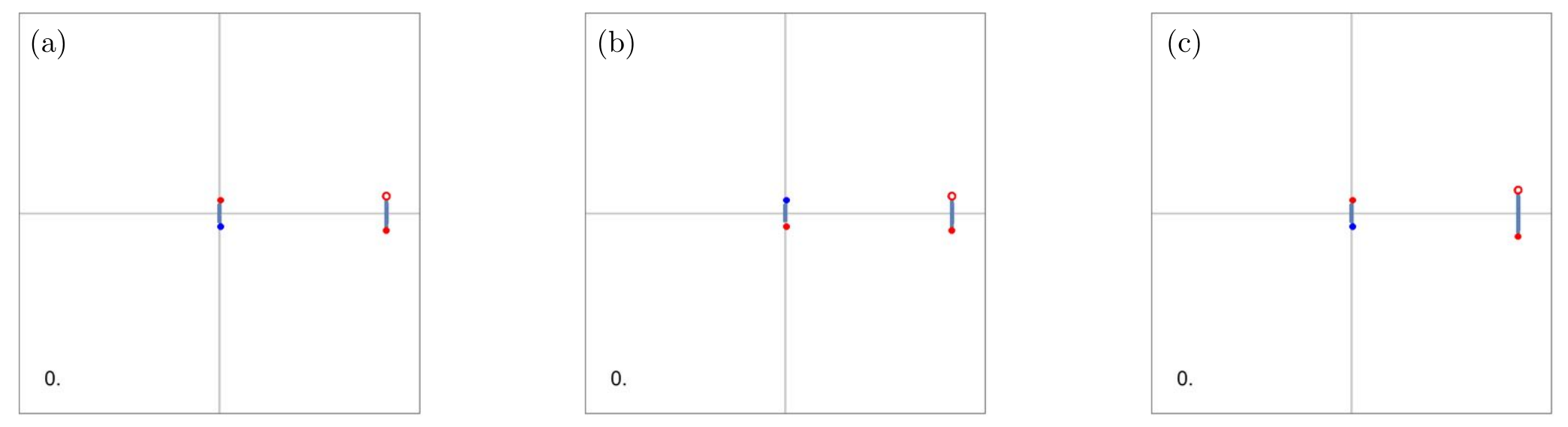}
\caption{The three initial configurations for a u meson and a baryon (ud) scattering. The displayed region is
$\tilde x_i \in [-60,60]$. The meson is placed at $(\tilde x_1, \tilde x_2) = (50,0)$ and the target baryon is at
the origin.}
\label{fig:initial_bm}
\end{center}
\end{figure*}

In this subsection, we exhibit two scattering simulations whose initial configurations are given in
Fig.~\ref{fig:initial_bm} (a) and (b). For both, we put the same meson as the one given in Fig.~\ref{fig:single_b_m}
at $(\tilde x_1,\tilde x_2) = (50,0)$. Similarly, the same baryon as the one given in Fig.~\ref{fig:single_b_m}
is located at the origin. The meson runs toward the target baryon rotating at the origin.
The difference between Fig.~\ref{fig:initial_bm} (a) and (b) is that the initial angles of the target baryon are
different by 180 degree, so that relative positions of the meson and baryon at the moment of collision are different.

\begin{figure*}
\begin{center}
\includegraphics[width=0.95\hsize]{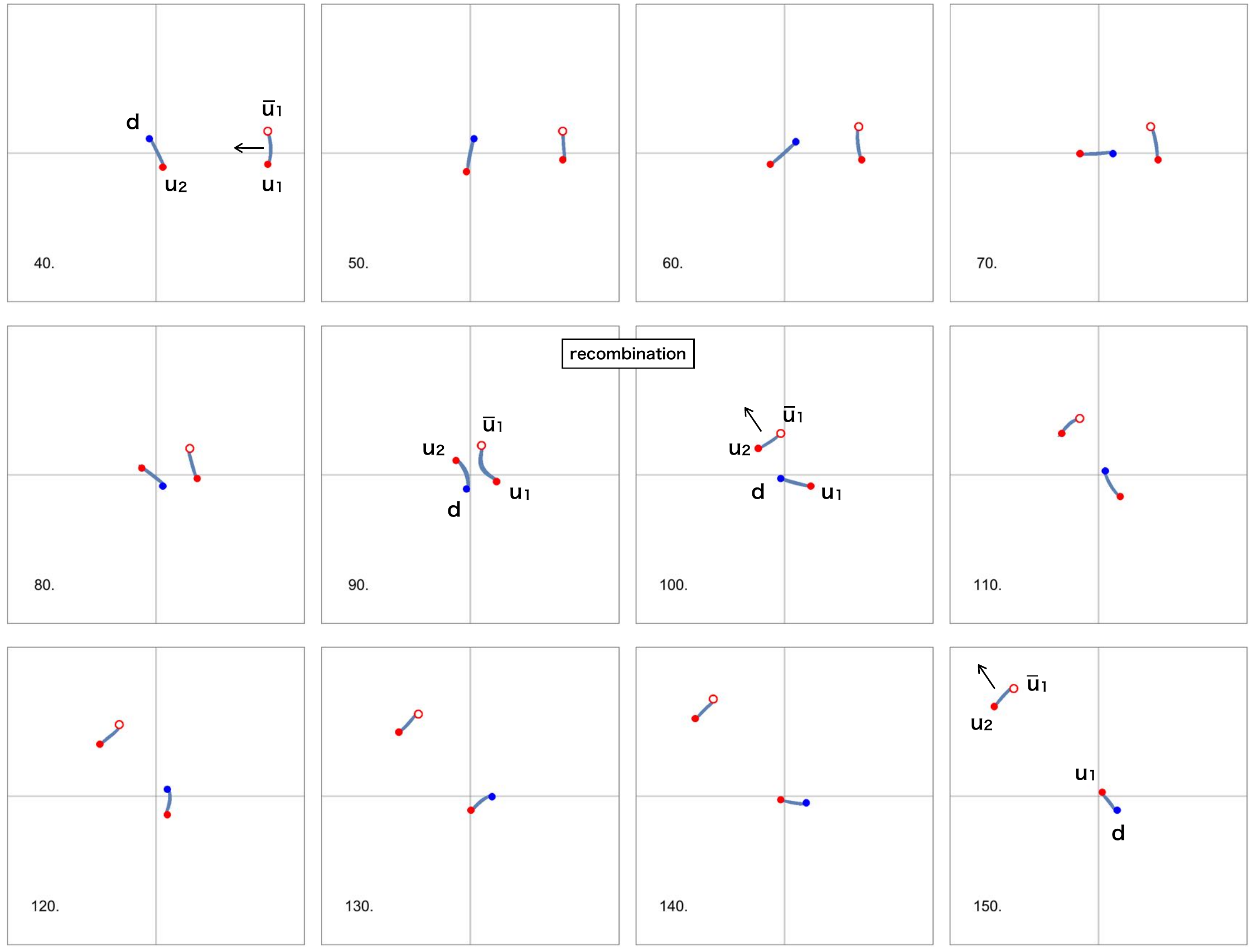}
\caption{The scattering of the meson $\bar{\rm u}_1{\rm u}_1$ and the baryon ${\rm u}_2{\rm d}$ with
the initial configuration given in Fig.~\ref{fig:initial_bm} (a). The collision accompanies the recombination
at about $\tilde t = 90$. The outgoing meson is $\bar{\rm u}_1{\rm u}_2$ and the new baryon ${\rm u}_1{\rm d}$ is
left near the origin. We show the snapshots from $\tilde t =40$ to $150$ with interval $\delta \tilde t = 10$, 
and the plot region is $\tilde x_{1,2} \in [-40,40]$.}
\label{fig:ss_KP_set01_ex04}
\end{center}
\end{figure*}

\begin{figure*}
\begin{center}
\includegraphics[width=0.75\hsize]{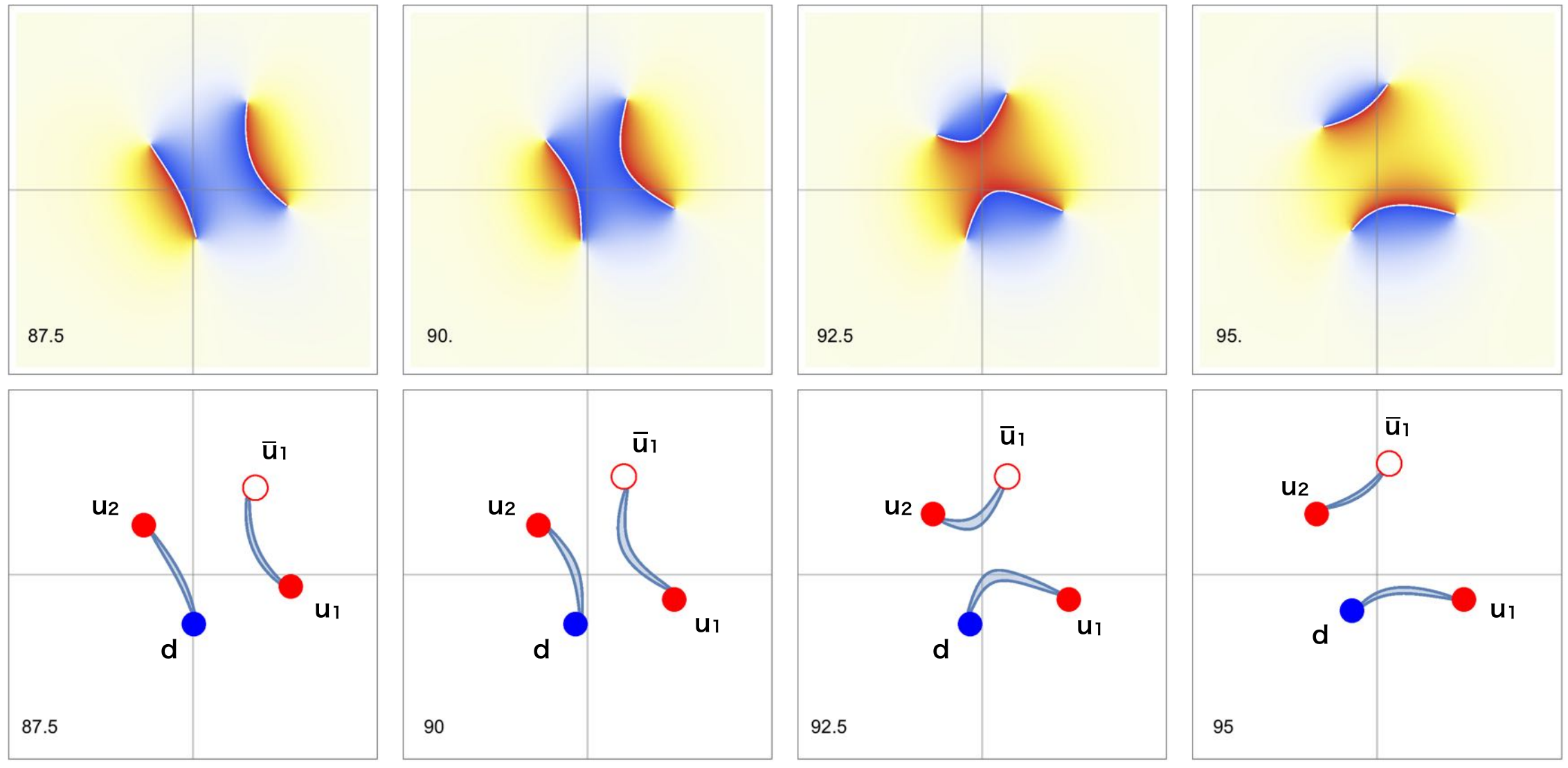}
\caption{Snap shots of Fig.~\ref{fig:ss_KP_set01_ex04} 
in close-up for $\tilde x_i \in [-15,15]$ at $\tilde t = 87.5, 90, 92.5, 95$ 
are shown.}
\label{fig:ss_DPKP_set01_ex04}
\end{center}
\end{figure*}

In Fig.~\ref{fig:ss_KP_set01_ex04},
the scattering from the initial configuration in Fig.~\ref{fig:initial_bm} (a) is shown.
The u meson approaches the rotating target baryon 
and they collide. After the collision, a u meson runs toward
a direction with a scattering angle about 45 degree while the baryon is left near the origin.
Thus, what happens is just $\bar{\rm u}{\rm u}\ + \ {\rm ud}\ \to \ \bar{\rm u}{\rm u}\ + \ {\rm ud}$, and one
might find nothing interesting for this process. However, as the case of the meson-meson scatterings, it is not
so simple. One of nontrivial phenomena is the recombination before and after the collision. 
As can be seen in the panels with the time stamps $\tilde t = 90$ and $100$, the meson and baryon exchange
the u constituent vortices. Therefore, both the meson and baryon after the collision are different from those before
the collision. If we express the initial meson as $\bar{\rm u}_1{\rm u}_1$ and the initial baryon as ${\rm u}_2{\rm d}$,
the process can be summarized as the following 
vortical reaction 
\beq
\bar{\rm u}_1{\rm u}_1 \quad + \quad {\rm u}_2{\rm d} 
\quad \to \quad 
\bar{\rm u}_2{\rm u}_1 \quad + \quad {\rm u}_1{\rm d}.
\label{eq:reaction_bm}
\eeq
To see this better, let us carefully look at the collision period. Fig.~\ref{fig:ss_DPKP_set01_ex04} shows  
snapshots in close-up for $\tilde x_i \in [-15,15]$ around the collision. We find that the snapshots are
qualitatively the same as those given in Fig.~\ref{fig:ss_DPKP_set06_ex01_umum} for the $\bar{\rm u}{\rm u}$-$\bar{\rm u}{\rm u}$
scattering. As before, the recombination takes place
together with partial annihilation of the SG and anti-SG solitons.

\begin{figure*}
\begin{center}
\includegraphics[width=0.95\hsize]{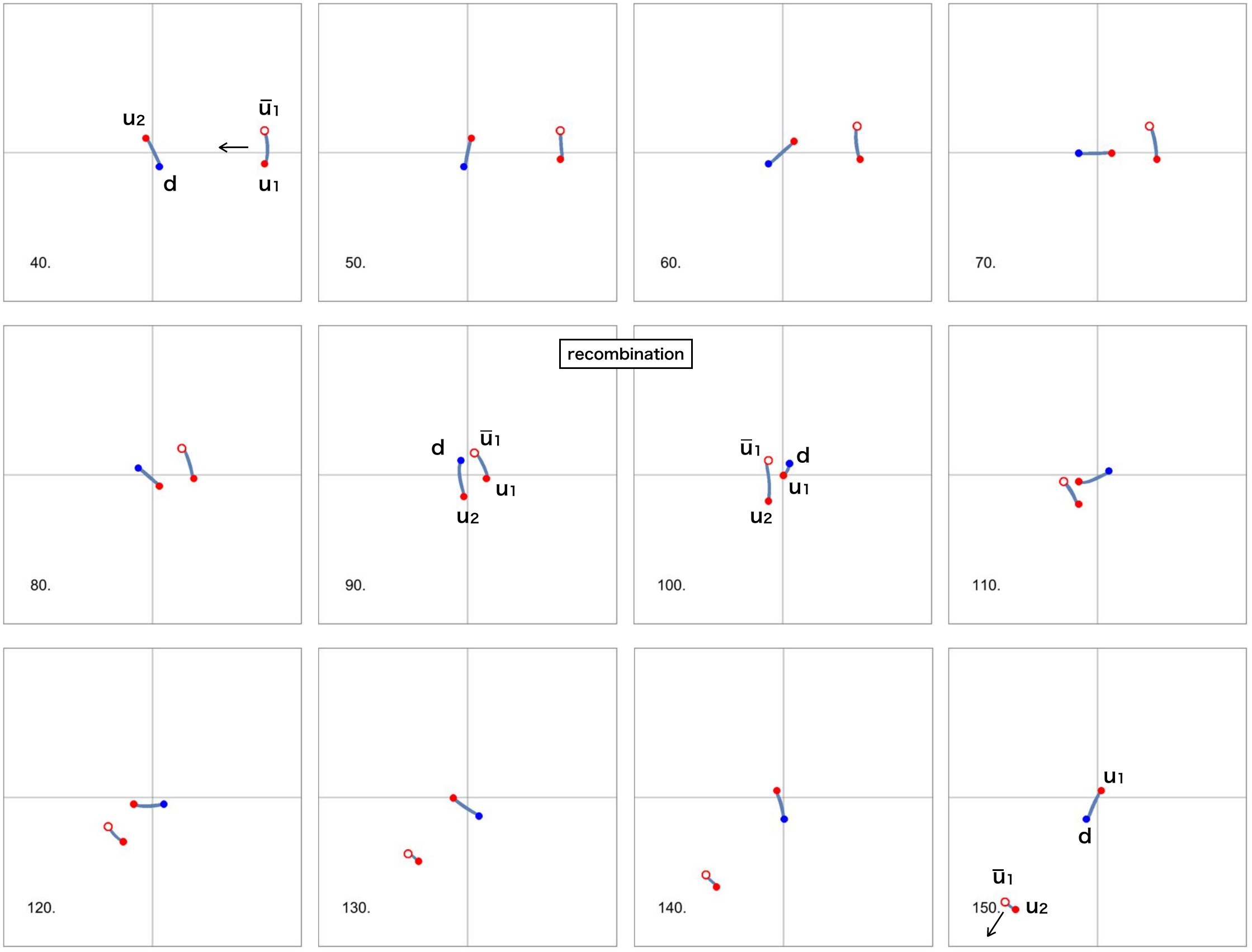}
\caption{The scattering of the meson $\bar{\rm u}_1{\rm u}_1$ and the baryon ${\rm u}_2{\rm d}$ with
the initial configuration given in Fig.~\ref{fig:initial_bm} (b). The collision accompanies the recombination
at about $\tilde t = 90$. The outgoing meson is $\bar{\rm u}_1{\rm u}_2$ and the new baryon ${\rm u}_1{\rm d}$ is
left near the origin. We show the snapshots from $\tilde t =40$ to $150$ with interval $\delta \tilde t = 10$, 
and the plot region is $\tilde x_{1,2} \in [-40,40]$.}
\label{fig:ss_KP_set02_ex04}
\end{center}
\end{figure*}

\begin{figure*}
\begin{center}
\includegraphics[width=0.75\hsize]{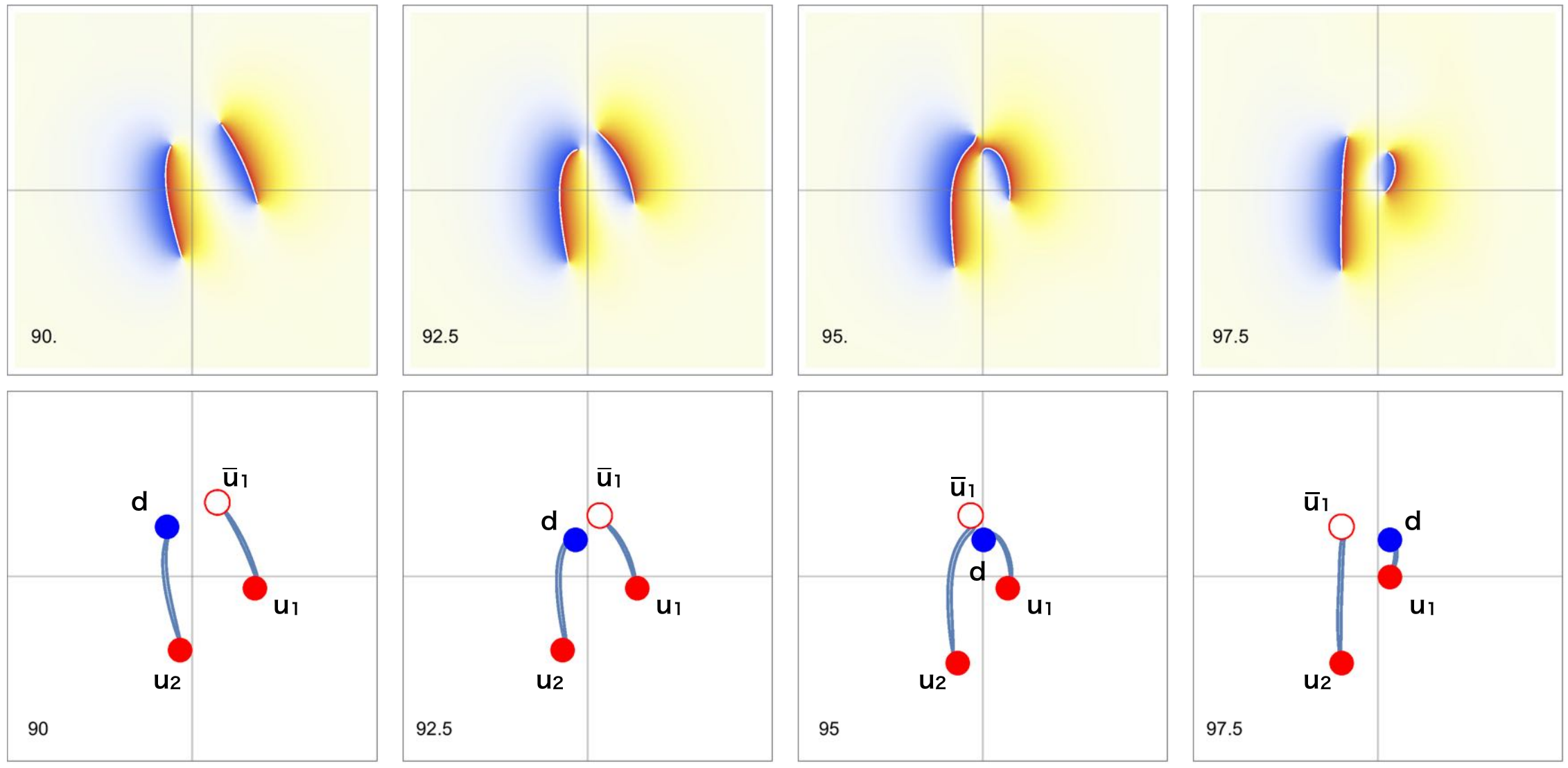}
\caption{Snap shots of Fig.~\ref{fig:ss_KP_set02_ex04} 
in close-up for $\tilde x_i \in [-15,15]$ at $\tilde t = 87.5, 90, 92.5, 95$ 
are shown.}
\label{fig:ss_DPKP_set02_ex04}
\end{center}
\end{figure*}

The other scattering from the initial configuration 
Fig.~\ref{fig:initial_bm} (b) is also shown in Fig.~\ref{fig:ss_KP_set02_ex04}.
The scattering goes almost the same as the previous one. The u meson collides with the target baryon,
and a u meson is scattered while a baryon is left near the origin. The scattering angle is now about $-45$ degree.
The difference of the scattering angles is just due to the relative position and angle of the meson and baryon at the moment of the collision.
Similarly to the first simulation, the meson and baryon experience the recombination once during the collision.
Therefore, the vortical reaction is the same as Eq.~(\ref{eq:reaction_bm}). Nevertheless, detail dynamical processes
of the recombination are not same.
To see the difference clearer, let us zoom in on the collision moment shown in Fig.~\ref{fig:ss_DPKP_set02_ex04}.
On contrary to the previous case, the collision is accompanied not by the SG and anti-SG solitons but the
SG and SG solitons. Hence, the recombination here is very similar to what we saw in Fig.~\ref{fig:ss_DPKP_set06_ex01_umdm}
for the $\bar{\rm u}{\rm u}$ and $\bar{\rm d}{\rm d}$ meson scattering. Since the two SG solitons repel each other, they
cannot be close. Only one set of the edges of the meson and baryon come across to exchange $\bar{\rm u}_1$ of $\bar{\rm u}_1{\rm u}_1$
and d of ${\rm u}_2{\rm d}$. Then, a new meson $\bar{\rm u}_1{\rm u}_2$ goes away and a new baryon ${\rm u}_1{\rm d}$
remains near the origin and keep rotating.
We have examined lots of collisions with various initial configuration, and have found that 
all of them are qualitatively the same.

\subsection{A vortical hadron jet}
In high energy physics, 
when hadrons collide with sufficiently high energy, they fragment into quarks or gluons.
However, obeying the color confinement in QCD, no color charged objects can exist alone.
Therefore, these fragments create new colored particles around them to form color neutral objects,
namely hadrons.
A bunch of the hadrons form a narrow beam which is called a hadron jet.
Since our vortices also obey the $U(1)_R$ confinement in BECs, which we have seen quite similar to
QCD, we expect that a hadron jet would be observed also in BECs.

\begin{figure*}[h]
\begin{center}
\includegraphics[width=0.95\hsize]{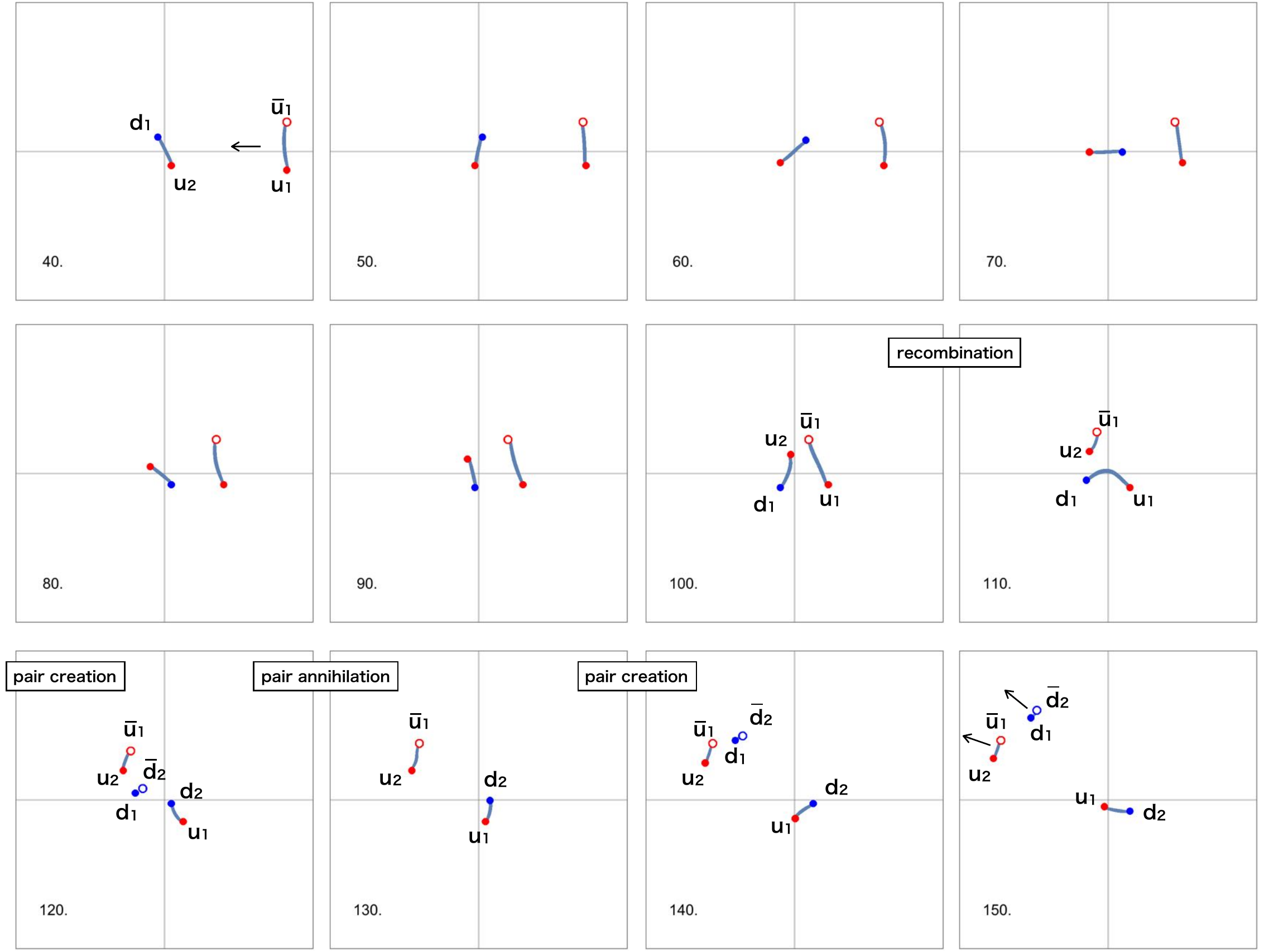}
\caption{The scattering of the meson $\bar{\rm u}_1{\rm u}_1$ and the baryon ${\rm u}_2{\rm d}$ with
the initial configuration given in Fig.~\ref{fig:initial_bm} (c). The collision accompanies the recombination, 
pair creation, and pair annihilation. A new meson $\bar{\rm d}_2{\rm d}_1$ is emitted, which we interpret a vortical
hadron jet. We show the snapshots from $\tilde t =40$ to $150$ with interval $\delta \tilde t = 10$, 
and the plot region is $\tilde x_{1,2} \in [-40,40]$.}
\label{fig:ss_KP_set01_ex03}
\end{center}
\end{figure*}

\begin{figure*}
\begin{center}
\includegraphics[width=0.75\hsize]{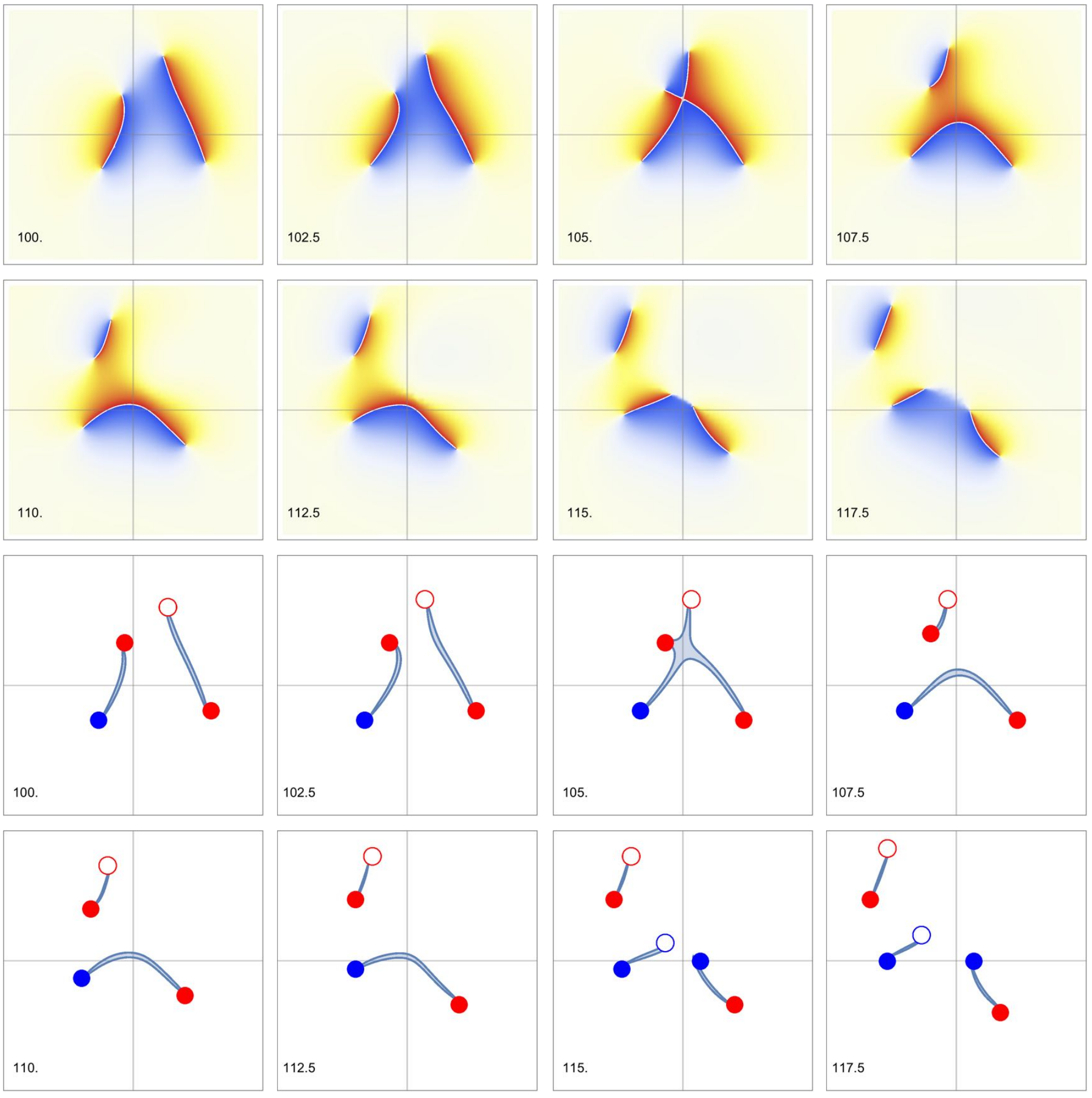}
\caption{Snapshots of Fig.~\ref{fig:ss_KP_set01_ex03} 
in close-up for $\tilde x_i \in [-15,15]$ from $\tilde t = 100$ to $117.5$ with interval $\delta \tilde t = 2.5$ 
are shown.}
\label{fig:ss_DPKP_set01_ex03}
\end{center}
\end{figure*}

Running speed of a vortical meson is determined by length of the meson. Namely, we cannot freely change
the running speed of mesons unlike relativistic particles in reality.
As shorter the meson is,  faster it runs. However, there is a threshold of the minimum length of the meson over which
mesons are unstable to be annihilated. Therefore, we cannot give a large kinetic energy to
mesons. Hence, it is not easy to set up a simulation with ultra high colliding energy in BECs. Indeed, as
we have seen in the previous subsections so far, the typical scatterings do not yield any additional new
hadrons. Then we change our strategy: instead of taking a shorter meson, we take a longer meson. Thus, we prepare
the third initial configuration given in Fig.~\ref{fig:initial_bm} (c) in which the baryon is the same as the one
of Fig.~\ref{fig:initial_bm} (a) but the meson is longer.
 It is in Fig.~\ref{fig:ss_KP_set01_ex03} 
how the collision of a longer meson and normal baryon occurs.
Since the longer meson runs slower, it takes more time to reach the target baryon at the origin. Thus, compared
with Fig.~\ref{fig:ss_KP_set01_ex04}, the baryon rotates slightly more so that the relative angle between
the meson and baryon is also slightly different. However, the first reaction is not affected by such a small difference.
Namely, the meson and baryon again experience the recombination accompanied by a partial annihilation of the SG and anti-SG solitons. 
While the corresponding vortical reaction process is the same as Eq.~(\ref{eq:reaction_bm}), the details of the collision
are shown in Fig.~\ref{fig:ss_DPKP_set01_ex03}. As before, a relatively short meson $\bar{\rm u}_1{\rm u}_2$
and a relatively long baryon ${\rm u}_1{\rm d}_1$ form during the collision, 
The former flies away with a smaller scattering angle. The latter baryon is very long and bended. Then 
it soon disintegrates into smaller hadrons. Indeed, it fragments by creating a ${\rm d}_2$ and $\bar{\rm d}_2$ 
in a certain point middle of the longer baryon. As a consequence, the third hadron, $\bar{\rm d}_1{\rm d}_2$ is emitted
toward the similar direction of the first meson $\bar{\rm u}_1{\rm u}_2$. We interpret the new hadron which appears
as a result of fragmentation of the confining SG soliton a vortical hadron jet, though the jet consists of only two
mesons $\bar{\rm u}_1{\rm u}_2$ and $\bar{\rm d}_2{\rm d}_1$. 
The vortical reaction for it is
\beq
{\rm d}_1{\rm u}_1
\quad \to \quad
\bar{\rm d}_2{\rm d}_1
\quad + \quad
{\rm u}_1{\rm d}_2.
\label{eq:baryon_fragment}
\eeq
The third meson $\bar{\rm d}_2{\rm d}_1$ is very short.
So it crawls under waves of $\Psi_i$ for a while and
soon emerges again, see the panels with $\tilde t = 130$ and $140$ of Fig.~\ref{fig:ss_KP_set01_ex03}.

\subsection{The Feynman diagrams}

\begin{figure*}[h]
\begin{center}
\includegraphics[width=0.5\hsize]{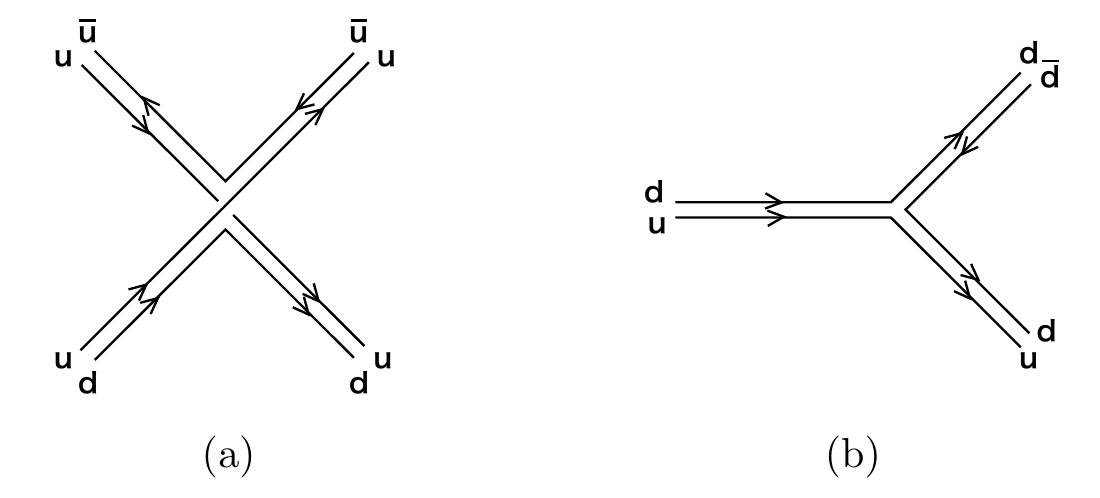}
\caption{(a) meson-meson-baryon-baryon vertex corresponding to the real process for Eq.~(\ref{eq:reaction_bm}). (b)
meson-meson-baryon vertex corresponds to the real process for Eq.~(\ref{eq:baryon_fragment}).}
\label{fig:twig_bm}
\end{center}
\end{figure*}

We have found the two new vortical reactions summarized in  Eqs.~(\ref{eq:reaction_bm}) and (\ref{eq:baryon_fragment}) 
through the meson-baryon scatterings. They can also be described by Feynman diagrams as before. The corresponding
diagrams are a meson-meson-baryon-baryon vertex and a meson-meson-baryon vertex shown in Fig.~\ref{fig:twig_bm}.
The former vertex is new in a sense that any of the discrete symmetries $F$, $T$, and $P$ cannot relate 
it to any of the previous diagrams shown so far. Although the latter vertex has not also been encountered yet, 
it can be obtained by flipping an external leg of 
the vertex (mbb4) of Fig.~\ref{fig:all_vertex}. However, as mentioned before, the presence 
(mbb4) of Fig.~\ref{fig:all_vertex} as a real process does not immediately mean that the vertex (b) of Fig.~\ref{fig:twig_bm}
indeed occurs as a real processes. Thus, we again put our emphasis on that we include 
the vertex (b) of Fig.~\ref{fig:twig_bm} into our diagram list because of the fact that we have found it in the simulation.

\section{A connection to the confinement problem in particle physics}
\label{sec:connection_to_QCD}

So far, we have investigated the topological objects in non-relativistic two component BECs in 2+1 dimensions.
At glance, these seem to be very far from relativistic particle physics in 3+1 dimensions. 
Nevertheless, the terminology (u, d, meson, baryon, and so on) 
and the description by Feynman diagrams
borrowed from QCD are surprisingly fit BECs.
Although a precise connection between QCD and BECs is not clear at all, let us try to give some hints for
unveiling it along with the observations by Son and Stephanov in  Ref.~\cite{Son:2001td}.

The key ingredient is a duality.
The vortices are particle-like topological defects in $2+1$ dimensions, and they are not elementary constituents of the
original models.
Nevertheless, it is known that a duality sometimes interchanges the defects and elementary constituents. 
A classic example is the duality between
sine-Gordon solitons and fermions in the massive Thirring model in $1+1$ dimensions \cite{Coleman:1974bu,Mandelstam:1975hb}. 
Another simplest example more relevant to this paper is particle-vortex duality
between the $XY$-model and the Abelian-Higgs model in $2+1$ dimensions \cite{Peskin:1977kp,Dasgupta:1981zz},
which gives insights for understanding fractional quantum Hall effect \cite{Lee:1989fw}. 
In the particle physics context, the dualities have been expected to be one of the powerful tools for us to understand much 
better non-perturbative dynamics in strongly coupled systems. 

On the other hand, one of the most important unsolved problems in modern high energy physics is the confinement of colors in QCD.
The quarks and gluons are elementary constituents of QCD, but we cannot observe them at low energy since they are
strongly confined to form hadrons. 
Widely accepted picture of  the confinement is that chromo-electric flux from a quark is squeezed to form a flux tube.
Then, interaction energy between (anti-)quarks is proportional
to the distance: it is confinement.
Though this picture of confinement is quite plausible, 
it is merely qualitative. 
Indeed, it is very difficult to prove analytically whether it occurs or not in real QCD.
Then, instead of QCD, many studies have been done for QCD-like theories. A brilliant milestone was put by Seiberg and Witten 
in supersymmetric $SU(2)$ Yang-Mills theory \cite{Seiberg:1994rs,Seiberg:1994aj}. 
They analytically showed condensation of the magnetic monopole at low energy ground state indeed takes place.

Another, rather old back in 70s, important remark was made by Polyakov \cite{Polyakov:1976fu}. He considered a compact $U(1)$
gauge theory in 2+1 dimensions which can be obtained as a low energy effective theory of the Georgi-Glashow model with
$SU(2)$ gauge field coupled to an adjoint scalar field $\phi_a$ ($a=1,2,3$). In the ground state, the adjoint scalar field develops a non-zero vacuum expectation value
$\phi_a = (0,0,v)$, and it breaks $SU(2)$ to its diagonal compact $U(1)$ subgroup. After integrating
all the massive fields, we are left with a free massless photon of the compact $U(1)$ group at low energy. 
Since the photon has only one polarization in three dimensions, it can be dualized to a periodic scalar field $\vartheta \in [0,2\pi)$, so-called
dual photon which is related to the original $U(1)$ gauge field $A_\mu$ by
\beq
F_{\mu\nu} = \frac{e^2}{4\pi} \varepsilon_{\mu\nu\rho}\p^\rho\vartheta,\qquad
F_{\mu\nu} = \p_\mu A_\nu - \p_\nu A_\mu,
\eeq
with $\mu,\nu = 0,1,2$ and $e$ is a $U(1)$ gauge coupling constant.
Under the duality relation, electric charges in the original theory are interchanged by  vortices in the dual theory as
\beq
\vartheta(z) = \sum_{a}q_a{\rm Im}\log(z-z_a)
\eeq
with $z = x + i y$. Here, $z_a$ are positions of vortices and 
$q_a = \pm 1$ are their charges.
The dual photon is  massless in perturbation theory but it gets mass by non-perturbative 
instanton ('t Hooft-Polyakov type monopole) effects
in the Georgi-Glashow model.
In the weakly coupled region, the instantons interact with the Coulomb force and behave as a dilute plasma. 
Then the Debye screening effect gives a non-zero
mass to the dual photon, which can be summarized in the following
low energy effective theory \cite{Polyakov:1976fu,Kogan:2002au,Antonov:2004kj,Shifman:2008ja}
\beq
{\cal L}_{\rm eff} = \frac{e^2}{32\pi^2}\p_\mu\vartheta\p^\mu\vartheta + c \xi \eta^3 \cos \vartheta,
\label{eq:Ldual}
\eeq
with $c$ being an undetermined parameter. The dimension full parameter $\eta$ is related to the mass $M_W$ of the massive
gauge bosons as $\eta^3 = M_W^{7/2}/e$. The so-called monopole fugacity $\xi$ 
is exponentially small as $\xi = \exp\left(-\frac{2\pi M_W}{e^2}\epsilon\right)$ where $\epsilon$ is a function of the ratio
of $M_W$ and the Higgs mass $M_\phi$, which is known to be of order one.
From Eq.~(\ref{eq:Ldual}), the dual photon mass reads
\beq
M_\vartheta^2 = \frac{16\pi^2 c \eta^3\xi}{e^2}.
\eeq
The non-perturbative instanton effects is responsible for another important phenomenon, the charge confinement.
Putting a prove charge in the Georgi-Glashow model corresponds to putting a vortex in the dual theory
as $\vartheta = {\rm Im}\log z = \theta$ ($z = r e^{i\theta}$). When we go around the vortex,
$\vartheta$ passes a potential peak once at $\theta = \pi$. 
Namely, a semi-infinite domain wall attaches at the vortex, and
it corresponds to the confining string attached to the prove charge in the original picture.
This is standard understanding of the confinement in the compact QED in 2+1 dimensions.

Now we are in a position to observe a relation between the dual theory (\ref{eq:Ldual}) and
the Gross-Pitaevskii equations (\ref{eq:gp1}) of the 2 component BECs. 
The Lagrangian for Eq.~(\ref{eq:gp1}) is given by
\beq
{\cal L}_{\rm GP} &=& \sum_i \left[- \frac{i\hbar}{2}\left(\Psi_i\dot \Psi_i^* - \dot\Psi_i \Psi_i^*\right) - \frac{\hbar^2}{2m}|\nabla\Psi_i|^2
+ \mu_i |\Psi_i|^2 - \frac{g_i}{2}|\Psi_i|^4\right] - g_{12} |\Psi_1\Psi_2|^2 - V_R,
\label{eq:LGP}
\eeq
with $V_{\rm R} = -  \hbar \omega \left(\Psi_1\Psi_2^* + \Psi_1^*\Psi_2\right)$. 
Then we truncate this by
substituting the expression of the condensates
$\Psi_i = (v + r_i)e^{i\theta_i}$ into Eq.~(\ref{eq:LGP}) 
and  by integrating out the amplitude modes $r_i$ \cite{Son:2001td,Eto:2017rfr}:
\beq
\tilde {\cal L}_{\rm GP} = \tilde {\cal L}_{S}  + \tilde {\cal L}_{R},
\label{eq:LET_BEC}
\eeq
with
\beq
\tilde {\cal L}_{S} &=& \frac{\hbar^2}{g+g_{12}}\dot\theta_S^2 - \frac{\hbar^2 v^2}{m} (\nabla \theta_S)^2,\\
\tilde {\cal L}_{R} &=& 
\frac{\hbar^2}{g-g_{12}}\dot\theta_R^2
- \frac{\hbar^2 v^2}{m} (\nabla \theta_R)^2
+ 2\hbar \omega v^2 \cos 2\theta_R,
\label{eq:L_R}
\eeq
where we have ignored constants and dealt with the Rabi term perturbatively. 
Here $\theta_S = (\theta_1+\theta_2)/2$ is a phonon 
and $\theta_R = (\theta_1-\theta_2)/2$ is known 
as the Leggett mode or phason.
Note that the potential term in Eq.~(\ref{eq:L_R}) is identical to $V_R$ given in Eq.~(\ref{eq:rabi_V}).
Now, we would like to identify ${\cal L}_{\rm eff}$ in Eq.~(\ref{eq:Ldual}) with
$\tilde {\cal L}_R$ in Eq.~(\ref{eq:L_R}). This can be achieved by making a correspondence as
\beq
\vartheta \quad &\Leftrightarrow& \quad 2 \theta_R,\\
\frac{e^2}{8\pi^2} \quad &\Leftrightarrow& \quad \frac{\hbar^2}{g-g_{12}},\\
c\xi\eta^3 \quad &\Leftrightarrow& \quad 2\hbar\omega v^2,
\eeq
and we have rescaled the spatial coordinate  as
$\tilde x_i = \sqrt{\frac{m}{(g-g_{12})v^2}}\,x_i$ in Eq.~(\ref{eq:L_R}).

The presence of the additional massless scalar field $\theta_S$ is 
a crucial difference between the Polyakov's dual theory (\ref{eq:Ldual})
and the low energy effective theory of the BECs in (\ref{eq:LET_BEC}).
We envisage a gauge theory whose gauge symmetry is broken like $U(1)_S \times SU(2) \to U(1)_S \times U(1)_R$
as an electric dual of the two-component BECs.
Thanks to the additional $\theta_S$, we have two different species of the vortices, namely u- and d-vortices,
and we can deal with not only the mesonic molecules but also the baryonic molecules unlike the original Polyakov's
model. Furthermore, we would like to emphasize that one of the nicest virtue of the two-component BECs is accessibility
to dynamical aspect of the confining phenomena as we have shown above.

Having these duality pictures in our minds, we ask ourself again what physical meaning of the
Feynman diagrams is. Although we have no rigorous arguments, the classical diagrams of the u- and d-vortices
would correspond to quantum Feynman diagrams of the elementary particles of the dual theory. For example, the classical
1-loop diagram in Fig.~\ref{fig:1loop_umdm} seems to be mapped onto the quantum 1-loop diagram in the dual theory.
If we could compute a quantum quantity from dynamics of some classical topological defects in very different theory, 
it is interesting. We have no further observations pushing the idea forward, so we leave it as an open question 
at this moment.

\begin{figure*}[h]
\begin{center}
\includegraphics[width=0.6\hsize]{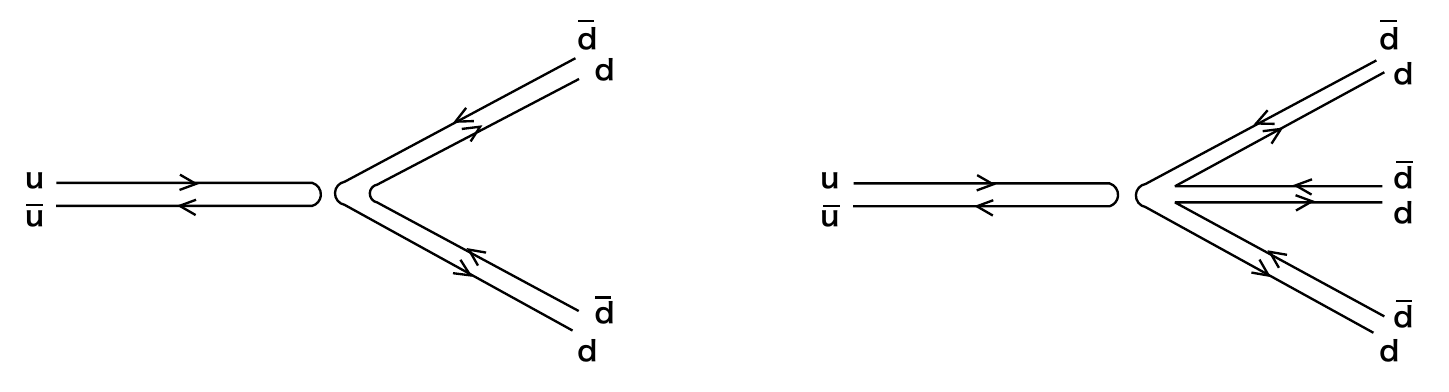}
\caption{Disconnected diagrams for $\bar{\rm u}{\rm u}$ decay.}
\label{fig:ozi}
\end{center}
\end{figure*}

Finally, as a possible clue for pinning down the connection between particle physics and two-component BECs, 
let us briefly mention the Okubo-Zweig-Iizuka (OZI) rule \cite{Okubo:1963fa,Zweig:1981pd,Iizuka:1966fk} 
which is a phenomenological rule of QCD found in 1960s \cite{Fritzsch:2018xbs}.
The OZI rule explains the narrow decay width of the vector meson. Kinematically, the decay
$\phi(\bar{\rm s}{\rm s}) \to \pi^+(\bar{\rm d}{\rm u}) + \pi^0(\bar{\rm d}{\rm d}) +
\pi^-(\bar{\rm u}{\rm d})$ dominates the other decay process 
$\phi(\bar{\rm s}{\rm s}) \to K^+(\bar{\rm s}{\rm u}) + K^-(\bar{\rm u}{\rm s})$ because the phase space of the 
former process is much lager than the latter. Nevertheless, the process $\phi \to 3\pi$ is highly suppressed
relative to $\phi \to K^+K^-$. The OZI rule in QCD is a phenomenological postulation that diagrams with
disconnected quark lines are suppressed relative to connected ones.
It seems natural for us to ask whether a similar rule holds for hadronic molecules in
two-component BECs or not. In order to answer this question, let us consider decay of a $\bar{\rm u}{\rm u}$ mesonic
molecule, and verify if the diagrams given in Fig.~\ref{fig:ozi} are observed or not. 
Throughout this work,
we have met the decay diagrams of $\bar{\rm u}{\rm u}$ three times. The first and the second ones are 
given in Fig.~\ref{fig:all_vertex} (mmm3) and (mbb1), respectively. The third one is the left-most diagram
of Fig.~\ref{fig:meson_disintegration}. None of them are disconnected diagrams. Moreover, not only these three diagrams
but also all the diagrams we encountered
so far are connected diagrams. Thus, we have found an empirical rule that diagrams with disconnected vortex lines
are highly suppressed relative to connected ones. Namely, the OZI rule seems to hold even in BECs.
The OZI rule in QCD is explained as follows. Disconnected quark lines are indeed connected by internal
gluon lines. High momentum transfer by the gluons makes the QCD coupling constant small, so that such channels
are highly suppressed. Thus, the QCD OZI rules is a consequence of a quantum effect.
On the other hand, the dynamics of vortical hadrons in BECs essentially obeyed the classical GP equations.
We expect that a counterpart of the gluon is the classical wave functions $\Psi_i$, and the classical OZI rule in BECs
corresponds to the quantum OZI rule in QCD via a duality. As we repeated, we do not have a precise connection 
between QCD and BECs.
This is an open question.

\section{Summary and discussion}
\label{sec:summary}

In this work, we have investigated dynamics of the vortex molecules in two-component BECs by numerically 
solving the Gross-Pitaevskii equations (\ref{eq:gp1}).
This paper is on the line of the previous works \cite{Son:2001td,Tylutki:2016mgy,Eto:2017rfr} focusing on
the confinement property of fractional vortices by the SG solitons. While the dynamical property of an individual molecule, such as 
precession and disintegration, were figured out in Refs.~\cite{Tylutki:2016mgy,Eto:2017rfr}, here we have studied 
the scattering and collision of  molecules.
Our numerical studies are twofolds: the meson-meson scattering and the meson-baryon scattering.\footnote{
For both the simulations, the non-relativistic nature of the GP equations is crucial since
it ensures that the molecule size is kept to be finite and constant even though the constituent vortices are
pulled by the SG soliton. In relativistic models, this cannot happen because the molecules soon shrink.
Especially, it is very difficult to prepare mesonic molecules which soon decay into radiations.
Cost we have to pay is that both the precession speed of a baryonic molecule and the translation speed of a
mesonic molecule are determined by the molecule size \cite{Eto:2017rfr}, but we can, nevertheless, make
use of such dynamical properties (the precession of baryons and the translation of mesons) for planning
scattering experiments of the mesons and baryons. Indeed, our numerical simulations provide quite rich and
interesting results as we have described above in details.
}

In Sec.~\ref{sec:meson-meson}, we have dealt with the meson-meson scattering of the same spices ($\bar{\rm u}$u-$\bar{\rm u}$u). Firstly, we have demonstrated
three simulations of the head-on collisions (zero impact parameter) by varying the incident angles, which are summarized in Fig.~\ref{fig:mm_scatterings}.
We have found that the mesons collide involving the vortical reaction given in Eq.~(\ref{eq:mm_01}). Namely,
the constituent vortices swap the partners. 
We also showed that the recombination can be understood as a collision of the SG and anti-SG solitons, and 
the swapping is nothing but the pair annihilation and creation 
of the confining SG solitons as can be seen in Fig.~\ref{fig:ss_DPKP_set06_ex01_umum}.
Of course, the details of the final states sensitively depend on the initial incident angles. 
The simulation with the initial relative angle $\pi$ happens to show
the right angle scattering of the two mesons, which is very common among relativistic topological solitons.
On the other hand, the scattering with smaller angle $\pi/2$ exhibits more complicated out-going state involving
the pair creation of a new meson and a subsequent pair annihilation of the original meson.
In principle, there are thousands of different scatterings depending on the model parameters, initial configurations,
and so on. Clearly, solving numerically the GP equations for each time is not a good strategy.
To avoid such infinitely high cost surveys and to get efficient outlook, we developed a useful description by
using the Feynman diagrams. A fractional vortex corresponds to a line with an arrow toward the time direction whereas
an anti-fractional vortex is expressed by a line with an arrow opposite to the time direction.
The elementary vertices found through the meson-meson scatterings with an aid of time reversal and the party
transformations are summarized in Fig.~\ref{fig:all_vertex}. These are building blocks of the real scattering processes.
We also have studied the meson-meson scatterings with non-zero impact parameters. It is found that 
the scattering angles depend on the impact parameter: the smaller impact parameter is, the lager the scattering
angle is. As increasing the impact parameter, the scattering angle reduces and it crosses zero and becomes negative. Then, it asymptotically becomes zero as the impact parameter is taken to be infinity.

In Sec.~\ref{sec:meson_meson_2}, we have turned to studying the meson-meson scattering of the different spices 
($\bar{\text{u}}\text{u}$-$\bar{\text{d}}\text{d}$). We have repeated the numerical simulations similar to those
in Sec.~\ref{sec:meson-meson}. The head-on collision with the 0 relative angle seems to be boring at glance:
the two mesons just pass through, see Fig.~\ref{fig:ss_KP_set06_ex01_umdm}. However, 
we have found that the scattering of the two SG solitons occurs. They bounce off, and 
the molecules interchange the SG solitons before and after the collision.
The head-on collisions with the smaller relative angles are more interesting. We have found that
the incoming u and d mesons are converted to the intermediate baryon and anti-baryon pair during the collision. 
The intermediate baryons rotate, and then they are reformed back into the mesons
at the second recombination. Forming the intermediate baryonic state results in 
the shift of the outgoing line from the ingoing one, see Fig.~\ref{fig:delay}.
As before, we have described the numerical simulations by using the Feynman diagrams, and divided them into
the elementary vertices. The newly found vertices are shown in Figs.~\ref{fig:1loop_umdm}.
In addition, we also have put the new vertices 
 in Figs.~\ref{fig:meson_disintegration} and \ref{fig:baryon_disintegration}, 
 corresponding to the disintegrations of the meson and baryon
found in Refs.~\cite{Tylutki:2016mgy,Eto:2017rfr}.
We also have studied the scatterings with non-zero impact parameters  but the results are not so dramatic as those
for the mesons of the same spices.

In Sec.~\ref{sec:meson_baryon}, we have studied the meson-baryon scatterings ($\bar{\rm u}$u and ud).
We have injected the meson toward the baryon staying at the origin.
A variety of the scatterings arises from the difference of the relative angles of the two molecules at the collisions.
We have examined two initial configurations (the molecules are initially parallel and anti-parallel) 
given in the left-most and the middle panels of Fig.~\ref{fig:initial_bm}. We have found that there are no qualitatively
large differences between the two cases. In the both cases, the meson and baryon swap their constituent u vortices,
and the new meson goes out while the new baryon stays at slightly shifted point from the original baryon point.
Much more interesting thing happens when we scatter a longer meson to the baryon. The typical initial configuration
is given in the right-most panel of Fig.~\ref{fig:initial_bm},
and the scattering goes as shown in Fig.~\ref{fig:ss_KP_set01_ex03}. The recombination takes place also in this case,
so that a longer and kink bend baryon is formed at first stage. However, such a long and bent molecule is unstable
and soon disintegrates into a set of shorter meson and baryon. As a result, the final state includes larger number
of molecules than the prepared one. This somehow resembles what happens in real Hadron collider experiments.
When we collide two hadrons in a Hadron collider with sufficiently large energy, thousands of hadrons come out
as a hadron jet. Although our final state consists of only a few hadrons, we exaggeratedly call it a vortical hadron jet
in the VHC (vortical hadron collider) experiment. It is surprising to us that the simple classical system (\ref{eq:gp1})
include such rich phenomena somehow common to QCD which needs highly quantum regimes.

In Sec.~\ref{sec:connection_to_QCD}, we have made supplementary comments to make a connection between QCD and BECs clearer.
We have compared the Polyakov's dual photon model in $2+1$ dimensions to the low energy effective theory based on the
two-component BECs. We have seen the latter is an extension of the former, and so we expect the confinement
of the u- and d-vortices studied here would shed some lights to the confinement of the elementary particles.
A good advantage using the BECs is that we can easily access to dynamical aspects of the confinement which is 
in general difficult. As a related topic, we also have pointed out that the similar rule to the OZI rule in QCD seems to
hold in the BECs.

To close this paper, let us list what we have not done in this work.
All numerical analysis in this paper have been done under the condition (\ref{eq:flavor_sym}).
Therefore, the u- and d-vortices have the same masses. The dynamics of the hadronic molecules 
for generic cases will be surely more complicated, but they might be more similar to QCD since
the quarks have different masses in nature. Furthermore, we considered the model with two condensates.
It is the reason why we have two different spices, the u- and d-vortices. If we include three or more
condensates, we can consider hadrons consisting of more than two constituent vortices \cite{Eto:2012rc,Eto:2013spa}.
It would be
especially interesting to study 
molecules such as a proton like molecule uud and a
neutron like molecule udd to simulate situations closer to QCD.
It might be also interesting to take into account the vortical hadrons at finite temperature and/or density.
In QCD, there exist several phases such as quark gluon plasma and color superconducting phase. Exploring the 
phase diagram of the vortex matter in BEC 
would shed some light on the phase diagram of real QCD, see Ref.~\cite{Eto:2013hoa}  for vortices in color superconductors in QCD.

\ \\

{\bf Acknowledgments}\ \ 

This work was supported by the 
Ministry of Education, Culture, 
Sports, Science (MEXT)-Supported Program for the Strategic Research 
Foundation at Private Universities ``Topological  Science'' 
(Grant  No.~S1511006) 
and JSPS KAKENHI Grant Numbers 16H03984.
The work of M.~E.~ is also supported in part 
by JSPS Grant-in-Aid for Scientific Research 
KAKENHI Grant No. JP19K03839, and
by MEXT KAKENHI Grant-in-Aid for 
Scientific Research on Innovative Areas
``Discrete Geometric Analysis for Materials Design'' No. JP17H06462 from the MEXT of Japan.
The work of M.~N.~is also supported 
in part by JSPS KAKENHI Grant Number 18H01217 
by 
a Grant-in-Aid for Scientific Research on Innovative Areas 
``Topological Materials Science'' 
(KAKENHI Grant No.~15H05855) from MEXT of Japan.

\newcommand{\J}[4]{{\sl #1} {\bf #2} (#3) #4}
\newcommand{\andJ}[3]{{\bf #1} (#2) #3}
\newcommand{\AP}{Ann.\ Phys.\ (N.Y.)}
\newcommand{\MPL}{Mod.\ Phys.\ Lett.}
\newcommand{\NP}{Nucl.\ Phys.}
\newcommand{\PL}{Phys.\ Lett.}
\newcommand{\PR}{ Phys.\ Rev.}
\newcommand{\PRL}{Phys.\ Rev.\ Lett.}
\newcommand{\PTP}{Prog.\ Theor.\ Phys.}
\newcommand{\hep}[1]{{\tt hep-th/{#1}}}


\end{document}